\documentclass[12pt,hidelinks]{UoKthesis}

\usepackage{UoKextentions}

\usepackage[titletoc]{appendix}

\usepackage{tikz-qtree}
\usepackage{graphicx}
\usepackage{cite}
\graphicspath{ {figures/} }
\usepackage{multirow}
\usepackage{lscape}
\usepackage{tikz}
\usepackage{amsmath}
\usepackage{booktabs}
\usepackage{url}
\usepackage[vlined,ruled,linesnumbered,lined]{algorithm2e}
\usepackage{float}
\usepackage{multirow}
\usepackage{rotating}
\usepackage{scalerel}
\usepackage[flushleft]{threeparttable}
\usepackage{balance}
\usepackage{amssymb}
\usepackage{amsmath}
\usepackage{tikz}
\usepackage{verbatim}
\usepackage{pgfplots}
\usepackage{threeparttable}
\usepackage{enumitem}
\usepackage{pgfgantt}
\usepackage{longtable}
\usetikzlibrary{patterns}
\usepackage{alltt}
\usepackage{fnbreak} 
\usepackage{relsize}
\usepackage{balance}
\usepackage[all]{xy}
\usepackage{tabularx}

\usepackage{hyperref}
\usepackage{breakcites}

\DeclareMathOperator*{\argmax}{argmax}

\usepackage[center]{caption}

\usepgfplotslibrary{clickable}
\usepgfplotslibrary{fillbetween}
\pgfplotsset{legend style={line width=1pt}}
\usepackage{lscape}
\usepackage{bibentry}

\usetikzlibrary{automata,positioning}
\setlist[enumerate]{label*=\arabic*.}

\usetikzlibrary{plotmarks}

\usepackage[font={small}]{caption}
\usepackage[labelformat=parens]{subcaption}

\usepackage[format=plain,
                       labelfont=it,
                        textfont=it]{caption}

\begin{document}
\normalsize
\title{Robust Deep Learning Frameworks For Acoustic Scene and Respiratory Sound Classification}
\author{Lam Dang Pham}
\subject{Computer Science}
\degree{PhD}

\begin{preface}

\section{Acknowledgements}

Firstly, I acknowledge with thanks my supervisors, Professor Ian McLoughlin and Professor Palaniappan Ramaswamy for their invaluable advice and guidance in my research. 
I am very much thankful to Professor Ian McLoughlin for spending time on discussing ideas and revising papers, and I am specially grateful for his kind care and help in my personal life. \\ \\
Secondly, I would like to thank Doctor Huy Phan, School of Electronic Engineering and Computer Science, Queen Mary University of London, and Professor Alfred Mertins, Institute for Signal Processing, University of Luebeck for their invaluable advice and supporting devices used to conduct experiments.  \\ \\
Finally, I am also thankful for my colleagues who have made the School and Computing in Medway a friendly and comfortable working environment.

\section{Abstract}

Although research on Acoustic Scene Classification (ASC) is very close to, or even overshadowed by different popular research areas known as Automatic Speech Recognition (ASR), Speaker Recognition (SR) or Image Processing (IP), this field potentially opens up several distinct and meaningful application areas based on environment context detection.
The challenges of ASC mainly come from different noise resources, various sounds in real-world environments, occurring as single sounds, continuous sounds or overlapping sounds.
In comparison to speech, sound scenes are more challenging mainly due to their being unstructured in form and closely similar to noise in certain contexts. 
Although a wide range of publications have focused on ASC recently, they show task-specific ways that either explore certain aspects of an ASC system or are evaluated on limited acoustic scene datasets. 

Therefore, the aim of this thesis is to contribute to the development of a robust framework to be applied for ASC, evaluated on various recently published datasets, and to achieve competitive performance compared to the state-of-the-art systems. 
%
To do this, a baseline model is firstly introduced. 
Next, extensive experiments on the baseline are conducted to identify key factors affecting final classification accuracy.
From the comprehensive analysis, a robust deep learning framework, namely the \textit{Encoder-Decoder} structure, is proposed to address three main factors that directly affect an ASC system. These factors comprise low-level input features, high-level feature extraction methodologies, and architectures for final classification.
Within the proposed framework, three spectrogram transformations, namely Constant Q Transform (CQT), gammatone filter (Gamma), and log-mel, are used to convert recorded audio signals into spectrogram representations that resemble two-dimensional images.
These three spectrograms used are referred to as low-level input features.
To extract high-level features from spectrograms, a novel \textit{Encoder} architecture, based on Convolutional Neural Networks, is proposed. 
In terms of the \textit{Decoder}, also referred as to the final classifier, various models such as Random Forest Classifier, Deep Neural Network and Mixture of Experts, are evaluated and structured to obtain the best performance. 

To further improve an ASC system's performance, a scheme of two-level hierarchical classification, replacing the role of \textit{Decoder} classification recently mentioned, is proposed.
This scheme is useful to transform an ASC task over all categories into multiple ASC sub-tasks,  each spanning fewer categories, in a divide-and-conquer strategy.
At the highest level of the proposed scheme, meta-categories of acoustic scene sounds showing similar characteristics are classified.
Next, categories within each meta-category are classified at the second level. 
Furthermore, an analysis of loss functions applied to different classifiers is conducted.
This analysis indicates that a combination of entropy loss and triplet loss is useful to enhance performance, especially with tasks that comprise fewer categories.  

Further exploring ASC in terms of potential application to the health services, this thesis also explores the 2017 Internal Conference on Biomedical Health Informatics (ICBHI) benchmark dataset of lung sounds. 
A deep-learning framework, based on our novel ASC approaches, is proposed to classify anomaly cycles and predict respiratory diseases.
The results obtained from these experiments show exceptional performance. 
This highlights the potential applications of using advanced ASC frameworks for early detection of auditory signals. 
In this case, signs of respiratory diseases, which could potentially be highly useful in future in directing treatment and preventing their spread.

\end{preface}

\chapter{Introduction} 
\label{c01}

\section{Motivation and State-of-the-art Approaches}    
\label{c01_mot_approach}

\subsection{Motivation}
\label{c01_motivation}
The role of audio in communication applications has become essential to modern human life,  comparable to the role played by images or text. 
A historical development of audio processing systems began with the invention of the phonograph in the 1870s, and ran to current high-tech systems, as summarised by Ian McLoughlin in his textbook~\cite{bk_ian_speech}. This highlights the important, but often hidden, role of audio-related research over the years. 
In terms of audio signal detection, there are four major fields of research including Automatic Speech Recognition (ASR), Speaker Recognition (SR) and non-speech research into Acoustic Event Detection (AED) and Acoustic Scene Classification (ASC).
As regards the first two research areas, human languages are the focus, and impressive achievements have been demonstrated from these research areas.
For example, the Siri tool from Apple and the Alexa tool from Amazon are two famous applications of ASR, while SR has become popular in security systems that consider speech biometrics as one of the necessary security layers, especially in modern telephone banking applications. 
Regarding the two remaining research areas, these are considered to be newly (or recently) emerging fields over the past few years. One of the pioneers, Lyon \cite{bk_lyon_17_human}, termed this ``Machine Hearing", described in terms of how accurately a machine could listen to and understand sounds in a real-world environment compared to humans. 

In terms of acoustic scenes in ASC, they refer to environmental sound occurrences such as the sounds heard \textit{in an office, on a train, in a car} or \textit{in a forest}. 
Detecting surrounding environments could be described by the terms ``scene detection", ``context detection", or ``sound scene classification''. 
By detecting the current location, devices could obtain useful information to enable them to respond appropriately or adjust certain functions, opening up a wide range of distinct applications. 
For instance, \cite{ravi2005_ap01, xiang2010_ap01} show the contribution of scene detection for enhancing the listening experience of users. 
Moreover, scene classifiers can support sound event detection when these sound events are mixed in real-world environments \cite{toni2013_ap01}. 
There are also significant applications in robotics, where the function of scene detection integrated into a robotics system was an early proposal made by Clarkson \emph{et al.}~\cite{brian1998_ap01}, followed by El-Maleh \emph{et al.} with mobile applications \cite{maleh1999_ap01}. 
Recent success in developing a distributed sensor-server system for acoustic scene classification was described by Jakob Abeber \emph{et al.}~ \cite{jakob2017_acoustic}, which promisingly opens up the likelihood of further practical applications in the near future.  

Despite the great potential for enabling a variety of applications, compared to the mature fields of automatic speech recognition (ASR) or speaker recognition (SR), ASC-based applications are still in their infancy due to the presence of several challenges. 
In particular, analyses of audio recordings readily reveals that sound events and sound scenes always exit simultaneously in real-world environments. 
For example, \textit{bird song} is usually heard in a park, a \textit{car horn} is usually outdoors; but a \textit{barking dog} could be outside or inside a house. If the background and foreground are referred to as noise and signal respectively, it is a fact that the signal-to-noise ratio exhibits extremely high variability due to the diverse range of environments and recording conditions. 
To complicate matters further, a lengthy sound event could be considered background in certain contexts and foreground in others. 
For instance, a \textit{pedestrian street} recording may have a generally quiet background, but with short vehicle engine foreground events, as traffic passes. 
However, a lengthy engine sound in an recording \textit{on a bus} would be considered a background sound rather than a foreground event.
Furthermore, both background and foreground contain true noise -- continuous, periodic or aperiodic acoustic signals that interferes with the understanding of the scene. 
Besides, other challenges may come from the available datasets for studying this area.
In particular, some datasets lack sufficient recorded data, contain unbalanced data (i.e. a large number of recordings of some classes with other classes having very little data), or even high-cross correlation among sound categories.
Recently, the issue of mismatched recording devices, something which often occurs in practical applications  has been raised as a new challenge for the ASC task (i.e. different devices record data in different classes, or data from some devices are used for training, but different devices are used in practice). 

The variabilities and difficulties mentioned make acoustic scene classification (ASC) particularly challenging.
To deal with such challenges, recent ASC publications have tended to focus on two main aspects of machine hearing, which are discussed in the following subsections.

\subsection{State-of-the-art Approaches}
The first aspect of machine hearing addressed by most state-of-the-art systems aims to solve the lack of discriminative information by exploiting various methods of low-level feature extraction.
In particular, it is notable that early publications used Mel Frequency Cepstral Coefficients (MFCC) parameters~\cite{bk_ian_speech} or combined MFCCs with temporal characteristics of audio sound such as loudness, average short-time energy, zero-crossing rate, spectral flux, or spectral centroid~\cite{dc_16_erik_dcase, dc_16_t04, dc_17_vaf_dcase_tp, dc_16_nico_dcase}.
More recently, ASC research has exploited spectrogram representations, and has made efforts to explore information from different recording channels~\cite{dc_18_oct_dcase, dc_18_yuma_tp}, different kinds of spectrogram~\cite{dc_18_tb01, dc_18_truc_icme, dc_18_hos_dcase, huy_lit_jr}, and different time resolutions of spectrogram input~\cite{dc_18_yuma_tp}.
By using multiple input features, recent ASC systems apply various ensemble methods such as majority voting, sum or product fusion~\cite{huy_lit_aes, dc_18_lam_aes} to fuse results obtained from discrete models. 
Although using multi-input features, combined with ensemble models, helps to achieve high performance, none of the publications has provided a comprehensive analysis of how to select optimum input features for different tasks.
Furthermore, ensemble models exhibit a high cost of computation, which may include significant redundancy. Due to the computation cost, almost all state-of-the-art systems were evaluated on limited size datasets. 

The second research trend found in state-of-the-art publications focuses on constructing and training powerful learning models, with the aim of obtaining high-performing high-level features.
For example, Lidy and Schindler~\cite{dc_16_lidy_dcase} proposed two parallel CNN-based models with different kernel sizes to learn from a CQT spectrogram input, capturing different regions of the spectrogram. 
Focusing on pooling layers, where high-level features are condensed, Zhao \emph{et al.}~\cite{dc_18_zhao_dcase, dc_18_zhao_ica} proposed an attention pooling layer that showed an effective improvement compared to conventional max or mean pooling layers.  
With the inspiration that different frequency bands in a spectrogram contain distinct features, Phaye \emph{et al.}~\cite{dc_18_phaye_ica} proposed a \textit{SubSpectralNet} network which is able to extract discriminative information from $30$ sub-spectrograms.
These examples of ASC systems recently mentioned all involve an end-to-end training process.
In such systems, values of the second to last layer are referred to as high-level features (or sometimes as embeddings), while the final layer, normally implemented as a softmax, performs the final classification. 
Another approach applies two different models which are trained separately.
While the first model is used to extract high-level features, the second model takes the role of the final classifier.

%
It can be seen that recent works have made efforts to improve the overall performance of ASC systems with more and more complex approaches.
However,  they mainly focus on specific aspects of an ASC system, and there are various issues that have not been deeply considered or directly addressed.
Firstly, although multiple low-level features have been shown to be effective at enhancing the performance of systems, no recent publication has addressed which low-level factors have the greatest influence on overall performance.
Furthermore, ensembles of multiple input features face a direct trade-off between system performance and extensive computation cost, essentially throwing computing power at the problem. 
This unfortunately makes the approaches incompatible with many real-time applications or platforms which are constrained in terms of computing power or applications which are constrained in terms of latency. 
A question arises as to whether there is an effective way to combine the most important low-level features, thus solve both the lack of refined input information and simultaneously avoid high computation costs.
Secondly, although systems using complicated deep-learning networks show effectiveness in extracting good high-level features, none of the publications investigates the role of the final classifier in exploring those high-level features. Most simply present a complete architecture without further experimentation or analysis.
It can be seen that these issues mentioned above may be grouped into three main topics of low-level feature input, high-level feature extraction, and output classification -- each of which affects ASC system performance in different ways.
Almost all state-of-the-art ASC systems are chosen or optimised in a task- specific way (i.e. they perform well for one task, but are not evaluated for others), and no consensus has emerged regarding an optimum choice for any of the three factors.
Finally, while the ability to perform early detection (i.e. low latency classification) is very useful for real-world applications such as human-robot interaction, noise reduction during calls or in  acoustic security systems, very few publications explore or even mention this~\cite{huy_early_2015, huy_early_2018, ivm_early_2018}.
These several factors motivate this research to develop an ASC system that targets the three most important issues mentioned above while providing a comprehensive analysis on the ability of early detection and exploration of the main factors involved in ensuring good performance.

\section{Contribution}
\label{c01_contribution}

During my PhD research in this field, I have made contributions in the five following areas;

\textbf{A comprehensive analysis of low-level features in ASC.} 
I have explored a variety of low-level features to tackle the lack of input information in ASC research. 
To understand how low-level features affect the classification result in an ASC system, this thesis firstly proposes a baseline system which is used to provide a comprehensive analysis.
The baseline uses spectrogram representations as low-level feature input and employs a C-DNN-based architecture (defined later in Chapter 3) for classification.
By using the baseline described, various low-level feature settings such as channel information, spectrogram type, time resolution, and data augmentation are evaluated.
This analysis helps to identify the most important low-level features and how they affect the final classification accuracy.
Part of this contribution was published in the Audio Engineering Society (AES) 2019 Conference ~\cite{dc_18_lam_aes} and the 20th INTERSPEECH 2019 Conference~\cite{dc_16_lam_int}. 

\textbf{A novel \textit{Encoder-Decoder} framework for Acoustic Scene Classification.}
It is a fact that condensed and discriminative high-level features directly affect the final classification accuracy in an ASC system.
Therefore, successfully developing a high-performing extractor is one of the most important aspects of building an effective ASC system.
This thesis contributes to ASC research by introducing a novel and robust architecture which we denote the \textit{Decoder-Encoder} framework. This enables a system to learn multiple low-level input features, from which the framework generates high-performing high-level features (specifically, this is the role of the \textit{Encoder}).
Furthermore, it then provides an analysis of various final classifiers models (the \textit{Decoder} function). 
This contribution was published in Journal of Digital Signal Processing~\cite{dc_all_lam_journal}.

\textbf{The ability of early detecting recording environments.}
Although early detection of recording environments promisingly opens a wide range of applications as introduced in Section \ref{c01_motivation}, a few of research~\cite{ivm_early_2018, huy_early_2018} mentioned and not many experiments have been conducted. 
This contribution shows an analysis of early detecting recording environments by using the novel \textit{Decoder-Encoder} framework recently mentioned.
This contribution was published in Journal of Digital Signal Processing~\cite{dc_all_lam_journal}.

\textbf{Two-level hierarchical classification scheme as an effective encoder in an ASC system.}
Further investigation to improve the final classifiers, referred to as \textit{decoders} mentioned under the description of the \textit{Decoder-Encoder} framework. This research contribution is to present a novel scheme of two-level hierarchical classification.
By exploiting cross-relation among environmental categories and using the triplet loss function for training, the scheme is able to enhance system performance.
This contribution was published in the International Joint Conference on Neural Networks (IJCNN) 2020~\cite{dc_18_lam_ijcnn}. 

\textbf{Application of the above for early prediction of respiratory disease.}
This thesis has already claimed that an effective ASC can enable important future applications, and in this work one such important application is explored. Up to this point, all evaluations have been done using public databases of audio clips that are presented as part of worldwide ASC challenges, particularly as part of Detection and Classification of Acoustic Scenes and Events (DCASE). These have real-world aspects, but are essentially artificially constructed datasets collected under controlled conditions.
This contribution now considers an important real-time application -- specifically the detection of respiratory diseases from lung sounds. This follows the 2017 Internal Conference on Biomedical Health Informatics (ICBHI) lung sound dataset and challenge.
The results obtained show the potential to apply the deep-learning frameworks developed in this research (and described in the early chapters of this thesis) to create advanced computational techniques for early detection of respiratory diseases. Furthermore, for these frameworks to be compatible with real time portable or wearable computational devices.
This contribution is published in the 42nd Annual International Conferences of the IEEE Engineering in Medicine and Biology Society~\cite{icbhi_lam_embc} and being considered for publication in IEEE Journal of Biomedical and Health Informatics~\cite{icbhi_lam_journal}, the 43th Annual International Conferences of the IEEE Engineering in Medicine and Biology Society~\cite{icbhi_lam_embc_02}

\section{Published and Preprint Papers}
\label{c01_publish}

This section shows published and preprint papers that are relevant to and contribute into the thesis.

\textbf{-First-author papers:}
\begin{enumerate}

\item \textbf{L. Pham}, I. McLoughlin, H. Phan, R. Palaniappan, and Y. Lang, “Bag-of-features models based on C-DNN network for acoustic scene classification”, in Audio Engineering Society Conference: 2019 AES International Conference on Audio Forensic (AES), 2019~\cite{dc_18_lam_aes}.

\item \textbf{L. Pham}, I. McLoughlin, H. Phan, and R. Palaniappan, “A robust framework for acoustic scene classification”, in Proc. Annual Conference of the International Speech Communication Association (INTERSPEECH), pp. 3634–3638, 2019~\cite{dc_16_lam_int}.

\item \textbf{L. Pham}, H. Phan, T. Nguyen, R. Palaniappan, A. Mertins, and I. McLoughlin, “Robust acoustic scene classification using a multi-spectrogram encoder-decoder framework”, Digital Signal Processing, vol. 110, 2021~\cite{dc_all_lam_journal}.

\item \textbf{L. Pham}, I. McLoughlin, H. Phan, R. Palaniappan, and A. Mertins, “Deep feature embed- ding and hierarchical classification for audio scene classification”, in Proc. International Joint Conference on Neural Networks (IJCNN), pp. 1-7, 2020~\cite{dc_18_lam_ijcnn}.

\item  \textbf{L. Pham}, I. McLoughlin, H. Phan, M. Tran, T. Nguyen, and R. Palaniappan, “Robust deep learning framework for predicting respiratory anomalies and diseases”, in Proc. 42nd Annual International Conferences of the IEEE Engineering in Medicine and Biology Society (EMBC), pp. 164-167, 2020~\cite{icbhi_lam_embc}.

\item \textbf{L. Pham}, H. Phan, A. Schindler, R. King, A. Mertins, and I. McLoughlin, “Inception- based network and multi-spectrogram ensemble applied for predicting res- piratory anomalies and lung diseases”, in Proc. 43nd Annual International Conferences of the IEEE Engineering in Medicine and Biology Society (EMBC), 2021~\cite{icbhi_lam_embc_02}.

\item \textbf{L. Pham}, H. Phan, R. Palaniappan, A. Mertins, and I. McLoughlin, “Cnn-moe based framework for classification of respiratory anomalies and lung disease detection”, IEEE Journal of Biomedical and Health Informatics, 2021~\cite{icbhi_lam_journal}.

\item \textbf{L. Pham}, A. Schindler, M. Schütz, J. Lampert, S. Schlarb, R. King “Deep Learning Frameworks Applied For Audio-Visual Scene Classification”, arXiv preprint arXiv:2106.06840, 2021~\cite{pham_dc_2021_1B}.

\item \textbf{L. Pham}, H. Tang, A. Jalali, A. Schindler, R. King “A Low-Compexity Deep Learning Framework For Acoustic Scene Classification”, arXiv preprint arXiv:2106.06838, 2021~\cite{pham_dc_2021_1A}.

\item \textbf{L. Pham}, C. Baume, Q. Kong, T. Hussain, W. Wang, M. Plumbley “An Audio-Based Deep Learning Framework For BBC Television Programme Classification”, in Proc. European Signal Processing Conference (EUSIPCO), 2021~\cite{pham_bbc}.

\end{enumerate}

\textbf{-Co-author papers related to the thesis:}

\begin{enumerate}

\item I. McLoughlin, Y. Song,  \textbf{L. Pham}, H. Pham, P. Ramaswamy, and L. Yue, “Early detection of continuous and partial audio events using CNN,” in Proc. Annual Conference of the International Speech Communication Association (INTERSPEECH), pp. 3314–3318, 2018~\cite{ivm_early_2018}.

\item  H. Phan, O. Y. Chen, P. Koch, \textbf{L. Pham}, I. Mcloughlin, A. Mertins, and M. D. Vos, “Unifying isolated and overlapping audio event detection with multi-label multi-task convolutional recurrent neural networks,” in Proc. IEEE International Conference on Acoustics, Speech and Signal Processing (ICASSP), pp. 51-55, 2019~\cite{huy_lit_aes}.

\item H. Phan, O. Y. Chen, P. Koch,  \textbf{L. Pham}, I. McLoughlin, A. Mertins, and M. De Vos, “Beyond equal-length snippets: How long is sufficient to recognize an audio scene?,” in Audio Engineering Society Conference: 2019 AES International Conference on Audio Forensic (AES), Jun 2019~\cite{huy_lit_aes}.

\item H. Phan, O. Y. Chen, \textbf{L. Pham}, P. Koch, M. De Vos, I. Mcloughlin, and A. Mertins, “Spatio-temporal attention pooling for audio scene classification,” in Proc. Annual Conference of the International Speech Communication Association (INTERSPEECH), pp. 3845–3849, 2019~\cite{huy_lit_int_02}.

\item D. Ngo, H. Hoang, A. Nguyen, T. Ly, and \textbf{L. Pham}, ``Sound context classification basing on join learning model and multi-spectrogram features'' arXiv preprint arXiv:2005.12779, 2020~\cite{dat_jour}.

\item  H. Phan, \textbf{L. Pham}, P. Koch, N. Duong, I. Mcloughlin, and A. Mertins, “On multitask loss function for audio event detection and localization,” in Proc. Detection and Classification of Acoustic Scenes and Events (DCASE) Workshop, pp. 160-164, 2020~\cite{huy_coau}.

\item  H. Phan, HL. Nguyen, OY. Chen, \textbf{L. Pham}, P. Koch, I. Mcloughlin, and A. Mertins, “Multi-view Audio and Music Classification”, in Proc. IEEE International Conference on Acoustics, Speech and Signal Processing (ICASSP), pp. 611-615, 2021~\cite{huy_coau_00}.

\end{enumerate}

\section{Organisation of This Thesis}
\label{c01_organisation}
These remaining chapters of this thesis are organised as follows. 

\textbf{Chapter 2} carries out a comprehensive literature review of ASC research, which covers state-of-the-art systems and recently published datasets.
From the detailed analyses, open issues related to the ASC task are raised and discussed. 

\textbf{Chapter 3} presents a baseline system applied for ASC. 
Using the baseline, the effect of different low-level features (and their settings) on classification accuracy is analysed, thus identifying low-level features which are able to perform well for various scenarios. When the most influencing low-level features are indicated, mixup data augmentation applied on these features is also evaluated (The first-author paper 1 mentioned in Section \ref{c01_publish} mainly contributes into this chapter).

Based on the comprehensive analysis of low-level features provided in Chapter 3, \textbf{Chapter 4} develops a novel \textit{Encoder-Decoder} framework applied for ASC. 
In particular, this chapter presents a novel \textit{Encoder} architecture that helps to learn multiple spectrograms simultaneously, thus extracts and combines high-performing high-level features. 
Furthermore, these high-level features extracted are explored by various \textit{Decoder} models, reports the final classification accuracy. 
The results obtained prove the proposed \textit{Encoder-Decoder} framework to be robust and general for ASC (The first-author papers 2 and 3 mentioned in Section \ref{c01_publish} mainly contribute into this chapter). 

\textbf{Chapter 5} proposes a scheme of two-level hierarchical classification.
The scheme is used to train and explore high-level features extracted from an \textit{Encoder} architecture mentioned in Chapter 3. 
The results obtained in this chapter indicate that the combination of the proposed scheme and a triplet loss function during training are useful to exploit the cross-correlation between environmental categories, which helps to improve accuracy (The first-author paper 4 mentioned in Section \ref{c01_publish} mainly contributes into this chapter). 

\textbf{Chapter 6} further explores ASC, but this time in the context of investigating a specific application of respiratory diseases detection.  
The extremely good results obtained from this system indicate the great potential for applying such deep-learning frameworks to  not only the early detection of lung-related diseases, but also to similar application areas (The first-author papers 5, 6 and 7 mentioned in Section \ref{c01_publish} mainly contribute into this chapter)..  

\textbf{Chapter 7} presents conclusion and future works.

\textbf{Appendix} where computation of spectrograms and network layers are described in detail.

\chapter{Literature Review}
\label{c02}

This chapter first defines the ASC task, then introduces some acoustic scene datasets which are popular in recent research literature.
Next, it analyses acoustic scene representations and classification algorithms.
From this analysis, this chapter then continues to identify issues with current ASC research --  which in turn for the main motivation behind this thesis.

\section{ASC Definition}
\label{c02_define}

\begin{figure}[thb]
    \centering
    \includegraphics[width=0.6\linewidth]{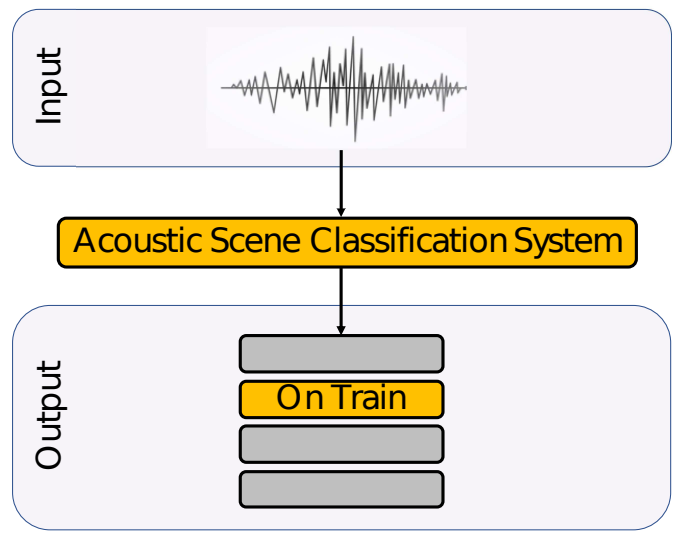}
    	\vspace{0.1cm}
	\caption{\textit{Task definition of Acoustic Scene Classification.}}
    \label{fig:c02_task_def}
\end{figure}

\begin{figure}[t]
    \centering
    \includegraphics[width=0.85\linewidth]{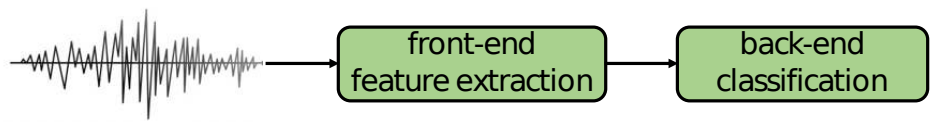}
    	\vspace{-0.2cm}
	\caption{\textit{A general system applied for Acoustic Scene Classification.}}
    \label{fig:c02_gen_sys}
\end{figure}

The ASC task aims to classify a recording into one or more predefined categories that characterise the environment in which it was recorded.
For example, a recording is classified into \textit{in caffe, on bus, in office} or \textit{on train}, as shown in Figure \ref{fig:c02_task_def}. 
%
%
A general system structure for performing ASC is described as Figure \ref{fig:c02_gen_sys}, showing a waveform analysed in two main steps; front-end feature extraction and back-end classification, respectively.
The purpose of the first step, front-end feature extraction, is to transform a segment of recorded audio into another form that contains compact information and is suitable for the subsequent stage of classification. 
A high-performed transformation generates well-presented features that benefit the back-end classifier.  
By using features extracted during the front-end extraction step, the back-end classifier aims to classify an environmental recording into certain predefined categories.

\begin{figure}[t]
    \centering
    \includegraphics[width=0.7\linewidth]{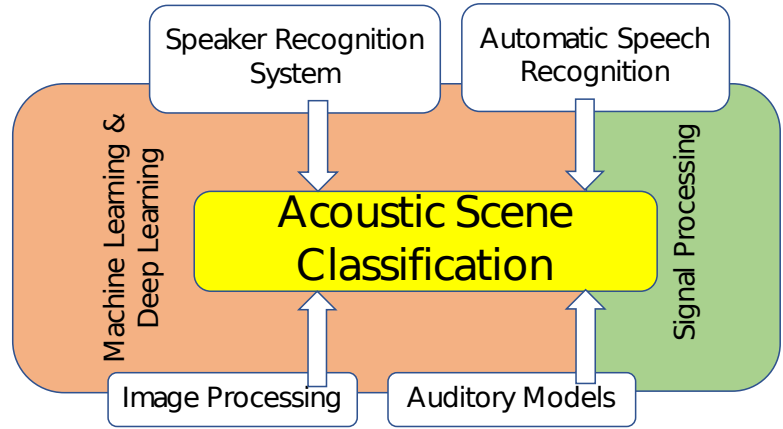}
    	\vspace{-0.1cm}
	\caption{\textit{Acoustic Scene Classification and relationship to overlapping research areas.}}
    \label{fig:c02_asc_tech}
\end{figure}

Techniques, which are generally applied to ASC systems, are diverse and have often been borrowed from different related research fields as shown in Figure \ref{fig:c02_asc_tech}. 
They focus on developing effective feature extraction techniques and robust models for the classification. 
%
%
\section{ASC Datasets}
\label{c02_datasets}
\begin{table}[h]
    \caption{\textit{The main Acoustic Scene Classification challenge datasets.}} 
     \vspace{-0.2cm}
    \centering
   \scalebox{0.8}{
    \begin{tabular}{ |l | c |c |c| c|}
        \hline
	      \textbf{ASC Dataset}                    &\textbf{Published In}          & \textbf{Classes}   &\textbf{Time Recorded} &\textbf{Segment Lenght}\\
	       \textbf{       (Name)}                                                              &\textbf{(Year)}                    & \textbf{(No.)}   &\textbf{(Hours)}  & \textbf{(Second)} \\

        \hline 
	      DCASE 2019 Task 1A~\cite{data_dc_19}    &2019          &10              &40.00             &10                 \\ 
              DCASE 2019 Task 1B~\cite{data_dc_19}    &2019          &10              &46.00             &10                 \\ 
              DCASE 2019 Task 1C~\cite{data_dc_19}    &2019          &10              &44.00             &10                 \\ 
              DCASE 2018 Task 1A~\cite{data_dc_18}    &2018          &10              &24.00             &10                 \\ 
              DCASE 2018 Task 1B~\cite{data_dc_18}    &2018          &10              &28.00             &10                 \\ 
              DCASE 2017 Task 1 ~\cite{data_dc_17}    &2017          &15              &17.50             &10                 \\ 
              DCASE 2016 Task 1 ~\cite{data_dc_16}    &2016          &15              &13.00             &30                 \\ 
	      AucoDer07~\cite{data_auco}              &2015          &4               &4.20              &not fixed          \\ 
              Litis-Rouen~\cite{data_litis}           &2014          &19              &25.51             &30                 \\ 
              DCASE 2013~\cite{data_dc_13}            &2013          &10              &0.83              &30                 \\ 
	      CASA 2010~\cite{data_casa}              &2010          &13              &8.88              &4                  \\ 
	      CASA 2009~\cite{data_casa}              &2009          &10              &18.88             &4                  \\ 
	      UEA-Series2~\cite{data_uea}             &2006          &10              &2.92              &not fixed          \\ 
	      UEA-Series1~\cite{data_uea}             &2006          &10              &0.66              &not fixed          \\ 
         \hline
    \end{tabular}
    }
    \label{table:c02_dataset} 
\end{table}
Table \ref{table:c02_dataset} lists the most prominent acoustic scene datasets which have been published as part of international challenges.
These datasets were recorded in real environments and released alongside the necessary meta information in Wave format (i.e. .wav files).  
The Litis-Rouen dataset~\cite{data_litis} shows the highest number of separate classes at 19, followed by DCASE 2016~\cite{data_dc_16} and DCASE 2017~\cite{data_dc_17} with 15 classes and CASA 2010~\cite{data_casa} with 13 classes.
The remaining dataset challenges have ten different environments, with the exception of AucoDer07~\cite{data_auco} which only comprises 4 separate classes.
Over time, as this research field has progressed, the recorded duration has increased from 0.66 hours for UAE-Series1~\cite{data_uea} to 46 hours for the recordings in DCASE 2019~\cite{data_dc_19}.
Recently, the IEEE AASP Challenge on Detection and Classification of Acoustic Scenes and Events (DCASE) has provided a diverse set of acoustic scene datasets,  motivates a lot of publications that have been evaluated with these datasets.
In this thesis, the analysed systems are mainly evaluated over Litis-Rouen~\cite{data_litis}  and DCASE 2016 Task 1~\cite{data_dc_16}, DCASE 2017 Task 1~\cite{data_dc_17}, DCASE 2018 Task 1A \& 1B~\cite{data_dc_18} and DCASE 2019~\cite{data_dc_19} Task 1A \& 1B datasets.
These datasets are independently evaluated (i.e. each dataset is separated into Train. and Eva. subsets for training and evaluating, respectively).

\section{Evaluation Metric}
\label{c02_metric}

As this thesis evaluates ASC datasets of Litis Rouen and EEE AASP Challenge on Detection and Classification of Acoustic Scenes and Events (DCASE) in years of 2016, 2017, 2018 and 2019, the evaluation metric of accuracy used in this thesis follows these challenges. 
In particular, if $C$ is considered as the number of audio segments which are correctly predicted, and the total number of audio segments is $T$, the classification accuracy (Acc.\%) mentioned in these challenges shares the similar computation as (note that the segment length evaluated depends on specific datasets),
\begin{equation}
    \label{eq:c03_mean_stratergy_patch}
    Acc. (\%) = 100\frac{C}{T}.
\end{equation}
As these datasets are slightly unbalanced and experimental results across categories are shown, other metrics such as Recall, Precision, or F1 score are not presented for ASC tasks in this thesis.

\section{ Acoustic Scene Representation}
\label{c02_representation}

According to the basic stages in a typical ASC system presented in Section \ref{c02_define}, the state-of-the-art systems applied to ASC use two main approaches for front-end feature extraction, namely one-dimensional frame-base measures or two-dimensional spectrogram representations respectively.
As there are various transformation used for generating spectrograms, mathematical definitions of transofrmation methods are described in detail in Section~\ref{c08} separately.  
The outputs from the front-end feature extraction are referred to as low-level features, and these are analysed in the following sections.  

\subsection{Frame-based Representations}
\label{c02_framebase}

Frame-based representations often utilise MFCC~\cite{bk_ian_speech}, and provide powerful feature extraction capabilities which are borrowed from the ASR community~\cite{data_dc_16}. 
To improve discrimination between MFCC frames, MFCCs are often combined with a wide range of temporal features such as loudness, probability of voicing, average short-time energy, sub-band energy, zero-crossing rate, spectral flux, or spectral centroid~\cite{dc_16_erik_dcase, dc_16_t04, dc_17_vaf_dcase_tp, dc_16_nico_dcase} (note that an MFCC frame is represented as a real valued vector, with the temporal features concatenated to that vector, effectively increasing the feature dimension for each frame).

Some systems first transform audio signals into MFCC spectrograms or log-mel spectrograms, then attempt to learn different aspects of those spectrograms to extract frame-based features. 
For instance, Nico \emph{et al.}~\cite{dc_16_nico_dcase} applied an Amplitude Modulation Filter Bank (AMFB) method to analyse and extract features from MFCC spectrogram before concatenating other temporal features such as flux, centroid spectral entropy.
Meanwhile, MultiScale-Kernel Fisher Discriminant Analysis (MSKFDA), coming from the emotion recognition field, was used by Erik~\cite{dc_16_erik_dcase} to provide multi-scale analysis over acoustic scene factors (combinations of  MFCCs and temporal features).
From log-mel spectrogram, Alain \emph{et al.}~\cite{data_litis} applied Non-negative Matrix Factorisation (NMF) techniques to extract condensed features. 
I-vector extraction, a powerful technique widely used in the SR research community~\cite{ivector_speaker}, has also recently been applied for ASC. 
Recent publications include various methods to extract i-vectors from MFCC spectrograms by using a Universal Background Model (UBM) model~\cite{dc_16_t01, dc_17_t05, dc_17_t09} or Gaussian Mixture Model (GMM)~\cite{dc_16_t06}. 
In an ASC system proposed by Abidin \emph{et al.}~\cite{dc_16_abi_jr}, frame-based features were extracted after many steps, via a complicated extractor. 
Firstly, auditory signals were transformed into a CQT spectrogram. 
Then, authors used Local Binary Pattern (LBP) techniques, borrowed from image texture extraction research, to extract a Time-Frequency Representation (TFR) from the CQT spectrogram.
The TFR was continuously solved by two different image process techniques.
The first method, using Histogram of Oriented Gradients (HOG), extracted HOG features from the TFR.
The second method, based on Local Binary Patterns (LBP), extracted histogram features located at linear zones of the TFR.
Eventually, two frame-based features, HOG and LBP, were concatenated before being fed into a Support Vector machine (SVM) model for classification.

Operating directly with audio signals, Song \emph{et al.}~\cite{dc_16_song_int} applied the auditory statistics of a cochlear filter model to extract discriminative features, operating without any spectrogram transformation step in their proposed system. 

\subsection{Spectrogram Representations}
\label{c02_spec}

Spectrogram images have higher resolution and contain richer information, in terms of both temporal and frequency dimensions, than general frame-based approaches. This thesis therefore explores a variety of spectrograms including short term Fourier transform (STFT), log-mel~\cite{dc_18_tb01, dc_18_zhao_ica, dc_18_phaye_ica}, MFCC~\cite{data_dc_17}, CQT~\cite{dc_16_lidy_dcase}, Gamma~\cite{huy_lit_jr, huy_lit_int} and scalogram~\cite{dc_17_zhao_jr}.

Further exploring spectrograms, publications applied various filters or image processing techniques for improving  spectrogram quality.
For instance, Truc \emph{et al.}~\cite{dc_18_tb01} applied a Nearest Neighbour Filter (NNF) on a log-mel spectrogram to generate a new NFF spectrogram.
By using a median-filtering harmonic percussive source separation over a log-mel spectrograms, Octave Mariotti \emph{et al.}~\cite{dc_18_oct_dcase} and Yuma \emph{et al.}~\cite{dc_18_yuma_tp} generated two spectrograms each of which focuses on either the time or the frequency resolution.
Yang \emph{et al.}~\cite{dc_17_kl_ica} used the Kullback-Leibler (KL) divergence scale to develop a KL filter bank.
Next they applied these filters on a log-mel spectrogram, generating a KL spectrogram that experimentally outperformed log-mel and CQT.
Waldekar and Saha~\cite{dc_16_wal_int} firstly generated a log-mel spectrogram. 
Then they applied a Haar wavelet on the log-mel spectrogram to generate new features named Mel-Frequency Discrete Wavelet Coefficients (MFDWC).

Combining image texture techniques known as Difference of Gaussians (DoG) and the Sobel edge detection operator~\cite{wu_sobel}, Wu \emph{et al.}~\cite{dc_17_wu_ica} applied these techniques to log-mel  spectrograms in order to enhance sound textures. 
Similarly, Park \emph{et al.}~\cite{dc_17_t06} extracted temporal energy density and energy variations for each frequency bin of a Gamma spectrogram by using a covariance matrix (COV) and double Fast Fourier Transform (FFT) image.

\subsection{Multiple Low-level Input Features and Data Augmentation to Address ASC Challenges} 
\label{c02_augmentation}
To deal with the ASC challenges mentioned in Section \ref{c01_motivation}, publications adopt two main approaches in terms of exploring low-level features.
The first approach considers that each low-level feature may capture distinct features of an audio signal.
Therefore, if multiple input features are used, it is effective at improving system performance.
Meanwhile, the second approach considers the use of data augmentation to tackle the issues related to datasets, such as lack of, or unbalanced nature of the data.

\textbf{Multiple low-level input features:}
For frame-base representations, MFCCs are often combined with temporal features as mentioned in Section \ref{c02_framebase}, or even with a variety of features such as perceptual linear prediction (PLP) coefficients, power nomalised cepstral coefficients (PNCC), robust compressive gamma-chirp filter-bank cepstral coefficients (RCGCC) or subspace projection cepstral coefficients (SPPCC)~\cite{dc_16_t03} that helps to achieve top-three system proposed in DCASE 2016 challenge.
For spectrogram representations, published papers show a diverse combination of log-mel and different types of spectrogram such as Mel-based Nearest Neighbour Filter (NNF) spectrogram~\cite{dc_18_tb01, dc_18_truc_icme}, CQT~\cite{dc_18_hos_dcase}, or MFCC and Gamma~\cite{huy_lit_jr}.
Testing a wavelet-transform derived spectrogram representation, Ren \emph{et al.}~\cite{dc_17_zhao_jr} compared results from STFT spectrograms and both \textit{Bump} and \textit{Morse} scalograms. 
They indicated that combination of STFT spectrogram and \textit{Bump} scalogram is useful to enhance the proposed ASC system. 

By exploiting channel information, Yuma \emph{et al.}~\cite{dc_18_yuma_tp} generated multi-spectrogram input from two channels, the average and difference of two channels, and explored separated harmonic and percussive spectrograms from each channel. 
By fusing results from channel information, the authors achieved the top-one score in DCASE 2018 challenge.
Some papers proposed combining both frame-based and spectrogram features such as i-vectors with an MFCC spectrogram in~\cite{dc_16_t01, dc_17_t05, dc_18_t09}.

\textbf{Data augmentation:}
To deal with the challenges causing by unbalanced classes within a dataset or the lack of representative data, some publications have proposed a variety of data augmentation methods. These can improve the robustness and enhance the learning ability of deep network models.
Early data augmentation methods combined the signals with multiple lengths of recorded audio~\cite{aug_t_length}.
This idea was improved by Salamon and Bello~\cite{aug_features} who provided an analysis of various data augmentation methods, including pitch shifting, time stretching, and the addition of background noise.
The research indicated that pitch shifting is useful for all types of experimental sound and a combination of all augmentation methods helps to improve the ASC system proposed.
Interestingly, Zang \emph{et al.}~\cite{lit_zang_int_02} proposed a sequence augmentation method.
Firstly, an audio signal was transformed into a STFT spectrogram. 
Next, a certain number of continuous STFT frames, referred to as the segment length $L$, were grouped as segments.
These segments thus were shuffled, re-arranged at different positions, and eventually were concatenated and generate a new sequence of STFF frames.  
By this way, the authors improved their proposed ASC system by 2\% and show that the proposed ASC system achieves the best performances with the segment length set to $L=64$.

Recently, mixing input data (mixup)~\cite{aug_mixup_i, aug_mixup_s01, aug_mixup_s02} and the application of Generative Adversarial Network (GAN) for data augmentation~\cite{dc_17_t01, dc_19_t01, dc_17_kl_ica} have become popular, and are shown to be effective.
The mixup method is easy to implement, which makes it popular in ASC and other systems.
Indeed, the top-eight highest performance systems for DCASE 2018 Task 1A challenge used this method, it was also used by almost all submitted systems for the DCASE 2019 Task 1A.

Using GAN to generate more fake data has been similarly shown to be effective in helping to improve system performance.
This is proved by systems in~\cite{dc_17_t01, dc_19_t01, dc_17_kl_ica}, achieving the highest scores in DCASE 2017 and 2019 challenges. 
However, systems using GAN for data augmentation~\cite{dc_17_t01, dc_19_t01, dc_17_kl_ica}  show competitive performance but are very complicated.
In particular, these systems need to configure and train a GAN network to be used to generate new data.
After training the GAN generator, they require a classified model to be trained as a filter to be able to select generated data which shows an appropriate distribution (SVM can be used).
Both old data and generated data need to then be shuffled, or interspersed, before being fed into a final classifier.

\section{Classification Algorithms}
\label{c02_classification}
It can be seen that the front-end feature extraction methods tend to fall under one of two main approaches, either frame-based or spectrogram representations.
Meanwhile the back-end classification methods are also divided into two main groups, analysed in some depth below.

\subsection{Machine Learning Models}
\label{c02_machinelearning}

\begin{figure}[!h]
    \centering
    \includegraphics[width=1.0\linewidth]{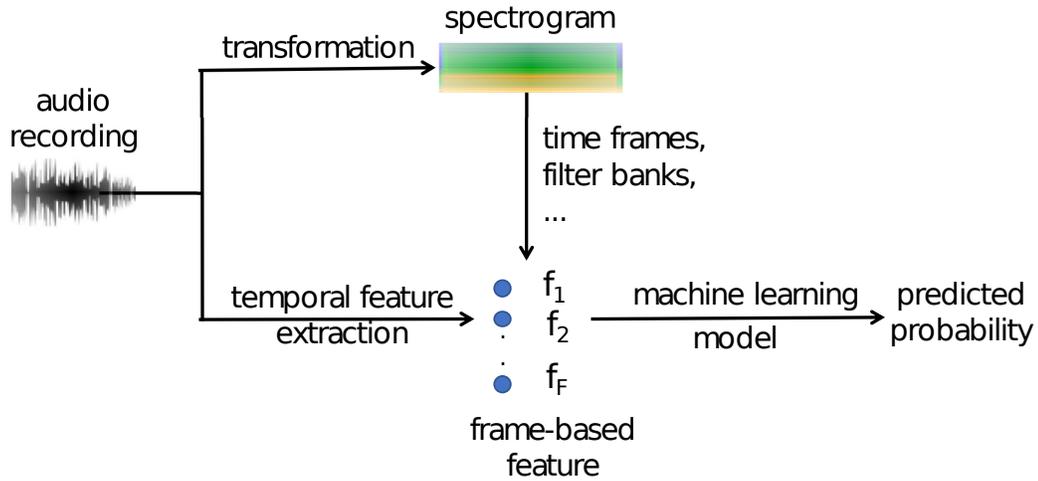}
    	\vspace{-0.7cm}
	\caption{\textit{ASC framework using frame-based feature and machine learning model}}
    \label{fig:c02_ml}
\end{figure}

The frame-based feature approaches are normally combined with traditional machine learning models as shown in Figure \ref{fig:c02_ml} and Table \ref{table:c03_sta_ml}.
For example, baseline of the DCASE 2016 challenge~\cite{data_dc_16} introduced MFCC feature extraction and Gaussian Mixture Model (GMM), which showed a very similar architecture to systems used in ASR research.
Similarly, Park \emph{et al.}~\cite{dc_16_t03} applied GMM models to evaluate various frame-based features such as MFCC, PLP, PNCC, RCGCC and SPCC.
Meanwhile, Support Vector Machine (SVM) was widely used with diverse types of frame-based input features such as MFCC~\cite{dc_16_t06, dc_16_ngoc_dcase}, auditory-summary-statistics features~\cite{dc_16_song_int}, HOG and LBP features in~\cite{dc_16_abi_jr}, MFDWC features in~\cite{dc_16_wal_int}.
Linear-based models have also been used, with Bisot \emph{et al.}~\cite{dc_16_t02} proposing a modified version of supervised dictionary model (TDL) to classify NMF features.
Meanwhile, Hamidn \emph{et al.}~\cite{dc_16_t01} used both Linear Discriminant Analysis (LDA)~\cite{lda_lin_model} and Within-Class Covariance Normalization (WCCN)~\cite{wccn_lin_model} to train i-vector features.
Recently, Multilayer-Perceptron-based (MLP-based) networks have been very widely used to train frame-base features.
For instance, MLP-based networks in~\cite{dc_16_ngoc_dcase, dc_16_xu_dcase},  ~\cite{dc_16_kong_dcase}, ~\cite{dc_17_t09} were used to learn MFCC, log-mel features, and a combination of MFCC and i-vector, respectively.
A variant of  MLP-based networks which introduces a dependency in time between frames, namely the Time Delay Neural Network (TDNN), was used in~\cite{dc_16_nico_dcase} to train AMFB features.
\begin{table}[t]
    \caption{\textit{The state-of-the-art frame-based ASC frameworks}} 
     \vspace{-0.1cm}
    \centering
    \scalebox{0.85}{
    \begin{tabular}{| l | c | c | } 
        \hline 
	    \textbf{Author}  & \textbf{Front-end} & \textbf{Back-end}  \\ [0.5ex] 
	               &  \textbf{Feature Extraction}      & \textbf{Classification} \\
        \hline 
        Mesaros \emph{et al.}~\cite{data_dc_16}                   & MFCC                      & GMM     \\
        Park \emph{et al.}~\cite{dc_16_t03}                    & MFCC, PLP, PNCC,            & GMM    \\
                             & RCGCC, SPCC            &     \\
        Elizalde \emph{et al.}~\cite{dc_16_t06}   & MFCC                      & SVM     \\
        Mafra \emph{et al.}~\cite{dc_16_ngoc_dcase}  & MFCC                      & SVM     \\
        Abidin \emph{et al.}~\cite{dc_16_abi_jr}                 & HOG, LBP                  & SVM     \\
        Waldekar \emph{et al.}~\cite{dc_16_wal_int}                & MFDWC                     & SVM      \\
        Bisot \emph{et al.}~\cite{dc_16_t02}                    & NMF                       & TDL     \\
        Eghbal-Zadeh \emph{et al.}~\cite{dc_16_t01}                    & i-Vector                  & LDA,WCCN      \\
        Mafra \emph{et al.}~\cite{dc_16_ngoc_dcase}   & MFCC     & MLP     \\
        Mika \emph{et al.}~\cite{dc_16_xu_dcase}   & MFCC     & MLP     \\
        Kong \emph{et al.}~\cite{dc_16_kong_dcase} & MFCC     & MLP \\
        Jee-Weon \emph{et al.}~\cite{dc_17_t09} & MFCC     & MLP \\
        Moritz \emph{et al.}~\cite{dc_16_nico_dcase}             & AMFB                      & TDNN     \\
        \hline 
    \end{tabular}
    }
    \label{table:c03_sta_ml} 
\end{table}
 
\subsection{Deep Learning Models}
\label{c02_deepleanring}

\begin{figure}[t]
    \centering
    \includegraphics[width=1.0\linewidth]{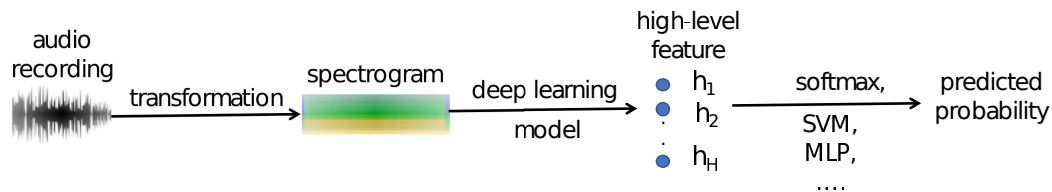}
    	\vspace{-0.7cm}
	\caption{\textit{ASC framework using spectrogram and deep learning model}}
    \label{fig:c02_dl}
\end{figure}

Regarding the second approach of using spectrogram representations, publications show a similar wide variety of back-end classification methods. 
In this case, most are using deep learning networks as shown in Figure \ref{fig:c02_dl} and Table \ref{table:c03_sta_dl}. 
Spectrogram features resemble two-dimensional images, and so to feed these into deep-learning models (note that the entire variable-length spectrogram for variable-length sound input is normally split into small overlapping or non-overlapping patches of equal size~\cite{dc_16_lam_int, lit_zang_int_02, dc_18_lam_aes, lit_zang_int, dc_18_tb01}).
Some systems obtain short spectrograms by adjusting hop size, and then feeding the entire spectrogram into back-end classifiers~\cite{data_dc_18, dc_19_dev_t04}.
Deep learning models applied to ASC can themselves be separated into three main categories: Multilayer Perceptron (MPL), Convolutional Neural Network (CNN) and Recurrent Neural Network (RNN) based architectures.
The deep learning framework used as the baseline of the DCASE 2017 challenge is an example of the first category of using MPL-based networks.
The input features of the MPL architecture systems need to be vectors, so the two-dimensional patches split from the much larger full spectrograms are flattened into vectors in these systems before being fed into the network~\cite{dc_16_kong_dcase}.
Although a wide range of deep learning network are applied to ASC, CNN-based architectures are now the most popular approach.
Indeed, there are a variety of ASC systems using CNN-based network such as Lenet~\cite{len_net}, VGG~\cite{vgg_net}, Resnet~\cite{res_net}, Capsule~\cite{cap_net}, etc. which have published recently.
Analysing the system characteristics submitted to the DCASE challenges over time,  while half of DCASE 2016 submissions used traditional machine learning models and the remaining system applied CNN-based networks, CNN-based architectures were used in  almost all DCASE 2017 systems.
In DCASE 2018 and DCASE 2019 challenges, all submitted systems that achieved higher performance than  the two challenge baselines either completely applied CNN-based networks or partly used them in their systems.
To further analyse the CNN-based networks that have been published, publications tend to make efforts to exploit certain aspects of the CNN networks.
For examples, Yang \emph{et al.}~\cite{dc_18_yang_dcase} proposed a complicated CNN-based architecture called the \emph{xception} network. 
This is inspired by the fact that a deep learning network trained by a wide range of feature scales and over separated channels can result in a very powerful model. 
Focusing on attention mechanisms, an attention-based pooling layer proposed by Zhao Ren \emph{et al.}~\cite{dc_18_zhao_ica, dc_18_zhao_dcase} helped to improve the quality of pooling layers compared with traditional pooling layers. 
Exploring different frequency bands in a spectrogram, Phaye \emph{et al.}~\cite{dc_18_phaye_ica} proposed a SubSpectralNet network which was able to extract discriminative information from 30 sub-spectrograms.
Recently, Song et al.~\cite{dc_18_hong_int} proposed a new way to handle distinct features in a sound scene recording; a deep learning model extracts a bag of features from a log-mel spectrogram, including both similar and distinct ones, from which a back-end network is exploited to enhance accuracy.

\begin{table}[t]
    \caption{\textit{The state-of-the-art spectrogram based ASC frameworks}} 
     \vspace{-0.1cm}
    \centering
    \scalebox{0.8}{
    \begin{tabular}{| l | c | c |} 
        \hline 
	    \textbf{Author}  & \textbf{Front-end} & \textbf{Back-end}  \\ [0.5ex] 
	               &  \textbf{Feature Extraction}      & \textbf{Classification} \\
        \hline 
        Kong \emph{et al.}~\cite{dc_16_kong_dcase}                  & MFCC               & MLP        \\
        Lecun \emph{et al.}~\cite{len_net}                           & log-Mel            & Lenet (CNN)     \\
        Simonyan \emph{et al.}~\cite{vgg_net}     & log-Mel            & VGGish (CNN)       \\
        Ren \emph{et al.}~\cite{dc_18_zhao_ica}     & log-Mel            & VGGish (CNN)       \\
        Zhao \emph{et al.}~\cite{dc_18_zhao_dcase}   & log-Mel            & VGGish (CNN)         \\
        Phaye \emph{et al.}~\cite{dc_18_phaye_ica}   & log-Mel            & VGGish (CNN)         \\
        He \emph{et al.}~\cite{res_net}                           & log-Mel            & Resnet (CNN)   \\
        Patrick \emph{et al.}~\cite{cap_net}                           & log-Mel            & Capsule (CNN)    \\
        Yang \emph{et al.}~\cite{dc_18_yang_dcase}                  & log-Mel            & x-Ception (CNN)  \\
        Zhang \emph{et al.}~\cite{lit_zang_int_02}                   & log-Mel            & LSTM (RNN)     \\
        Zhang \emph{et al.}~\cite{lit_zang_int_03}                   & log-Mel            & LSTM, attention (RNN)  \\
        Zhang \emph{et al.}~\cite{lit_zang_int}                      & log-Mel            & LSTM,temporal transformer (RNN)   \\
        Phan \emph{et al.}~\cite{huy_lit_int_02}            & log-Mel               & GRU (RNN)     \\
        Phan \emph{et al.}~\cite{huy_lit_int}        & log-Mel                & GRU, attention (RNN)    \\
        Phan \emph{et al.}~\cite{huy_lit_aes}           & log-Mel                & GRU, CNN (CNN \& RNN)     \\
        \hline 

    \end{tabular}
    }
    \label{table:c03_sta_dl} 
\end{table}
RNN-based networks are very powerful methods able to learn sequences across time series in addition to spectral relationships. Zang et al.~\cite{lit_zang_int, lit_zang_int_02, lit_zang_int_03} provided a deep analysis of the application of Long Short-term Memory (LSTM), a kind of RNN network, for ASC.
In particular, the authors not only evaluated a single LSTM~\cite{lit_zang_int_02} but also conducted extensive experiments on combinations of the LSTM with other techniques, such as an attention scheme~\cite{lit_zang_int_03} or a temporal transformer layer~\cite{lit_zang_int}. 
Another example showed to be effective in exploiting RNN-based network was published by Huy \emph{et al.}~\cite{huy_lit_aes, huy_lit_int, huy_lit_int_02}.
Instead of using LSTM-based RNN, Huy \emph{et al.} proposed using a Gate Recurrent Unit (GRU) based RNN~\cite{huy_lit_int_02}.
Then they further improved the model by applying an attention scheme~\cite{huy_lit_int} or combining this with CNN-based architectures~\cite{huy_lit_aes}.
Although RNN-based networks prove to be very powerful approaches in Acoustic Event Detection (AED) due to their effectiveness at capturing time sequence, they show poor performance when applied to ASC compared with CNN approaches.
Indeed, not a single RNN-based model was submitted to the recent DCASE 2019 ASC challenge, and furthermore, when RNN-based architectures are used in different contexts, they are normally combined with a CNN network to improve system performance~\cite{huy_lit_aes}.

\subsection{High-level Features}
\label{c02_high_level_feature}

It can be seen that deep-learning-based systems using spectrogram representations have complicated architectures~\cite{dc_18_yang_dcase, dc_18_phaye_ica, dc_18_zhao_ica, huy_lit_aes, huy_lit_int}.
A deeper analysis of the kind of deep learning networks used in these systems, shows that they belong to two main groups divided by the number of training processes used.
Systems which only use one training process are called \textit{end-to-end} learning systems.
In these \textit{end-to-end} systems, the network architecture is separated into two main parts.
While the first part helps to transfer low-level features (spectrogram representation) to high-level features, which contain condensed and discriminative information, the second part takes the role of classification from those condensed features.
In particular, high-level features are normally referred to as the values of the next-to-last layer, and the final layer (normally using Softmax) is referred to as the classification part~\cite{dc_18_yang_dcase, dc_18_phaye_ica, dc_18_zhao_ica}.
In order to gain high-performed high-level features, a variety of complicated architectures have been proposed.
For examples, Truc \emph{et al.}~\cite{dc_18_tb01} applied two parallel CNNs to learn from two type of spectrograms, then concatenated outputs of the CNNs to generate high-level features.
Similarly, Lidy \emph{et al.}~\cite{dc_16_lidy_dcase} used two parallel CNNs, each of which used different kernel sizes to capture different regions of a CQT spectrogram. 
Meanwhile, Soo \emph{et al.}~\cite{dc_16_t09} used both CNN and RNN to capture spatial and time sequence features.
Normally, high-level features extracted by CNN or RNN based structures are concatenated before feeding into fully-connected layers (i.e. Multilayer perceptron), referred to as the final classifier. \\ 
Inspired by the idea that if results of a first training process used to model low-level features are transferred into a second model to aggregate those features, it can improve classification accuracy without requiring an unduly complex single network architecture, the second group of ASC systems use two, or even more, different learning models.
While the first model is again used to extract high-level features (note that these high-level features are also called embedded features or embeddings in some papers), the second model aims to explore high-level features, reporting final classification accuracy.
An early system from this trend was described in~\cite{huy_lit_acm}.
In that system, the authors applied Random Forests (RF) to train from low-level Gamma spectrograms, converting the output into another form of features called labelled tree embedded (LTE) features. 
These LTE features were then classified by a SVM model.
To further explore LTE features, authors conducted various experiments on the second model by using both CNN and SVM~\cite{huy_lit_jr} or RNN~\cite{huy_lit_int}.
Other examples were shown in~\cite{huy_lit_int_02, huy_lit_aes}.
In these systems, deep-learning frameworks of either parallel~\cite{huy_lit_aes} or continuous~\cite{huy_lit_int_02} combinations of CNN and RNN were used to extract high-level features. 
Then, a SVM model was used as the final classifier.
This trend includes a variety of high-level features such as x-vectors extracted from CNNs ~\cite{dc_18_hos_dcase}, feature maps from C-NN networks~\cite{dc_17_chen_ica, dc_16_ars_eus}, deep-scalogram representations from CNN~\cite{dc_17_zhao_jr}.
Recently, transfer learning technique~\cite{dc_16_mun_ica}, a variant of this basic approach, has been widely applied.

\subsection{Ensemble Models}
\label{c02_ensemble}

As mentioned in Section \ref{c02_augmentation}, recent publications reporting high performance have tended to explore multiple low-level input features to deal with the lack of sufficient training information.
Systems using multi-input features often use separated learning models, then fuse the models' results to obtain a final classification accuracy. 
In general, these fusion methods are separated into three main groups, namely \textit{max-fusion, mean-fusion} and \textit{multiplication-fusion} methods, each of which is discussed in~\cite{huy_lit_aes}. 
Although these systems prove to have a high cost of computation as well as a high volume of training parameters, they are able to achieve competitive results.
Indeed, the top-three performing systems in the DCASE challenges~\cite{dc_19_t03, dc_16_t01, dc_16_t03, dc_17_t01, dc_17_t02, dc_17_t03, dc_18_dof_tp, dc_18_yuma_tp} and on the Litis-Rouen dataset~\cite{huy_lit_int_02, lit_zang_int}, use a variety of fusion methods.
In particular, these high-performing systems use a spectrogram representation for low-level input features.

\section{Open Issues}
\label{c02_issue}

Several of the existing works mentioned above have described the open issues that this thesis is going to analyse and investigate. 
Firstly, while multiple low-level input features such as spectrograms, channels, frequency and time resolution, etc. have been explored for use on ASC challenges, there has been much less comprehensive analysis to identify the most effective low-level feature.
Furthermore, using multiple input features usually combined with ensemble models, has a very high computational cost -- effectively throwing computational power at the problem. It would clearly be better to more precisely identify optimum features (and their characteristic settings) rather than blindly combining a large set of multiple features.
In terms of back-end classification, although a wide range of deep learning frameworks have been proposed, they mainly exploit specific features (i.e. are highly feature-specific) and are almost always evaluated over a very limited set of datasets. The danger there is of building locally-optimum systems which work well on one challenge, but perform poorly on others.
As ASC challenges mainly come from various sound events and scenes inside environmental recording datasets, focusing on specific aspects of an ASC system easily causes overfitting issues, especially troublesome when a model is evaluated over different datasets to those it is trained for.
Further analysis of back-end classification models shows that, although some high-performance high-level features have been recently defined, no current publications adequately explore the relationship between high-level features which have been extracted from different low-level features.
Moreover, very few published papers explore the nature and the effect of those high-level feature characteristics.
Finally, although environmental sounds have high cross-correlation due to the presence of similar types and degrees of background noise, few publications provide an analysis of this aspect of systems.

\chapter{Low-Level Feature Analysis}
\label{c03}
This chapter aims to provide a comprehensive analysis of low-level features in an ASC system and identify how these features affect the final classification accuracy.
To this end, a wide range of low-level features such as channel information, spectrogram types and various image patch sizes, etc., are evaluated.
These experiments are conducted over a C-DNN-based model, referred to as the proposed baseline, and evaluated using the DCASE 2018 Task 1A, 1B dataset.
After indicating the most influencing low-level features, C-DNN baseline's performance is evaluated again with mixup data augmentation.

\section{High-level Architecture}
\label{c03_high_level_architecture}
\begin{figure}[t]
    \centering
    \includegraphics[width=\linewidth]{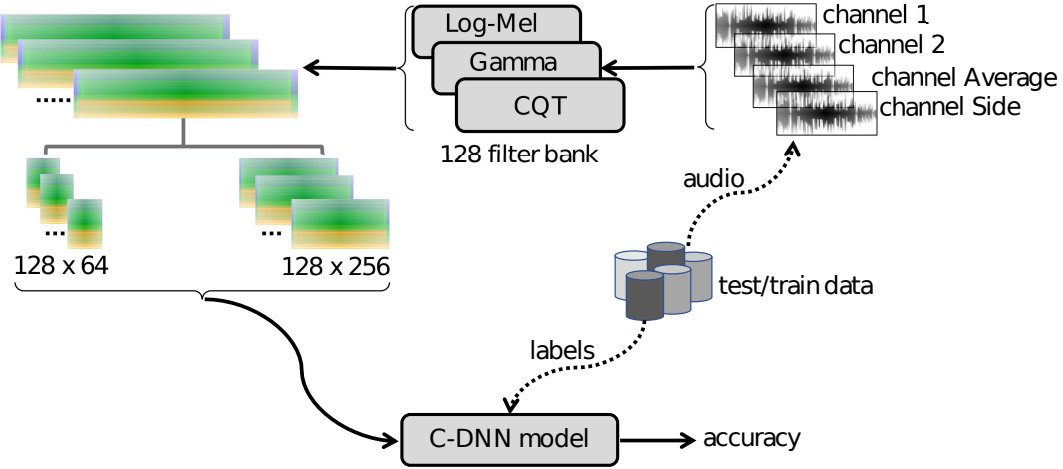}
	\caption{\textit{The high-level architecture of proposed baseline.}}
    \label{fig:c03_baseline}
\end{figure}

Starting with a spectrogram representation as the low-level feature, a general system architecture for ASC is presented in Figure \ref{fig:c03_baseline}.
It can be seen that the entire ASC system is separated into two main processes. 
The first process (top half) has the role of transforming the selected audio channels into one or more types of spectrogram, and then splitting the full spectrogram into smaller patches of different sizes to form a bag-of-features. In this case, the patches are non-overlapping and of predefined size. 

The second process (bottom half) is a machine learning model which trains the input patches extracted by the first process and performs classification. In this case, the model is a CNN combined with a DNN (abbreviated to C-DNN). The output of the machine learning model is the reported classification accuracy. 
\section{Low-level Feature Analysis}
\label{c03_feature_analysis}
Due to ASC challenges discussed previously in Section \ref{c02_augmentation}, it is known that an ensemble of multiple low-level features or models can promisingly enhance the classification accuracy in current state-of-the-art systems.
This motivates the analysis and exploration of the effects on classification accuracy of different bags-of-features.
Specifically, it drives to explore into three groups as shown in Table~\ref{table:c03_bag_of_feature}.

The first exploration focuses on the effect of different channels (since the data sets include two-channel recordings namely Left and Right). 
The possibilities are using the first channel alone (Left), the second channel alone (Right), and the space information of two channels obtained by Average and Side, where Average means (Left+Right)/2 and Side means (Left-Right).
\begin{table}[t]
    \caption{\textit{Bag-of-feature analysis settings for ASC.}} 
    \vspace{-0.2cm}
    \centering
       \scalebox{0.85}{

    \begin{tabular}{|l |c |} 
        \hline 
            \textbf{Channels}         &Channel 1, Channel 2,\\
	                              & Average, and Side of two channels  \\
	           \hline                                               
             \textbf{Patch sizes}      & 0.37\,s (64 bins), 0.74\,s (128 bins), \\ 
                                       & 1.11\,s (192 bins), 1.48\,s (256 bins)\\          
             \hline 
             \textbf{Spectrograms}     & log-mel, Gamma, and CQT \\
                                                       & (each with 128 filters) \\     
        \hline 

    \end{tabular}
    }
    \label{table:c03_bag_of_feature} 
\end{table}

The second exploration is for different patch sizes, in terms of the time duration that they cover. 
The number of frequency bins used for the analysis is fixed at 128, but a different numbers of time bins are evaluated, specifically 64, 128, 192, and 256 -- giving rise to the four different patch durations shown in Table~\ref{table:c03_bag_of_feature}.

The third exploration considers the use of three alternative spectrogram transformations, namely log-mel, gammatone (Gamma) and Constant Q Transform (CQT). Since these spectrograms are derived from different auditory models, it is plausible that they can each contribute distinct features for classification. By ensuring that parameters such as window size, hop size and patch number are fixed, the number of data items fed into the back-end learning model will be equal for different types of spectrogram, hence any difference in classification accuracy is due to the characteristics of each spectrogram, not due to different learning requirements. In this thesis, while log-mel and CQT are generated by using a popular toolbox namely Librosa~\cite{librosa_tool}, gammatone-like spectrogram toolbox~\cite{auditory2009_tool} is used to generate Gamma. The methods of transforming  a selection of audio signal into a spectrogram are presented in detail in Chapter \ref{c08}. 

Furthermore, this chapter also analyses how mixup data augmentation, a simple-implemented and popular method of data augmentation mentioned in Chapter \ref{c02}, affects to the classification accuracy. Description and settings for mixup data augmentation are presented in detail in next Section \ref{c03_effect_of_mixup}.
\section{Propose a Baseline} 
\label{c03_baseline}
\begin{table}[h]
    \caption{\textit{The primary operating parameters of the proposed baseline for \\ DCASE 2018 challenge, alongside the parameters from the challenge baseline.}} 
    \vspace{-0.1cm}
    \centering
    \scalebox{0.85}{
    \begin{tabular}{| l |c |c |} 
        \hline 
         \textbf{System Setting}                     &\textbf{Proposed C-DNN}     & \textbf{DCASE 2018}\\	                              
			                                                  &\textbf{baseline}         & \textbf{baseline}   \\
        \hline 
	    Window size       &0.044s           &0.04s   \\
	    Hop size          &12.5\%             &50\%    \\
	    Spectrogram method            &log-mel              &log-mel \\
	    Filter number                  &128      &40\\
	    Spectrogram size &$128{\times}1728$         &   $40{\times}500$   \\
	    Like-image features        &13 patches ($128{\times}128$)             &entire spectrogram \\
       \hline 
       \hline 

	    Deep learning model     &C-DNN               &C-DNN \\
	    	        & (Lenet-7)               &(The best model form DCASE 2016) \\

	      \hline 
    \end{tabular}
    }
    \label{table:c03_para_feature} 
\end{table}
%
To analyse different bags-of-features, a baseline architecture is first established, configured using the parameters set out on the left hand side in Table \ref{table:c03_para_feature}.
Additionally, this chapter also compares overall performance against the standard DCASE 2018 baseline system of~\cite{data_dc_18}.
The proposed C-DNN parameters are listed in Table \ref{table:c03_cdnn_conf}.

Given by Table \ref{table:c03_para_feature}, it can be seen that the proposed baseline architecture uses a window size of 0.044\,s (albeit extracted from a higher dimension spectrogram) and a smaller hop size of 12.5\% compared with a window size at 0.04s and a hop size of 50\% in the DCASE 2018 baseline, since the baseline's purpose aims to extract additional useful information from the input spectrogram. 
Regarding the spectrogram, the proposed architecture uses a log-mel filter with 128 bands which is much larger than the 40 Mel filter bands of the DCASE 2018 baseline. 
Based on these parameters, the proposed spectrogram, which are generated from from 10-second segments with sample rate of 48,000 Hz, has a bigger size in both time and frequency bins of 128$\times$1728 compared with 40$\times$500 in the DCASE 2018 baseline. 
Then, the entire spectrogram of 128$\times$1728 is split into 13 non-overlapping patches of $128\times128$ before feeding into a back-end classification. \\

As regards learning models, the DCASE 2018 baseline reuses the architecture of the top ranked submission of DCASE 2016~\cite{dc_16_t05}.
\begin{table}[t]
    \caption{\textit{The proposed C-DNN network configuration  \\ (the upper part is for CNN and the lower part is for DNN).}} 
    \vspace{-0.1cm}
    \centering
    \scalebox{0.85}{
    \begin{tabular}{|l |c |c|} 
        \hline 
	    \textbf{Layers}    &  \textbf{Output shape}  &   \textbf{Kernel/Drop Ratio}\\
       \hline 
        Input                 &$128{\times}128{\times}1$      &-    \\
        Convolutional 1         &$128{\times}128{\times}32$     &[$3{\times}3$] $@$ 32  \\
	ReLU  1                &$128{\times}128{\times}32$     &-     \\
	Batch normalization  1          &$128{\times}128{\times}32$     &-     \\
        Average pooling 1        &$64{\times}64{\times}32$       &[$2{\times}2$]  \\
	Dropout 1             &$64{\times}64{\times}32$       &10\% \\
       \hline    
        Convolutional 2         &$64{\times}64{\times}64$       &[$3{\times}3$]  $@$ 64 \\
	ReLU  2               &$64{\times}64{\times}64$       &-     \\
		Batch normalization  2          &$64{\times}64{\times}64$       &-     \\
        Average pooling 2        &$32{\times}32{\times}64$       &[$2{\times}2]$  \\
   	Dropout 2             &$32{\times}32{\times}64$       &15\% \\
       \hline 
        Convolutional 3         &$32{\times}32{\times}128$      &[$3{\times}3$]  $@$ 128 \\
   	ReLU 3                &$32{\times}32{\times}128$      &-     \\
   	Batch normalization  3          &$32{\times}32{\times}128$      &-     \\
        Average pooling 3        &$16{\times}16{\times}128$      &[$2{\times}2$]  \\
    	Dropout 3             &$16{\times}16{\times}128$      &20\% \\
       \hline 
        Convolutional 4         &$16{\times}16{\times}256$      &[$3{\times}3$]  $@$ 256 \\
   	ReLU        4         &$16{\times}16{\times}256$      &-     \\
   	Batch normalization  4          &$16{\times}16{\times}256$      &-     \\
	Global average pooling 4   &256                            &-     \\	     
   	Dropout 4             &256                            &25\% \\
        \hline
       \hline 
	Fully connected 5     &512    &-          \\
	ReLU 5                  &512    &-          \\
    	Drop out 5              &512    &30\%      \\
    	       \hline    

	Fully connected 6     &1024   &-          \\
	ReLU 6                  &1024   &-          \\
    	Dropout 6              &1024   &35\%      \\
    	       \hline    

	Fully connected 7     &10     &-          \\
	Softmax 7               &10     &-          \\
        \hline 
    \end{tabular}
    }
    \label{table:c03_cdnn_conf} 
\end{table}
However the DCASE 2016 system only had two convolution blocks (convolutional, batch normalization, rectify linear unit, and dropout layers) followed by one dense layer and an output layer with Softmax classification.

The new baseline is much more complicated, with four convolutional blocks based on Lenet-7~\cite{len_net}.
In particular, the C-DNN proposed presents four convolutional blocks, as configured in the upper part of Table \ref{table:c03_cdnn_conf}, each of which includes a convolutional layer followed by a rectify linear unit (ReLU), a batch normalization, an average pooling, and a dropout layer. 
At the final convolution block, instead of using a average pooling layer, a global average pooling layer is applied to enhance the accuracy based on the idea of considering the contribution of all channels output to the final convolution block as a bag-of-features and reducing noise.

The classification operation is shown in the lower part of Table \ref{table:c03_cdnn_conf} (note that while blocks are separated by single line, the role of classification in the lower parts is separated by a double line), handled by fully connected, ReLU and dropout layers. At the final layer, a Softmax function is used for classifying into different ten scene contexts.

\section{Experimental Setting} 
\label{c03_setup}

\subsection{Dataset}
\label{c03_dataset}
Experiments in this chapter are conducted using the development set (Dev. set) of DCASE 2018 Task 1A and 1B~\cite{data_dc_18}.
The audio files in these datasets are all wave file format recordings with a sample rate of 48000\,Hz and have a 10-second duration. 

As DCASE 2018 Task 1A dataset, all of the audio files were recorded by the same device, denoted `device A' (Soundman OKM II Klassik/studio A3 electret microphone and a Zoom F8 audio recorder), and are grouped into ten different categories, with one category label per recording. 
The data is unbalanced so that the number of recordings per category is slightly uneven. 
This can be seen in the left hand side of Table~\ref{table:c03_dataset} which identifies the categories and the number of 10-second audio files that each categories contains.
In total, the task includes 8640 audio files.  
Using the DCASE 2018 suggested test/train split~\cite{data_dc_18}, recordings of the development set are separated into a Training subset (6122 audio files) and a Test subset (2518 audio files).
\begin{table}[t]
    \caption{\textit{Development set of DCASE 2018 Task \\ 1A and 1B datasets are split into Test/Training subsets}} 
     \vspace{-0.1cm}
    \centering
    \scalebox{0.85}{
    \begin{tabular}{| l | c c | c c |} 
        \hline 
	    \textbf{Categories}  & \textbf{Training subset} & \textbf{Test subset} & \textbf{Training subset} & \textbf{Test subset} \\ [0.5ex] 
	               &  \textbf{(1A)}      & \textbf{(1A)} & \textbf{(1B)} & \textbf{(1B)} \\
        \hline 
        Airport                  & 599               & 265   &707                &301   \\
        Bus                      & 622               & 242   &730                &278   \\
        Metro                    & 603               & 261   &711                &297   \\
        Metro Station            & 605               & 259   &713                &295   \\
        Park                     & 622               & 242   &730                &278   \\
        Public Station           & 648               & 216   &756                &252   \\
        Shopping Mall            & 585               & 279   &693                &315   \\
        Pedestrian Street        & 617               & 247   &725                &283   \\
        Traffic Street           & 618               & 246   &726                &282   \\
        Tram                     & 603               & 261   &711                &297   \\
        \hline 
        \textbf{Total files}              & \textbf{6122}              & \textbf{2518}  &\textbf{7202}               &\textbf{2878}  \\
        \hline 

    \end{tabular}
    }
    \label{table:c03_dataset} 
\end{table}

DCASE 2018 Task 1B reuses all audio files from Task 1A, but extends that with additional recordings obtained from two other recording devices named B and C (e.g. recorded from a variety of smart phones and cameras).
However, it should be noted that the number of recordings made by devices B and C is much smaller than that of device A; in total just 4 hours for devices B and C compared to 24 hours for device A. 

While performance evaluation for subtask 1A is based on classification accuracy for device A recordings, scoring for subtask 1B is only based on the classification accuracy assessed for devices B and C. It thus tests how well a system, trained mainly with recordings from one device, performs on recordings made on other devices (which is a common real world scenario).

Like DCASE 2018 Task 1A, the subtask 1B dataset is split into test/train portions~\cite{data_dc_18}. Recordings within the development set are separated into a Training subset (7202 audio files) and a Test subset (2878 audio files), shown in two right hand columns of Table~\ref{table:c03_dataset}.

\subsection{Setting Hyperparametes and Training Process} 
\label{c03_hyper}
To train the baseline, cross entropy, defined in Equation (\ref{eq:c03_loss_func}), is minimised to tune the parameters, denoted as \(\Theta\).

\begin{equation}
    \label{eq:c03_loss_func}
    LOSS_{EN}(\Theta) = -\sum_{c=1}^{C}y_{c}\log \left\lbrace\hat{y}_{c}(\Theta)\right\rbrace  +  \frac{\lambda}{2}||\Theta||_{2}^{2},
\end{equation}

where \(LOSS_{EN}(\Theta)\) is the entropy loss function for all parameters \(\Theta\) of the C-DNN model, $\lambda$ denotes the $\ell_2$-norm regularization coefficient set to 0.001, $C$ is number of sound scene categories classified, \(y_{c}\) and \(\hat{y}_{c}\) are ground truth and predicted result for class $c$ respectively, in one-hot format.

The C-DNN baseline is built in the Tensorflow framework, set with epoch number, batch size and initial learning rate of 100, 100 and 0.0001 respectively, and using the Adam method for learning rate optimisation~\cite{kingma2014adam}.
Trainable parameters are initialised randomly with a normal distribution, having mean and variance set to 0 and 0.1, respectively.

\subsection{Ensemble Method}
\label{c03_ensemble}

As the back-end classification models returns the predicted probability of single patch, predicted probability of entire spectrogram is computed by taking the average of all patches' probabilities (this is similar to the \textit{mean-fusion} method mentioned in~\cite{huy_lit_aes}).
If $\mathbf{p^{n}} = (p_{1}^{n}, p_{2}^{n},..., p_{C}^{n})$,  with $C$ being the category number and the \(n^{th}\) out of \(N\) patches fed into learning model, are considered as the probability of a test sound scene instance, then the average classification probability is denoted as  $\mathbf{\bar{p}} = (\bar{p}_{1}, \bar{p}_{2}, ..., \bar{p}_{C})$ where,

\begin{equation}
    \label{eq:c03_mean_stratergy_patch}
    \bar{p}_{c} = \frac{1}{N}\sum_{n=1}^{N}p_{c}^{n}  
\end{equation}

and the predicted label from the C-DNN is determined using,

\begin{equation}
    \label{eq:c03_label_determine}
    \hat{y} = \argmax_{c \in \{1,2,\ldots,C\}}\bar{p}_c
\end{equation}

As regards an ensemble of channels, spectrograms or patch sizes, this \textit{mean-fusion} method can also be applied in the same way.
If $\mathbf{p^{m.n}} = (p_{1}^{m.n}, p_{2}^{m.n},..., p_{C}^{m.n})$,  with $C$ being the class number, the \(n^{th}\) out of \(N\) patches fed into learning model, and the \(m^{th}\) out of \(M\) channels, spectrograms, or patch sizes, are considered as the probability of a test sound scene instance. The mean classification probability is then denoted as $\mathbf{\bar{p}} = (\bar{p}_{1}, \bar{p}_{2}, ..., \bar{p}_{C})$ where,

\begin{equation}
    \label{eq:c03_mean_stratergy_file}
    \bar{p}_{c} = \frac{1}{M.N}\sum_{m=1}^{M}\sum_{n=1}^{N}p_{c}^{m.n}  
\end{equation}

and similarly the predicted label is determined as in Equation (\ref{eq:c03_label_determine}).
sddsss
\section{Experimental Results}
\label{c03_results}

To evaluate how low-level features such as channel, patch size, type of spectrogram, and ensembles of these features affect ASC performance, the baseline with using channel 1, patch size of $128\times128$, and log-Mel spectrogram proposed in Section \ref{c03_baseline} is firstly compared with DCASE baseline. 
This comparison is described in the next Section \ref{c03_baseline_compare}.
Then, ensembles of channels, sizes of patches, and types of spectrograms are conducted and compared to individual models (i.e. an individual model receives only one channel input, and use one type of patch size and spectrogram), which are shown in Section \ref{c03_channel_compare}, \ref{c03_size_compare}, and \ref{c03_spec_compare}, respectively.
The comparison shows how ensembles help to improve performance, compare with individual model and DCASE baseline.
Next, these ensembles are compared together, indicating the most effective low-level feature in Section \ref{c03_compare}.
Finally, the effect of data augmenation is evaluated in Section \ref{c03_effect_of_mixup}.

\subsection{Baseline Comparison}
\label{c03_baseline_compare}
 \begin{table*}[tbh!]
     \caption{\textit{Performance comparison (percentage category classification accuracy - \%) \\ between DCASE 2018 Task 1A baseline and the proposed C-DNN baseline.}} 
    \vspace{-0.1cm}
    \centering
    \scalebox{0.85}{
    \begin{tabular}{|l |c |c |} 
        \hline 
        \textbf{Categories}                    &\textbf{DCASE 2018}        &\textbf{Proposed C-DNN}      \\
                                                          &\textbf{baseline (\%)}        &\textbf{baseline (\%)}      \\                                  
        \hline 
	Airport                  & 72.9             &56.2         \\
	Bus                      & 62.9             &66.1        \\
	Metro                    & 51.2             &39.1         \\
        Metro  Station           & 55.4             &67.6        \\
        Park                     & 79.1             &80.6          \\
        Public  Square           & 40.4             &64.8        \\
        Shopping  Mall           & 49.6             &87.8         \\
	Street Pedestrian       & 50.0             &46.6          \\
        Street Traffic        & 80.5             &79.3         \\
        Tram                     & 55.1             &72.8          \\
        \hline 
	    \textbf{Average}              & \textbf{59.7}             &\textbf{66.2}         \\
        \hline 
    \end{tabular}    
    }
    \label{table:c03_baseline_comparison} 
\end{table*}
The accuracy of every category reported by the DCASE 2018 Task 1A baseline and by the proposed C-DNN network is displayed in Table \ref{table:c03_baseline_comparison}.
In general, the C-DNN baseline improves average accuracy by 6.5\%, compared to 59.7\% of DCASE 2018, but not every category improves. 

In terms of each category performance, \textit{Tram, Shopping Mall, Public Square} and \textit{Metro Station} show significant improvements, increasing by 17.7\%, 38.2\%, 24.4\%, and 12.2\%, respectively.
Performances on \textit{Airport, Metro}, by contrast, decrease to 56.2\% and 39.1\%, compared to 72.9\% and 51.2\% of DCASE 2018.
\begin{figure}[t]
    \centering
    \includegraphics[width=0.68\linewidth]{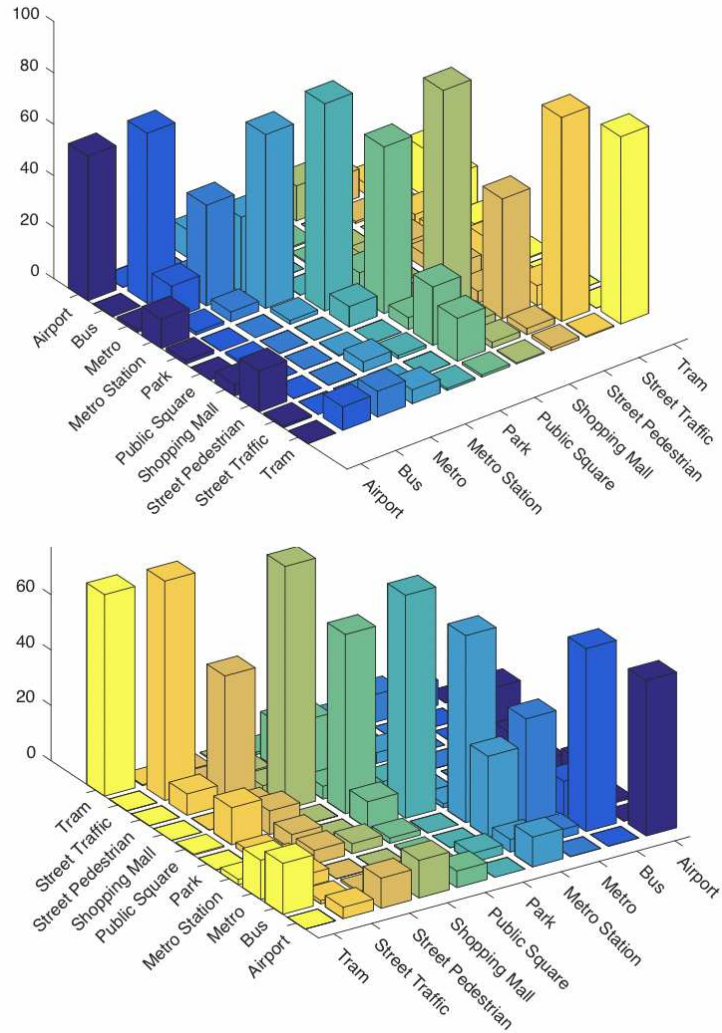}
    	\vspace{-0.3cm}
	\caption{\textit{Confusion matrix plot (displayed from two sides) \\ of C-DNN baseline performance per category.}}
    \label{fig:c03_confusion_matrix}
\end{figure}
The accuracy of the remaining categories is similar.
To further analyse the C-DNN baseline performance, 10 categories are divided into three meta categories, 
and re-compute the performance on 3 meta categories rather than on 10 categories.
Specifically, grouping the categories is as follows - \textit{vehicle (Bus, Metro, Tram), indoor (Airport, Metro Station, Shopping Mall)} and \textit{outdoor (Park, Public Square, Street Pedestrian, Street Traffic)}. The classification accuracy over each meta category is 85.8\%, 81.5\% and 91.5\% for \textit{indoor, vehicle} and \textit{outdoor}, respectively, with an averaging of 86.3\% over three meta categories. 
The relatively high performance indicates that environmental sounds show high-cross correlation, with a majority of incorrectly recognized samples dropping into the same meta categories.  
Indeed, the confusion matrix result of the C-DNN baseline, shown in Figure \ref{fig:c03_confusion_matrix} and Table \ref{table:c03_inacc}, indicates that incorrect cases in the same meta categories are larger than those across different meta categories. These are in fact 20.7\% and 13.1\%  on average, respectively.
This interesting discovery motivates a further comprehensive analysis on cross correlation among both individual and meta categories that will be presented in  Chapter \ref{c05}.
\begin{table}[tb]
    \caption{\textit{Accuracy (Acc. \%) and Inaccuracy (Inacc. \%) \\ across individual and meta categories.}} 
    \vspace{-0.1cm}
    \centering
    \scalebox{0.85}{
    \begin{tabular}{|l |c |c |c |} 
        \hline 
        \textbf{Categories}         &\textbf{Acc.}        &\textbf{Inacc. inside}    &\textbf{Inacc. outside}    \\ [0.25ex] 
        &\textbf{(\%)}   &\textbf{meta category (\%)} &\textbf{meta category (\%)} \\
        \hline 
	Airport                   &56.2   &24.4  &19.4   \\
	Bus                       &66.1   &21.5  &12.4   \\
	Metro                     &39.1   &27.0  &33.9   \\
        Metro  Station            &67.6   &14.9  &17.5   \\
        Park                      &80.6   &15.0  &4.4    \\
        Public  Square            &64.8   &29.8  &5.4    \\
        Shopping  Mall            &87.8   &5.0   &7.2    \\
	Street Pedestrian         &46.6   &31.6  &21.8   \\
        Street Traffic            &79.3   &19.1  &1.6   \\
        Tram                      &72.8   &19.1  &8.1   \\
        \hline 
	    \textbf{Average}      &\textbf{66.2} &\textbf{20.7} &\textbf{13.1}         \\
        \hline 
    \end{tabular}    
    }
    \label{table:c03_inacc} 
\end{table}

\subsection{Bag-of-channel Ensembles}
\label{c03_channel_compare}
\begin{figure}[th!]
    \centering
    \includegraphics[width=0.85\linewidth]{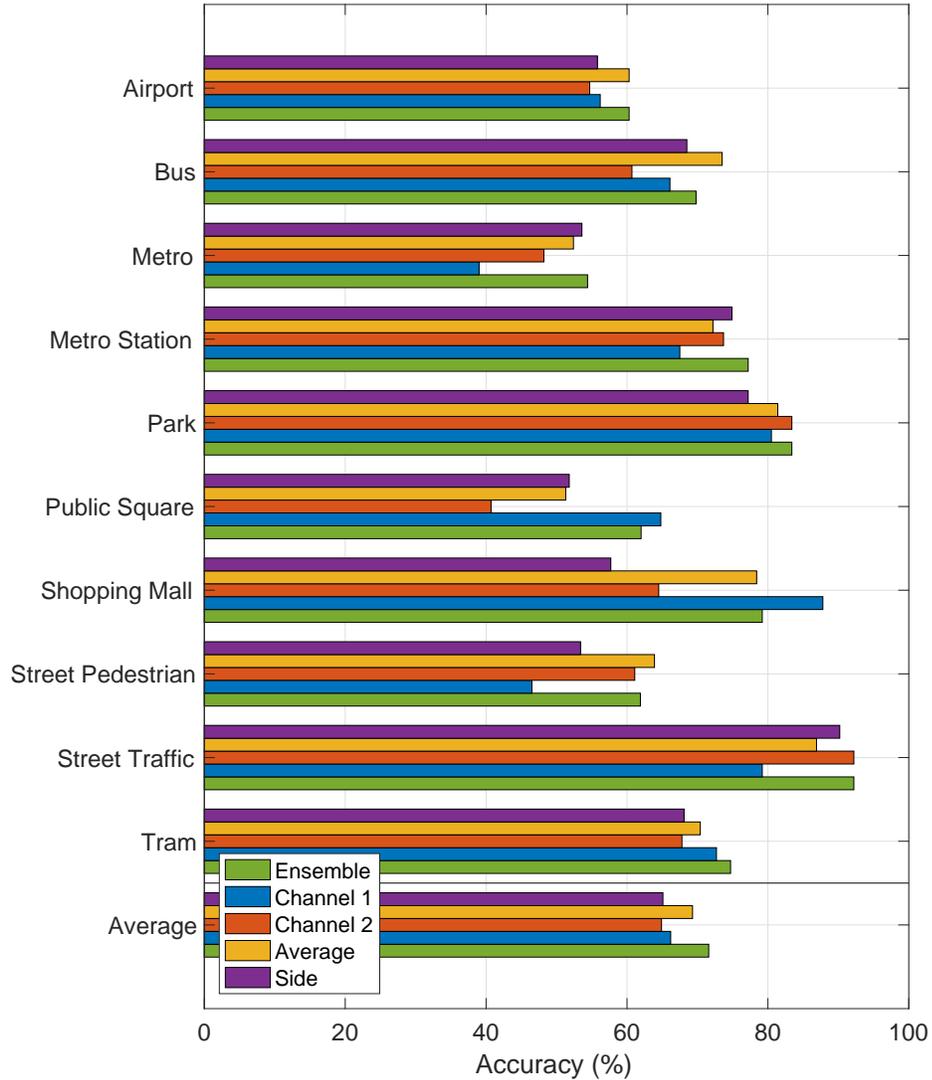}
    	\vspace{-0.3cm}
	\caption{\textit{Class-wise performance with channel effect.}}
    \label{fig:c03_channel_effect}
\end{figure}
Working from the proposed C-DNN baseline, the effect on  classification accuracy was analysed when using the four different channel arrangements (Note that computing the ensemble of different channels is mentioned in Section \ref{c03_ensemble}).
Results are presented in Figure \ref{fig:c03_channel_effect} and reveal that Channel 1 (Left), Channel 2 (Right), their 
Side, and Average differ more widely at the output of the C-DNN baseline with the highest score of 69.3\% obtained from the average of the two channels.  
Ensemble models exploiting different channels help to improve accuracy by 11.9\% better than the 59.7\% achieved by the DCASE 2018 baseline. 

\subsection{Bag-of-feature-size Ensembles}
\label{c03_size_compare}
\begin{figure}[h]
    \centering
    \includegraphics[width=0.85\linewidth]{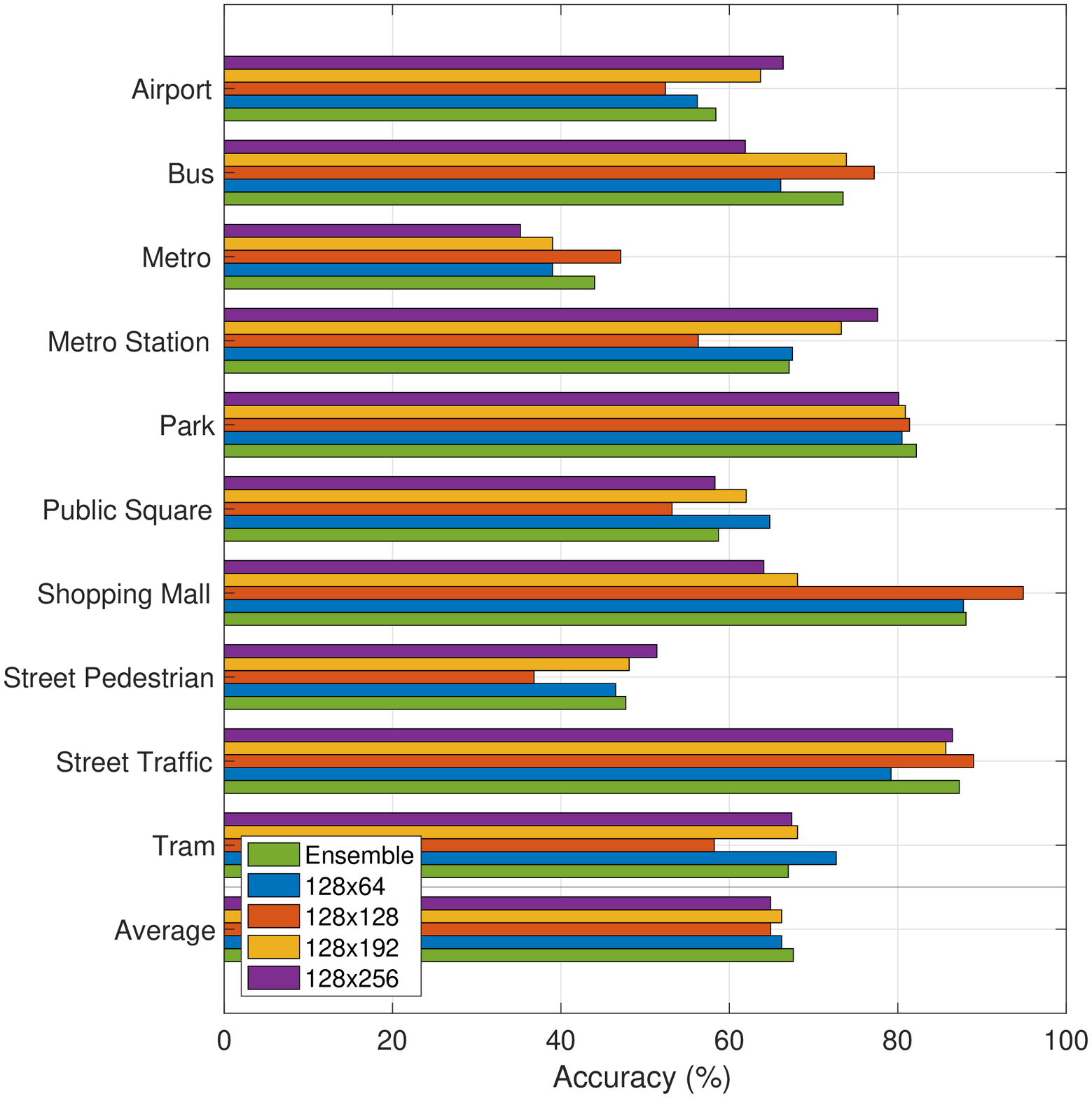}
    	\vspace{-0.5cm}
	\caption{\textit{ Class-wise performance with patch size effect.}}
    \label{fig:c03_patch_effect}
\end{figure}
This section presents an experiment to determine whether ensembles of different patch sizes could improve accuracy as shown in Figure \ref{fig:c03_patch_effect} (Note that computing the ensemble of different patch sizes is mentioned in Section \ref{c03_ensemble}). 
In general, the C-DNN performance results in different sizes are similar and are not significantly improved.
Ensemble result over all patch sizes slightly improve -- by  2.1\% and 8.6\% compared to the C-DNN baseline and DCASE 2018, respectively, which shows poorer performance than the channel ensemble.  

\subsection{Bag-of-spectrogram Ensembles}
\label{c03_spec_compare}
\begin{figure}[h]
    \centering
    \includegraphics[width=0.85\linewidth]{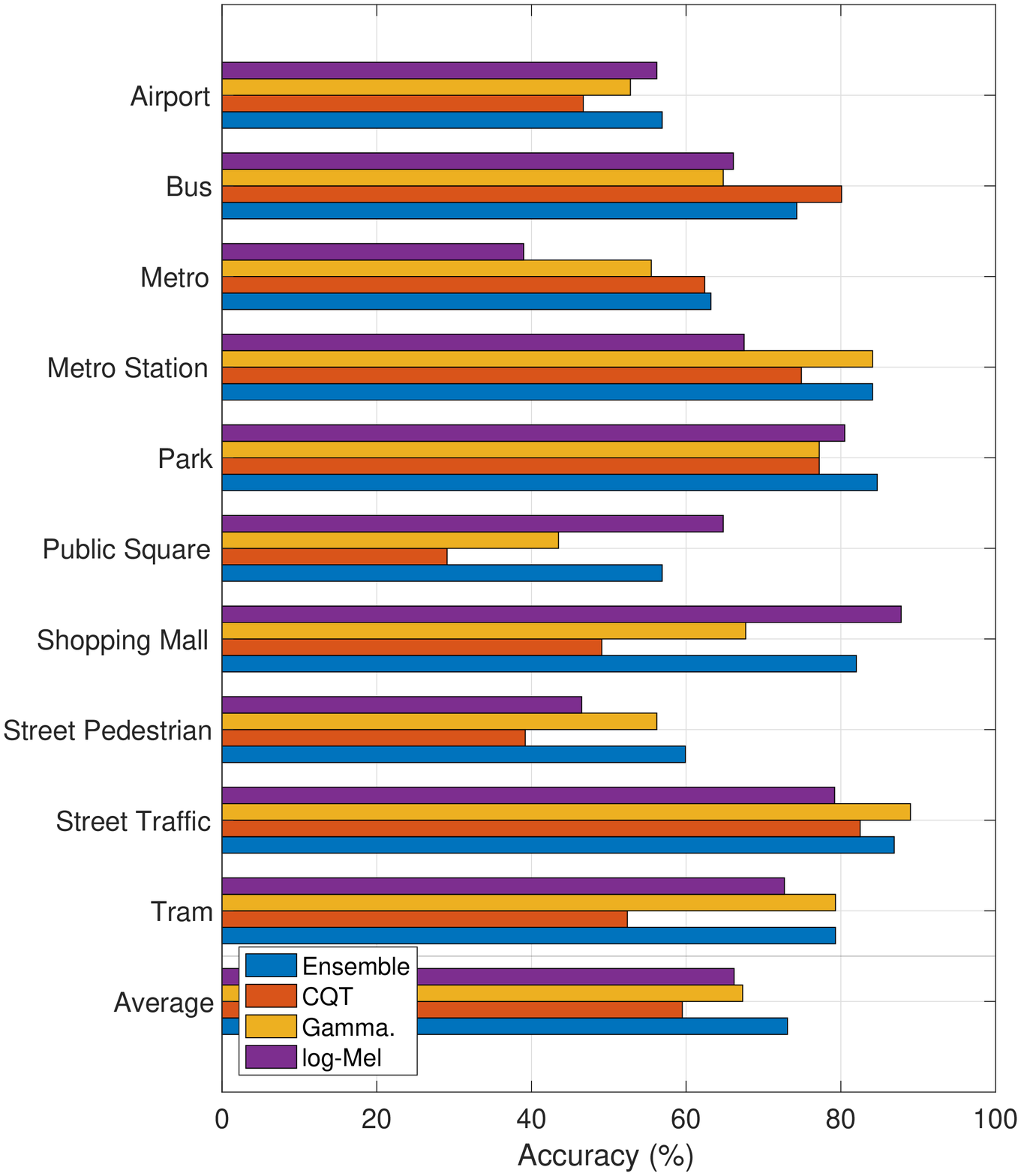}
    \vspace{-0.7cm}
	\caption{\textit{Class-wise performance with spectrogram effect.}}
    \label{fig:c03_spec_effect}
\end{figure}
Next, the effect of using the three spectrogram transformation types listed in Table~~\ref{table:c03_bag_of_feature} for the classification of patches sized 128$\times$128 (in the proposed baseline), is evaluated (Note that computing the ensemble of different types of spectrogram is mentioned in Section \ref{c03_ensemble}).
The results, shown in Figure \ref{fig:c03_spec_effect}, indicate that the best C-DNN result among the three spectrogram types is 67.3\%, for the Gamma spectrogram. 
While log-mel results are competitive to Gamma, CQT performance tends to be significantly poorer than the others. 
However, CQT performs well with \textit{vehicle}-related categories.
Noticeably, an ensemble among spectrograms helps to improve the overall accuracy 13.4\% more than that of DCASE 2018 baseline, and on average 7.0\% better than the highest single-spectrogram performance. 

\subsection{Comparison of bag-of-feature Ensembles}
\label{c03_compare}
%
 \begin{table*}[tbh!]
     \caption{\textit{Performance comparison (percentage accuracy - \%) between\\ DCASE 2018 Task 1A baseline and bag-of-feature ensemble of C-DNN proposed.}} 
   \vspace{-0.1cm}
    \centering
    \scalebox{0.85}{
    \begin{tabular}{|l |c |c |c |c |} 
        \hline 
        \textbf{Categories}                    &\textbf{DCASE 2018}        &\textbf{Spectrogram}        &\textbf{Patch Size}    &\textbf{Channel}  \\
                          &\textbf{baseline (\%)}        &\textbf{Ensemble (\%)}        &\textbf{Ensemble (\%)}    &\textbf{Ensemble (\%)}  \\ [0.25ex] 

        \hline 
	Airport                  & 72.9             &57.0 & 64.9 & 60.3          \\
	Bus                      & 62.9             &74.4 & 69.8 & 69.8         \\
	Metro                    & 51.2             &63.2 & 42.9 & 54.4          \\
        Metro  Station           & 55.4             &84.2 & 73.3 & 77.2         \\
        Park                     & 79.1             &84.7 & 80.5 & 83.4           \\
        Public  Square           & 40.4             &56.9 & 60.1 & 62.0         \\
        Shopping  Mall           & 49.6             &82.1 & 82.4 & 79.2          \\
	Street Pedestrian        & 50.0             &59.9 & 51.4 & 61.9           \\
        Street Traffic           & 80.5             &87.0 & 86.5 & 92.2          \\
        Tram                     & 55.1             &79.3 & 70.1 & 74.7           \\
        \hline                                      
	    \textbf{Average}              & \textbf{59.7}             &\textbf{73.1}  &\textbf{68.3} &\textbf{71.6}        \\
        \hline 
    \end{tabular}    }
    \label{table:c03_ensemble_comp} 
\end{table*}

 \begin{table*}[tbh!]
     \caption{\textit{Performance comparison (percentage accuracy - \%) between \\ DCASE 2018 Task 1B baseline with the use of spectrogram ensemble of the C-DNN.}} 
    \vspace{-0.1cm}
    \centering
    \scalebox{0.85}{
    \begin{tabular}{|l |c |c |c |c |} 
        \hline 
        \textbf{Categories}                    &\textbf{DCASE 2018}        &\textbf{Ens. of Spec.}        &\textbf{DCASE 2018}    &\textbf{Ens. of Spec.}  \\ [0.25ex] 
                                                          &\textbf{Dev. A (\%)}        &\textbf{Dev. A (\%)}        &\textbf{Dev. B\&C (\%)}    &\textbf{Dev. B\&C (\%)}  \\ [0.25ex] 
        \hline 
	Airport                  &73.4  &68.7 &72.5  &52.8           \\
	Bus                      &56.7  &70.7 &78.3  &88.9          \\
	Metro                    &46.6  &70.9 &20.6  &27.8           \\
        Metro  Station           &52.9  &83.4 &32.8  &61.1          \\
        Park                     &80.8  &82.2 &59.2  &83.3            \\
        Public  Square           &37.9  &52.8 &24.7  &55.6          \\
        Shopping  Mall           &46.4  &67.4 &61.1  &80.6           \\
	Street Pedestrian        &55.5  &64.0 &20.8  &52.8            \\
        Street Traffic           &82.5  &90.2 &66.4  &83.3           \\
        Tram                     &56.5  &78.9 &19.7  &27.8            \\
        \hline                                      
	    \textbf{Average}              & \textbf{58.9}             &\textbf{72.9}  &\textbf{45.6} &\textbf{61.4}        \\
        \hline 
    \end{tabular}    }
    \label{table:c03_task1b} 
\end{table*}
Comparing the performance of bag-of-feature ensembles with DCASE 2018 baseline for Task 1A, as shown in Table \ref{table:c03_ensemble_comp}, the spectrogram ensemble achieves the highest scores, followed by the channel ensemble and then the patch size ensemble.
Noticeably, ensemble methods are clearly able to improve on the baseline performance for almost all categories, with the exception of \textit{Airport}. Specially, \textit{Tram, Shopping Mall} and \textit{Metro Stations} categories show significant improvements when applying ensembles.
For both DCASE 2018 baseline and ensemble methods, \textit{Public Square} achieves the lowest scores.
By contrast, \textit{Street Traffic} achieves the highest scores for all of the models evaluated.

\begin{figure}[t!]
    \centering
    \includegraphics[width=0.85\linewidth]{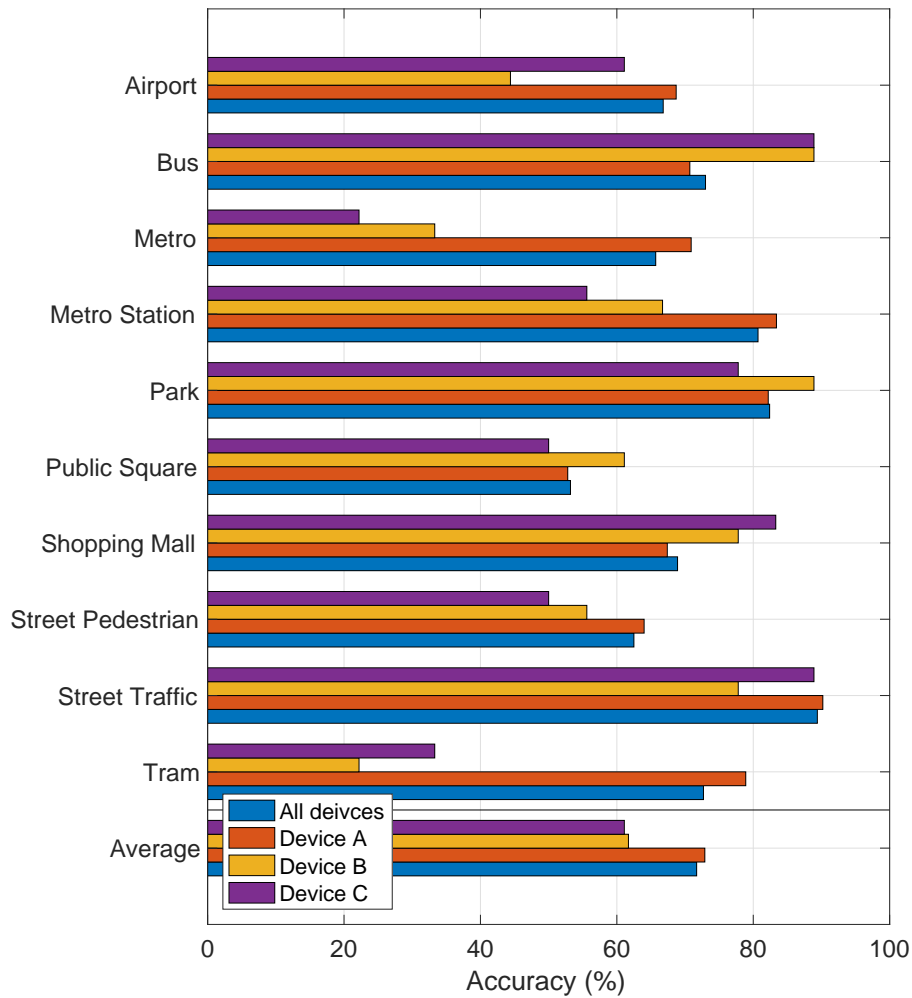}
    	\vspace{-0.2cm}
	\caption{\textit{Performance comparison (Accuracy \%) among Devices over DCASE 2018 Task 1B with spectrogram ensemble.}}
    \label{fig:c03_device}
\end{figure}

Moving on to the DCASE 2018 Task 1B which addresses the issue of mismatched recording devices, it is noted that only one channel is provided and thus channel ensemble is not possible (even though it performed well for Task 1A).
The performance of the spectrogram ensemble method is now explored in DCASE 2018 Task 1B, achieving the results shown in Figure \ref{fig:c03_device} and Table \ref{table:c03_task1b}.

From results shown in Figure \ref{fig:c03_device}, the classification performances on devices B and C are poorer than A due to unbalanced data and mismatch recorded devices, reporting an average of 72.9\%, 61.7\% and 61.1\% for devices A, B, and C, respectively.
Compared to the DCASE 2018 baseline, results in Table \ref{table:c03_task1b} show a significant improvement, increasing accuracy of B\&C by 15.8\% (note that DCASE 2018 Task 1B challenge only evaluates accuracy on device B\&C).

\subsection{Effect of Mixup Data Augmentation}
\label{c03_effect_of_mixup}
\begin{figure*}[h]
    \centering
    \includegraphics[width=\linewidth]{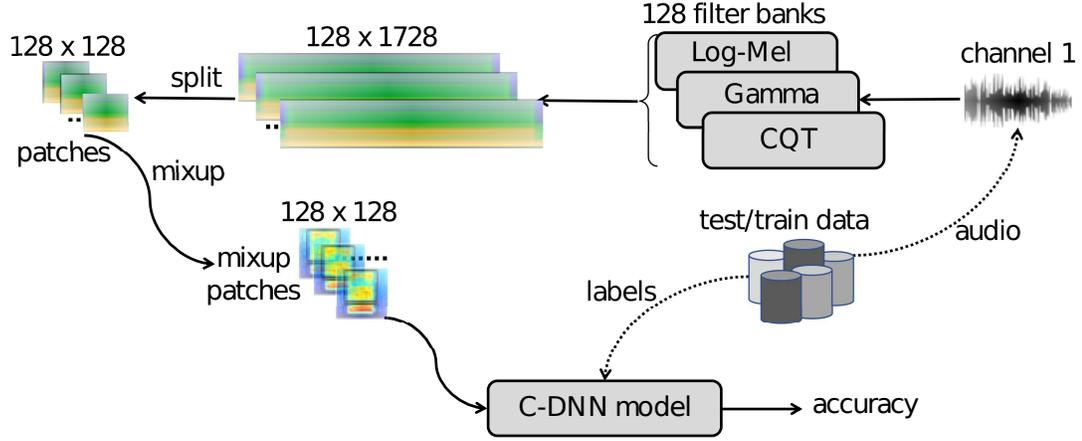}
    	\vspace{-0.5cm}
	\caption{\textit{The baseline system architecture with mixup data augmentation.}}
    \label{fig:c03_mixup}
\end{figure*}
As mentioned in Section \ref{c02_augmentation}, applying multiple input features and data augmentation are two main approaches to deal with ASC challenges in terms of low-level features.
It is fact that the recently comprehensive analysis of bag-of-features has proven that ensemble of spectrograms is effective to improve an ASC system's performance, outperform channel and time resolution features.
In this section,  the effect of augmentation affects on classification accuracy is evaluated.
In particular, Figure \ref{fig:c03_mixup} describes how to apply mixup data augmentation in the baseline system proposed.
Firstly, only channel 1 is used to transform into three types of spectrogram (log-mel, Gamma, and CQT). 
The entire spectrograms are thus split into non-overlapping image patches of $128\times128$. 
These two steps with setting parameters such as the filter number, window size or hop size, etc. are the same as those used in experiments of bag-of-features mentioned in Table \ref{table:c03_para_feature}.
Next, mixup data augmentation~\cite{aug_mixup_s01, aug_mixup_s02} is applied on the image patches. 
Let consider $\mathbf{X}_1$ and $\mathbf{X}_2$ as two image patches randomly selected from the set of original image patches with their labels $\mathbf{y}_1$ and $\mathbf{y}_2$, respectively, mixup data augmentation helps to generate new image patches as Equations below,
\begin{align}
    \mathbf{X}_{\text{mp1}} &= \alpha\mathbf{X}_{1} + (1-\alpha)\mathbf{X}_{2}, \label{eq:mix_up_x1} \\
    \mathbf{X}_{\text{mp2}} &= (1-\alpha)\mathbf{X}_{1} + \alpha\mathbf{X}_{2},     \label{eq:mix_up_x2} \\
    \mathbf{y}_{\text{mp1}} &= \alpha\mathbf{y}_{1} + (1-\alpha)\mathbf{y}_{2}, \label{eq:mix_up_y} \\
    \mathbf{y}_{\text{mp2}} &= (1-\alpha)\mathbf{y}_{1} + \alpha\mathbf{y}_{2}. \label{eq:mix_up_y}
\end{align}
where $\alpha$ is drawn from both Uniform or Beta Distribution, $\mathbf{X}_{\text{mp1}}$ and $\mathbf{X}_{\text{mp2}}$ are two new image patches resulted by mixing $\mathbf{X}_1$ and $\mathbf{X}_2$ with a random mixing coefficient $\alpha$. Similarly, $\mathbf{y}_{\text{mp1}}$ and $\mathbf{y}_{\text{mp2}}$ are two new labels resulted by mixing $\mathbf{y}_1$ and $\mathbf{y}_2$.
After mixup, old data and generated data from mixup data augmentation are shuffled and fed into C-DNN baseline proposed, double batch size and consider learning time.
Because of using multiple-spectrogram input features (log-mel, Gamma, and CQT), effect of mixup data augmentation on individual and ensemble spectrogram is evaluated.
\begin{table}[t!]
    \caption{\textit{Effect of mixup data augmentation on individual and \\ ensemble spectrograms, evaluated on C-DNN baseline and DCASE 2018 Task 1A \\ and 1B dataset (only devices B \& C in  DCASE 2018 Task 1B dataset is reported).}} 
    \vspace{-0.1cm}
    \centering
    \scalebox{0.85}{
    \begin{tabular}{|c |c |c |c |c|}
       \hline
	   \textbf{C-DNN} 	           &\textbf{Task 1A (\%)}   &\textbf{Task 1A (\%)}   &\textbf{Task 1B (\%)}   &\textbf{Task 1B (\%)}          \\ 
			      &(w/o mixup)                    &(w/ mixup)                    &(w/o mixup)           &(w/ mixup)                   \\ 
       \hline                                                                                         
	    Gamma            &67.3              &68.3               &55.5             &58.9                      \\
	    log-mel        &66.2              &67.8               &56.4              &59.4                     \\
	    CQT            &59.5              &60.2               &51.7              &51.4                      \\
       \hline                                                                                         
	    \textbf{Ensemble}       &\textbf{73.1}     &\textbf{74.0}      &\textbf{61.4}     &\textbf{66.9}             \\
       \hline 
    \end{tabular}    
    }
    \label{table:c03_mixup_res} 
\end{table}

\begin{table}[t]
    \caption{\textit{Performance comparison of the proposed system \\ (multiple-spectrogram low-level features, mixup data augmentation, \\ C-DNN ensemble) to top-ten DCASE 2018 challenge.} }
   \vspace{-0.1cm}
    \centering
    \scalebox{0.85}{
    \begin{tabular}{|l |c | l |c |} 
    \hline

           \textbf{DCASAE 2018 1A}              &\textbf{Acc. (\%)}        &\textbf{DCASE 2018 1B}                  &\textbf{Acc. (\%)}  \\
    \hline
           Li~\cite{dc_18_t10}           &$72.9$               &Baseline~\cite{dc_18_bsl}         &$45.6$           \\ 
           Jung~\cite{dc_18_t09}         &$73.5$               &Li~\cite{dc_18_tb07}              &$51.7$           \\ 
           Hao~\cite{dc_18_t08}          &$73.6$               &Tchorz~\cite{dc_18_tb06}          &$53.9$           \\ 
           Christian~\cite{dc_18_t07}    &$74.7$               &Kong~\cite{dc_18_tb05}            &$57.5$           \\ 
           Zhang~\cite{dc_18_t06}        &$75.3$               &Wang~\cite{dc_18_tb04}            &$57.5$           \\ 
           Li~\cite{dc_18_t05}           &$76.6$               &Waldekar~\cite{dc_18_tb03}        &$57.8$           \\ 
           Dang~\cite{dc_18_t04}         &$76.7$               &Zhao~\cite{dc_18_zhao_dcase}      &$58.3$           \\ 
           Octave~\cite{dc_18_oct_dcase} &$78.4$               &Truc~\cite{dc_18_tb01}            &$63.6$           \\ 
           Yang~\cite{dc_18_yang_dcase}  &$79.8$               &                                  &                 \\ 
           Golubkov~\cite{dc_18_t01}     &$\textbf{80.1}$      &                                  &                 \\ 
    \hline
	    Proposed system               &$74.0$ &Proposed system  & $\textbf{66.9}$ \\
	        \hline

    \end{tabular}  }  
    \label{table:c03_compare_challenge} 
\end{table}

By applying mixup data augmentation technique, the new labels $\mathbf{y}_{\text{mp1}}$ and $\mathbf{y}_{\text{mp2}}$ of the two mixup patches are no longer one-hot labels, Kullback-Leibler (KL) divergence loss~\cite{kl_loss} rather than the standard cross-entropy loss is used as shown in Equation below,
\begin{equation}
    \label{eq:loss_func}
    LOSS_{KL}(\Theta) = \sum_{c=1}^{C}y_{c}\log \left(  \frac{y_{c}}{\hat{y}_{c}(\Theta)} \right) +  \frac{\lambda}{2}||\Theta||_{2}^{2},
\end{equation}
where \(LOSS_{KL}(\Theta)\) is KL loss function, $\Theta$ denotes the trainable network parameters and $\lambda$ denotes the $\ell_2$-norm regularization coefficient, set to 0.001.
 \(y_{c}\) and \(\hat{y}_{c}\) denote the ground truth and the network output at category $c$, respectively.
Other hyper parameters are same as those mentioned in Section \ref{c03_setup}.
Obtained experimental results conducted on DCASE 2018 Task 1A and 1B datasets are shown in Table \ref{table:c03_mixup_res}.
It can be seen that performance of all experimental systems is improved on both datasets. 
In particular, mixup data augmentation helps to improve by 0.9\% with ensemble of spectrograms on DCASE 2018 Task 1A.
Noticeably, this technique is very effective for DCASE 2018 Task 1B when it shows an improvement of 5.5\%.

Compare the best results obtained (74.0\% and 66.9\% for DCASE 2018 Task 1A and 1B, respectively) to the top-ten DCASE 2018 challenge as shown in Table \ref{table:c03_compare_challenge}, it can be seen that C-DNN baseline model with multi-spectrogram input and using mixup data augmentation stands on the top-eight position as regards DCASE 2018 Task 1A, and outperforms DCASE 2018 Task 1B challenge.

\section{Conclusion}
\label{Conclusion}

From these experimental results obtained from the DCASE 2018 Task 1A and 1B datasets, there is a clear indication that using different spectrograms, coming from different auditory models, is effectively to improve classification accuracy. 
This improvement can also be achieved on different ASC tasks (assessed by comparing the classification performance on matched and mismatched recording devices). 
Furthermore, applying mixup data augmentation on image patches is effective to enforce learning ability of the back-end learning model, thus improve the accuracy.

\chapter{A novel Encoder-Decoder Framework}
\label{c04}

Comprehensive analysis provided in  Chapter \ref{c03} indicates that the combination of using three spectrograms of log-mel, gammatonegram (Gamma) and CQT as low-level features, along with mixup data augmentation, is effective at improving the performance of an ASC system.  
However, the results from deep learning models used in Chapter \ref{c03} show some issues of concern. 
Firstly, although an ensemble of multi-spectrogram input is useful to enhance ASC system performance, this method fuses the predicted probability of single models learned from individual spectrograms, but does not explore the interrelation between those spectrograms.
Secondly, although both single models using individual spectrograms and multi-spectrogram ensemble models were proposed in Chapter \ref{c03}, and these outperform DCASE 2018 baselines, their performance is not competitive with the most recent, state-of-the-art systems.
Furthermore, because the ASC systems proposed in Chapter \ref{c03} were only evaluated on two datasets (DCASE 2018 Task 1A and 1B), there is still not enough evidence to conclude whether the proposed learning models perform well or not in general, or whether they only work on a restricted task.

Motivated by these issues, this chapter aims to improve the back-end classification process, proposes a novel deep learning model called the \textit{Encoder-Decoder} framework, then evaluates it over a wide variety of published datasets including Litis Rouen~\cite{data_litis}, DCASE 2016 Task 1~\cite{data_dc_16}, DCASE 2017 Task 1~\cite{data_dc_17}, DCASE 2018 Task 1A~\cite{data_dc_18}, 1B, and DCASE 2019 Task 1A, 1B~\cite{data_dc_19} -- all of which will be summarised below.

Conducted experiments below obtain very good results; competitive with the best single-task systems, and far better than any previously published multi-task methods.

\section{High-level Architecture}
\label{c04_encoder_decoder}
\begin{figure*}[tb]
    \centering
    \includegraphics[width=\linewidth]{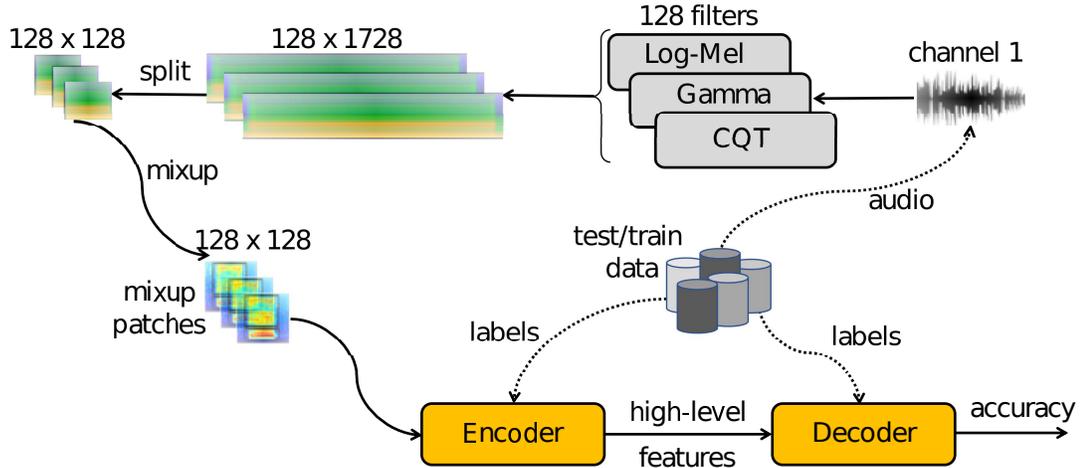}
    	\vspace{-0.7cm}
	\caption{\textit{High-level architecture of an ASC system \\ using the proposed \textit{Encoder-Decoder} framework.}}
    \label{fig:c04_architecture}
\end{figure*}

An overall ASC system using the proposed \textit{Encoder-Decoder} framework is illustrated as Figure \ref{fig:c04_architecture}.
As the comprehensive analysis of various low-level features in Chapter \ref{c03} identified the best settings to deal with ASC challenges, the \textit{Encoder-Decoder} framework only uses one channel 1 and one patch size based on that, but applies it to three spectrograms (log-mel, Gamma, and CQT) for low-level feature extraction.
Firstly, the recorded audio signal from one channel (Channel 1 - Left) is transformed into three types of spectrogram (log-mel, Gamma, and CQT) with 128 filters each.
Next, the entire spectrograms are sliced into non-overlapping patches of $128{\times}128$ before applying mixup data augmentation~\cite{aug_mixup_s01, aug_mixup_s02} as mentioned in Section~\ref{c03_effect_of_mixup}.
Patches after mixup are then fed into the \textit{Encoder} model to start the first training process.
This training process helps to map low-level features to high-level features which are vectors containing discriminative and condensed multi-dimensional information.
In  other words, the role of the \textit{Encoder} is to be a high-level feature extractor.
Next, the high-level features are extracted patch-by-patch, mixup data augmentation is again applied, and the resulting augmented dataset is used to train the \textit{Decoder} model.
The \textit{Decoder} model has the responsibility to perform final classification, and reports the classification accuracy at its output.
\section{\textit{Encoder-Decoder} Network Configuration}
\label{c04_encoder_decoder}

\subsection{\textit{Encoder} as High-level Feature Extractor}
\label{c04_encoder}
\begin{figure}[htb]
    \centering
    \includegraphics[width=0.85\linewidth]{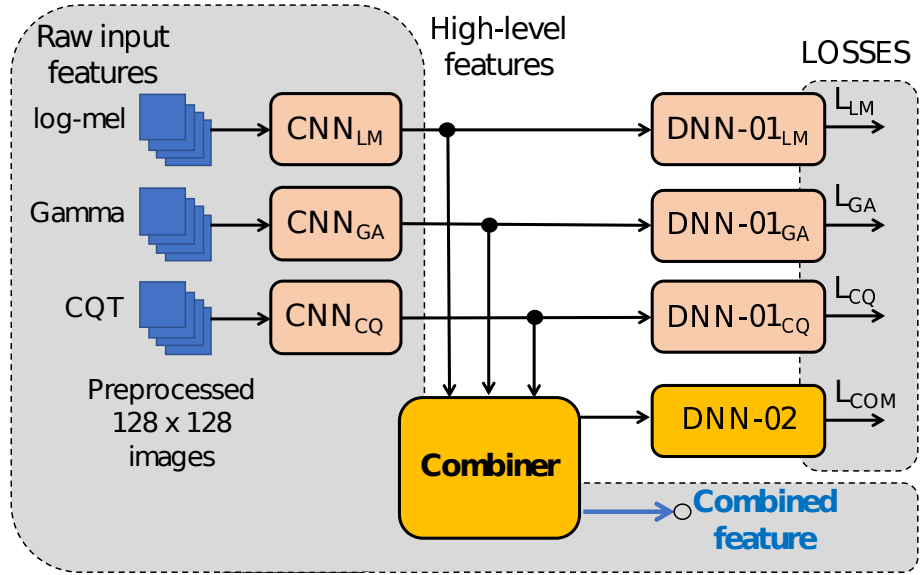}
    	\vspace{-0.1cm}
	\caption{\textit{High-level feature extraction from the \textit{Encoder} network.}}
    \label{fig:c04_encoder}
\end{figure}
%
\begin{table}[htb]
    \caption{\textit{Encoder network structures of the CNN (top),\\ DNN-01 (middle) and DNN-02 (bottom).}} 
        	\vspace{-0.2cm}
    \centering
    \scalebox{0.85}{

    \begin{tabular}{|l |c|} 
        \hline 
            \textbf{Encoder network architecture}   &  \textbf{Output}  \\
        \hline 
        \textbf{CNN$_{LM/GA/CQ}$} shares the same architecture & \\
         Input layer (image patch) & $128{\times}128$          \\
         BN - Cv [$3{\times}3$] $@ 32$ - ReLU - BN - AP [$2{\times}2$] - Dr ($10\%$)      & $64{\times}64{\times}32$\\
         BN - Cv [$3{\times}3$] $@ 64$ - ReLU - BN - AP [$2{\times}2$] - Dr ($15\%$)      & $32{\times}32{\times}64$\\
         BN - Cv [$3{\times}3$] $@ 128$ - ReLU - BN - Dr ($20\%$)      & $32{\times}32{\times}128$ \\
         BN - Cv [$3{\times}3$] $@ 128$ - ReLU - BN - AP [$2{\times}2$] - Dr ($20\%$)       & $16{\times}16{\times}128$\\
         BN - Cv [$3{\times}3$] $@ 256$ - ReLU - BN  - Dr ($25\%$)      & $16{\times}16{\times}256$ \\
         BN - Cv [$3{\times}3$] $@ 256$ - ReLU - BN -  GAP - Dr ($25\%$) & $256$ \\           
         \hline 
          \textbf{DNN$-$01$_{LM/GA/CQ}$} shares the same architecture & \\
         Input layer (vector) & $256$ \\
         FC  - Softmax   &  C         \\
         \hline
         \textbf{DNN-02} &  \\
         Input layer (vector) & $256$ \\
         FC - ReLU - Dr ($30\%$)        &  $512$       \\
         FC - ReLU - Dr ($30\%$)        &  $1024$    \\
         FC - Softmax   &  C        \\
       \hline 
    \end{tabular}
    }
    \label{table:c04_CDNN} 
\end{table}
The architecture of the \textit{Encoder} network, performing high-level feature extraction, is shown in Figure \ref{fig:c04_encoder}.
Three types of image patches of size $128{\times}128$ pixels (i.e. three types of image patches from CQT, Gamma, and log-Mel repectively), after mixup, are fed into the three parallel networks each of which comprises a CNN and a DNN-01 block, like the VGG-7 architecture~\cite{vgg_net}.
Subscripts LM, GA, and CQ are used to denote the three paths, as shown in Figure~\ref{fig:c04_encoder}, referring to the kind of spectrogram of log-Mel, Gamma, and CQT, respectively (e.g. CNN$_{LM}$ and DNN$-$01$_{LM}$ blogs in Figure \ref{fig:c04_encoder} are used for learning log-Mel input only).

The architecture of the CNN$_{LM/GA/CQ}$ and DNN$-$01$_{LM/GA/CQ}$ blocks are described in the upper and middle sections of Table \ref{table:c04_CDNN}, resepective.
As regards the CNN$_{LM/GA/CQ}$, they share the same architecture, which comprises six layers employing sub-blocks of batch normalization (BN), convolutional (Cv [kenel size] $@$ kernel number), rectified linear units (ReLU), average pooling (AP [kernel size]), global average pooling (GAP), dropout (Dr(\%)).
The DNN$-$01$_{LM/GA/CQ}$ blocks also share the same architecture, which performs fully connected (FC), and Softmax layers, with dimensions given in Table \ref{table:c04_CDNN}.
The number of categories within the given dataset is denoted by ``C''; this depends on the particular evaluation task.

The three parallel networks, each of which is configured to contain a CNN$_{LM/GA/CQ}$ and DNN$-$01$_{LM/GA/CQ}$, are used to learn and extract high-level features from one type of spectrogram for each.
While the structures of these three CNN$_{LM/GA/CQ}$ as well as the three DNN$-$01$_{LM/GA/CQ}$ blocks are identical, they will contain very different weights (trainable parameters) after training due to their different spectrogram input.

The output of each of the CNN$_{LM/GA/CQ}$ block shown in the upper part of Table \ref{table:c04_CDNN} is a 256-dimensional vector.
The vector extracted from each individual spectrogram is referred as to a high-level feature.
To combine the three high-level features, which are 256-dimensional vectors independently extracted from three parralell network streams, into a single combined feature, the ``Combiner'' block, as shown in  Figure~\ref{fig:c04_encoder}, is proposed.
There are three methods to combine the high-level features, which are evaluated.
The vector outputs of the CNN blocks are denoted as $\mathbf{x_{LM/GA/CQ}}$ [$x_{1}, x_{2}, ..., x_{256}$]. 
The first combination method, called ``sum-comb'', is the unweighted sum of the three vectors. i.e. the individual vectors contribute equally to the combined high-level feature,
\begin{equation}
    \label{eq:sum_com}
     \mathbf{x_{sum-comb}} =  \mathbf{x_{LM}} + \mathbf{x_{GA}} + \mathbf{x_{CQ}}
\end{equation}
The second method, which is called ``max-comb'', obtains $\mathbf{x_{max-comb}}[x_{1}, x_{2}, ..., x_{256}]$ by selecting the element-wise maximum of the three vectors across the dimensions as in Equation (\ref{eq:max_com}). 
The motivation is to pick the most important (highest magnitude) feature from among the three high level feature vectors,
\begin{eqnarray}
    \label{eq:max_com}
     \mathbf{x_{max-comb}}[x_{i}] =   max(\mathbf{x_{LM}}[x_{i}], \mathbf{x_{GA}}[x_{i}], \mathbf{x_{CQ}}[x_{i}]) ~~~  for  ~~ 1 \leq i \leq 256 
\end{eqnarray}
For the final method, it is assumed that elements of three vectors have a linear relationship across dimensions. Then, a simple data-driven combination method called ``lin-comb'' is proposed by employing a fully connected layer trained to weight and combine the three high level features, as in
\begin{eqnarray}
    \label{eq:lin_com}
     \mathbf{x_{lin-comb}} =  Relu \left\{ \mathbf{x_{LM}w_{LM}} + \mathbf{x_{GA}w_{GA} + x_{CQ}w_{CQ} + w_{bias}}  \right\}
\end{eqnarray}
where $\mathbf{w_{LM/GA/CQ/bias}}$[$w_{1}, w_{2}, ..., w_{256}$] are the trained parameters. 
The combined high level feature vector from the output of the ``Combiner'' block is then fed into DNN-02, with the structure shown in the lower part of Table \ref{table:c04_CDNN}. 
Note that the combined high level feature vectors, like the individual high level vectors, have a dimension of 256 -- meaning that the higher layer classifier of the decoder can be set for evaluation with either individual or combined feature input, without changing its structure or complexity.

Regarding training loss, four loss functions to train the encoder network are defined; three to optimize individual spectrograms, and the final one for their combination.
Eventually, the overall loss function $LOSSES$ is computed as
\begin{equation}
    \label{eq:loss_func}
	LOSSES =  \alpha(L_{LM} + L_{GA} + L_{CQ}) + {\beta}L_{com}
\end{equation}
and \(L_{LM}, L_{GA}, L_{CQ}$ and $L_{com}\) are individual losses from the log-mel, Gamma and CQT spectrograms, and their combinations. These are depicted from Figure~\ref{fig:c04_encoder} and will be defined in Section~\ref{c03_feature_analysis}.
The balancing parameters \(\alpha\) and \(\beta\) focus on learning particular features or combinations and are set to $1/3$ and $1.0$ here, making the contributions from each spectrogram equal.

After training the \textit{Encoder} network, the combined feature (i.e. the combined feature is also the input of DNN-02 block in Figure \ref{fig:c04_encoder}) is extracted. Then, it is fed into the \textit{Decoder} described below for classification.

\subsection{\textit{Decoders} for Back-end Classification}
\label{c04_decoder}
\begin{table}[th]
    \caption{\textit{MLP-based architecture of \textit{Decoder} network.}} 
        	\vspace{-0.2cm}
    \centering
    \scalebox{0.85}{

    \begin{tabular}{|l |c |} 
        \hline 
            \textbf{Network architecture}   &  \textbf{Output}  \\
        \hline 
         Input layer (vector) & $256$ \\
         FC - ReLU - Dr ($30\%$)        &  $512$       \\
         FC - ReLU - Dr ($30\%$)        &  $1024$    \\
         FC - ReLU - Dr ($30\%$)        &  $1024$    \\
         FC - Softmax   &  C        \\
       \hline 
    \end{tabular}
    }
    \label{table:c04_MLP} 
\end{table}
As regards the baseline architecture proposed in Chapter \ref{c03}, fully connected layers with the final Softmax layer takes the role of classification.
In the \textit{Encoder-Decoder} framework proposed in this chapter, the \textit{Decoder} is responsible for this role and receives combined high-level feature vectors extracted from \textit{Encoder} as its input (note that mixup data augmentation is applied on these feature before feeding into \textit{Decoder} network during training).
There are three types of \textit{Decoder} evaluated: A random forest classification (RFC) with classifier, a Multilayer Perceptron (MLP), and a Mixture of Experts (MoE), described below, 

\textbf{a) Random Forest Classification (RFC \textit{Decoder})}: 
A regression forest~\cite{rd_fr} is a type of ensemble model, comprising multiple regression trees. 
The role of each tree is to map the complex input space defined by the high level features from the encoder network, into a continuous class-dimension output space. 
Its nonlinear mapping is achieved by dividing the large original input space into smaller sub-distributions. 
Individual trees are trained using a subset randomly drawn from the original training set. 
By using many trees (e.g. $100$), the structure is effective at tackling overfitting issues that can occur with single trees.
Additionally, the regressor structure benefits from the continuous mixed-class training labels provided by employing mixup.
Eventually, the decoded output spaces are classified as in our previous work~\cite{huy_lit_acm} by average pooling the output over all trees. 

\textbf{b) Multilayer Perceptron Network  (MLP \textit{Decoder})}:
The proposed MLP \textit{Decoder} comprises four fully connected dense blocks as shown in Table \ref{table:c04_MLP}.
Comparing the MLP \textit{Decoder} architecture to DNN-02 block used in \textit{Encoder}, one more fully connected layer with 1024 nodes is added -- this is to handle the additional complexity of the input information.

\begin{figure}[t]
    \centering
    \includegraphics[width=0.85\linewidth]{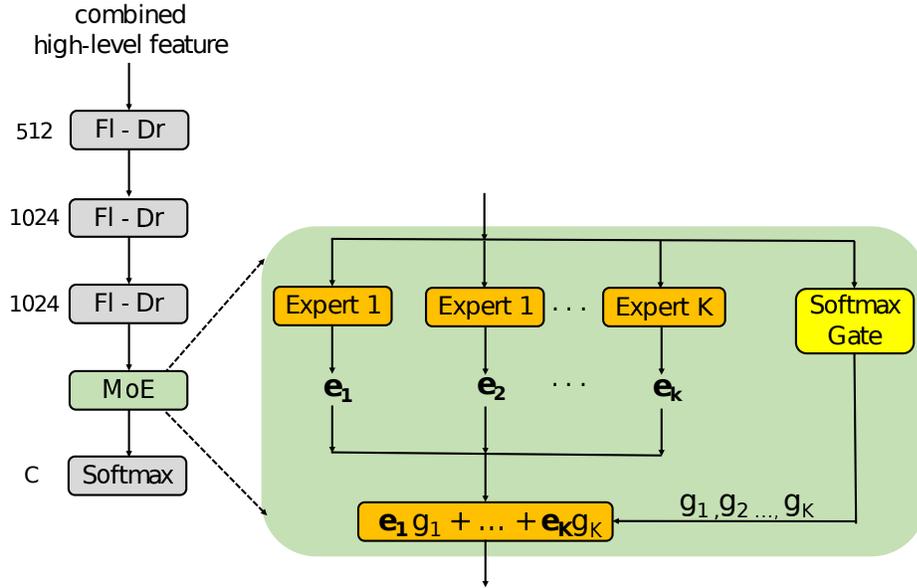}
    	\vspace{-0.3cm}
	\caption{\textit{Proposed mixture of experts (MoE) as back-end decoder network.}}
    \label{fig:c04_moe}
        	\vspace{-2mm}
\end{figure}

\textbf{c) Mixture of Experts (MoE \textit{Decoder})}: MoE is a machine learning technique that divides the problem spaces into homogeneous regions by using an array of different trained (but in this case identical structure) models, referred to as experts~\cite{moe}.
A conventional MoE architecture comprises many experts and incorporates a gate network to decide which expert is applied in which input region.
The MoE technique is used to classify the combined high-level features, as shown in Figure \ref{fig:c04_moe}.
Specifically, the 256-dimensional input vector goes through three dense layers with dropout, having 512, 1024, and 1024 hidden nodes, respectively, matching MLP \textit{Decoder} in the number of hidden units.
The output enters the MoE layer, which is explained in Figure \ref{fig:c04_moe}.
The combined result from the experts is gated before passing through a Softmax layer to determine the final $C$ class scores.
Each MoE expert comprises a dense block with a Relu activation function. Its input dimension is 1024 and its output size is $C$.
The gate network is implemented as a Softmax gate --  an additional fully connected layer with Softmax activation function and a gating dimension equal to the number of experts.

If  $\mathbf{e_{1}, e_{2}}, \dots  \mathbf{e_{K}} \in \mathbb{R^{C}}$ is considered as the output vectors of the $K$ experts, and  $g_{1}, g_{2}, \dots , g_{K}$ as the outputs of the gate network where $g_k \in [0, 1], \sum_{k=1}^K{g_k}=1$
The predicted output is then defined as,
\begin{equation}
    \label{eq:moe}
    \mathbf{\hat{y}} = softmax \left\{ \sum_{k=1}^{K} \mathbf{e_{k}}g_{k} \right \}.
\end{equation}
\section{Experiment Setup}
\label{c04_setup}

\subsection{Dataset}
\label{c04_dataset}

To clearly demonstrate the general performance of the proposed systems, five different ASC tasks are used for the evaluation.
While it is relatively easy to perform well in one challenge, it is considerably more difficult to do so for all -- this helps to explore one of the hypothesised strengths of this proposed combined-spectrogram approach, that it can be more generic.
Four of the datasets used are derived from annual DCASE challenges (DCASE 2016 Task 1, DCASE 2017 Task 1, DCASE 2018 Task 1A and 1B, DCASE 2019 Task 1A and 1B), whereas the fifth is the extensive LITIS Rouen dataset. 
Each is described below. 

\textbf{DCASE 2016 Task 1A and DCASE 2017 Task 1A}: 
Firstly, DCASE 2016 Task 1 dataset~\cite{data_dc_16} as shown in Table \ref{table:DCASE2016_dataset} were recorded at a sample frequency at 44100 \,Hz with a 30-second recording duration for every audio file. 
The data is subdivided into two sets; a development set (Dev. Set) and an evaluation set (Eva. Set), one for training and another for evaluating, with 15 categories as described in detailed in Table \ref{table:DCASE2016_dataset}. 
In total, the development and evaluation sets comprise 13 hours of data.
As regards DCASE 2017 Task 1 dataset as shown in Table \ref{table:DCASE2016_dataset} ~\cite{data_dc_17}, it reuses all DCASE 2016 dataset. 
In particular, each 30-second segment from the DCASE 2016 dataset was split into three 10-second segments used in DCASE 2017 dataset.
Besides, more 10-second audio segments were recorded and included, which create a total of 17.5 hours for both development and evaluation sets.
Similar to DCASE 2016 settings, while the development set (Dev. Set) is used to train the model, the evaluation set (Eva. Set) is for evaluating.
Both DCASE 2016 and DCASE 2017 contain balanced data, and each challenge has 15 categories. 
\begin{table}[t]
    \caption{\textit{Development and Evaluation Sets of \\ DCASE 2016 Task 1 and DCASE 2017 Task 1 Datasets.}} 
            	\vspace{-0.2cm}
    \centering
    \scalebox{0.85}{
    \begin{tabular}{|l |c |c |c |c|} 
        \hline
                \textbf{Categories}                                        &\textbf{DCASE 2016}      & \textbf{DCASE 2016}  &\textbf{DCASE 2017}      & \textbf{DCASE 2017} \\ [0.5ex] 

                                               &\textbf{Dev. Set}      & \textbf{Eva. Set}  &\textbf{Dev. Set}      & \textbf{Eva. Set} \\ [0.5ex] 
        \hline
        Beach   &78   &26  &312   &108 \\
        Bus   &78   &26 &312   &108\\
        Cafe/Restaurant   &78   &26 &312   &108\\
        Car   &78   &26 &312   &108\\
        City center   &78   &26 &312   &108\\
        Forest Path   &78   &26 &312   &108\\
        Grocery Store   &78   &26 &312   &108\\
        Home   &78   &26 &312   &108\\
        Library   &78   &26 &312   &108\\
        Metro station   &78   &26 &312   &108\\
        Office   &78   &26 &312   &108\\
        Park   &78   &26 &312   &108\\
        Residential area   &78   &26 &312   &108\\
        Train   &78   &26 &312   &108\\
        Tram   &78   &26 &312   &108\\
        \hline
    \end{tabular}    
    }
    \label{table:DCASE2016_dataset} 
\end{table}
\begin{table}[t]
    \caption{\textit{Development and Evaluation set of DCASE 2019 Task 1A and 1B datasets.}} 
            	\vspace{-0.2cm}
    \centering
    \scalebox{0.85}{
    \begin{tabular}{|l |c |c |c |c|} 
        \hline 
	    \textbf{Categories}  & \textbf{Dev. Set} & \textbf{Eva. Set} & \textbf{Dev. Set} & \textbf{Eva. Set} \\ [0.5ex] 
	               &  \textbf{(1A)}      & \textbf{(1A)} & \textbf{(1B)} & \textbf{(1B)} \\
        \hline 
        Airport                  &911                &421    &1019	                &529   \\
        Bus                      &928                &415    &1036	                &523   \\
        Metro                    &902                &433    &1010	                &541   \\
        Metro Station            &897                &435    &1005	                &543   \\
        Park                     &946                &386    &1054	                &494   \\
        Public Station           &945                &387    &1053	                &495   \\
        Shopping Mall            &896                &441    &1004	                &549   \\
        Pedestrian Street        &924                &429    &1032	                &537   \\
        Traffic Street           &942                &402    &1050	                &510   \\
        Tram                     &894                &436    &1002	                &544   \\
        \hline 
        \textbf{Total files}              & \textbf{9185}              & \textbf{4185}  &\textbf{10265}               &\textbf{5265}  \\
        \hline 

    \end{tabular} }
    \label{table:DCASE2019_dataset} 
\end{table}

\textbf{DCASE 2018 Task 1A, 1B and DCASE 2019 Task 1A, 1B}: 
As DCASE 2018 Task 1A and 1B challenges~\cite{data_dc_18} have not released their evaluation sets, only the development sets are explored in this Chapter.
Description and setting evaluation for DCASE 2018 Task 1A and 1B development set are similar and were introduced in previously Chapter \ref{c03}.
Regarding DCASE 2019 Task 1A and 1B datasets~\cite{data_dc_19}, these reuses DCASE 2018 Task 1A and Task 1B data, but incorporates additional audio segments.
Therefore, the recording files in DCASE 2018 and DCASE 2019 challenges have similar formats as well as the same number of categories, as shown in Table \ref{table:DCASE2019_dataset}.
The proposed \textit{Encoder-Decoder} framework described above was submitted to compete in the DCASE 2019 challenge, so this thesis also reports the results over the evaluation set via the DCASE 2019 competition, even through this evaluation dataset has not been released publicly yet. 

\textbf{Litis-Rouen dataset:}
This extensive dataset~\cite{data_litis}, as shown in Table \ref{table:Litis-Rouen_dataset}, comprises 19 urban scene classes with 3026 segments, divided into 20 training/testing splits. 
The audio was recorded at a sample rate of 22050\,Hz, with each segment duration of 30 seconds.  
Following the mandated settings, the dataset is separated and organised for 20 times cross validation, reporting the final classification accuracy by averaging over the 20 testing folds.

\begin{table}[tb]
    \caption{\textit{Litis-Rouen Dataset.}} 
       	\vspace{-0.2cm}
    \centering
    \scalebox{0.85}{    
    \begin{tabular}{|l |c|} 
        \hline 
	    \textbf{Categories}  & \textbf{Segment No.} \\ [0.5ex] 
        \hline 
          Plane	                  &  192                   \\
          Busy Street        	  &  143                    \\
          Café                    &  120                   \\
          Car	                  &  243                   \\
          Hall Gate                & 269                     \\
          Kid Game                 & 145                    \\
          Market	          &  276                   \\
          Metro Pari               & 139                   \\
          Metro Rouen        	  &  249                  \\
          Pedestriant Street	  &  122                  \\
          Plane	                 &   23                    \\
          Pool	                  &  155                   \\
          Quite Street	         &   90                   \\
          Restaurant	          &  133                    \\
          Shop	                  &  203                   \\
          Student Hall        	 &   88                   \\
          Train High Speed	   & 147                   \\
          Train Normal      	  &  164                  \\
          Tube	                  &  125                  \\
	  \hline 
        \textbf{Total files}              & \textbf{3026}  \\
        \hline 

    \end{tabular} 
    }
    \label{table:Litis-Rouen_dataset} 
\end{table}

\subsection{Setting Hyperparametes and Training Process}
\label{ssec:c04_hyper}
The Tensorflow framework is used, and the Kullback-Leibler (KL) divergence loss~\cite{kl_loss} as in Equation (\ref{eq:c04_loss_func}) is applied to all of the proposed networks. This is a common loss function for training ASC systems, typically obtaining good performance.
\begin{align}
    \label{eq:c04_loss_func}
    LOSS_{KL}(\Theta) = \sum_{c=1}^{C}y_{c}\log \left\{ \frac{y_{c}}{\hat{y}_{c}(\Theta)} \right\}  +  \frac{\lambda}{2}||\Theta||_{2}^{2},
\end{align}
where \(LOSS_{KL}(\Theta)\) is the KL loss function, $\Theta$ denotes the trainable network parameters and $\lambda$ denote the $\ell_2$-norm regularization coefficient, set to 0.001. \(C\) is the class number.
$y_{c}$ and $\hat{y}_{c}$  denote the ground truth and network output at class $c$, respectively.
Experiments use the Adam optimiser~\cite{kingma2014adam} to adjust learning rate, with a batch size of $50$. 
Results were obtained after $100$ epochs (in practice only a small degree of performance was lost by not continuing beyond this, but it helped significantly to reduce the duration of experiments).
Trainable parameters are initialised by a Normal Distribution with mean and variance set to 0 and 0.1, respectively.
As aforementioned, mixup data augmentation is applied to enhance the training processes.
As regards the training process on the \textit{Encoder}, each of the raw $128{\times}128$ dimensional feature was repeated twice by including same-dimension Beta and Uniform Distribution mixup images of the same dimension.
It is similar when training the \textit{Decoders} where mixup is applied on the high-level feature vectors prior to the final classifier. 
In each case, both original and generated mixup data are used in the training processes to improve performance, at the cost of increasing the training time.

\section{Experimental Results and Comparison}
\label{c04_results}
In this section the performance of the \textit{Encoder} network is firstly analysed to specifically understand the contribution made by different spectrogram types, as well as their combinations. 
The performance of the decoder, thus, is evaluated to assess different back-end classifiers, then the overall performance is compared to a range of state-of-the art methods.
\subsection{Performance of Each Spectrogram by Class}
\label{c04_result_spec}

\begin{figure}[t]
    \centering
    \includegraphics[width=0.85\linewidth]{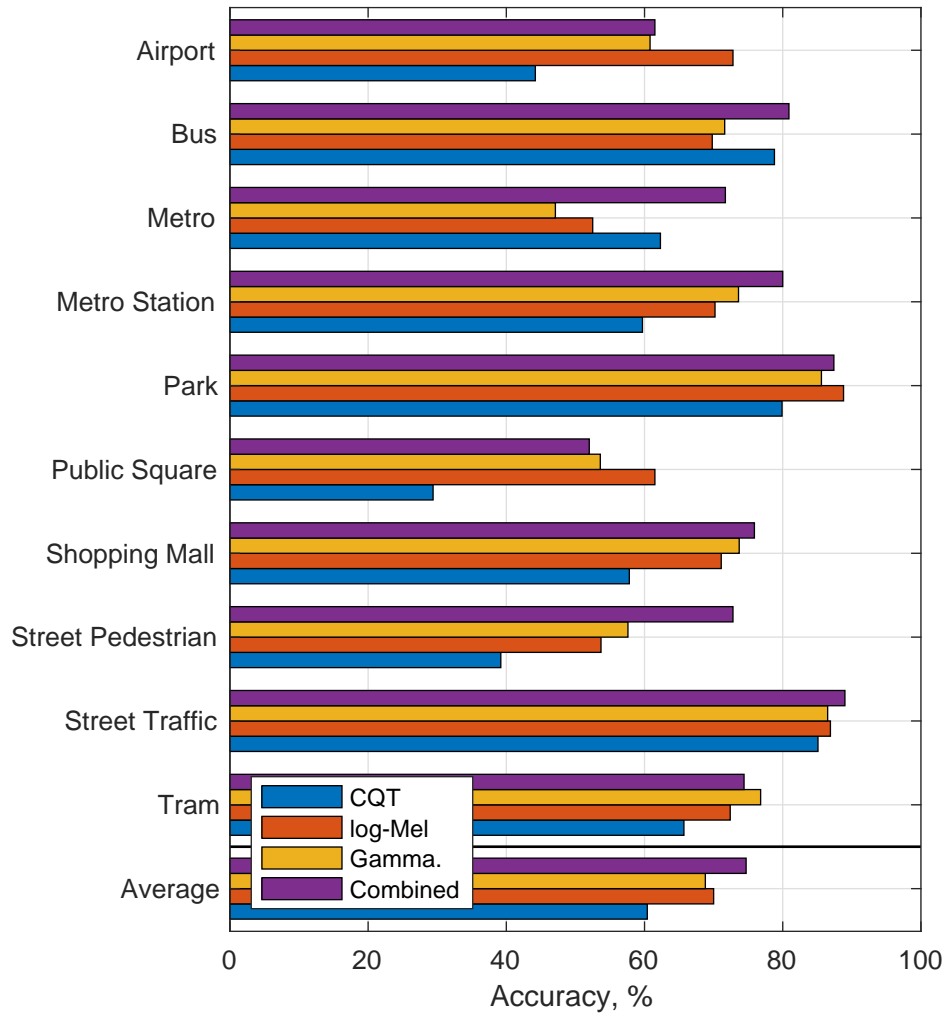}
    	\vspace{-0.1cm}
	\caption{\textit{Performance comparison of different spectrograms types, \\ and their combination, for the DCASE 2018 Task 1B Dev. set.}}
	    	\vspace{-0.2cm}
    \label{fig:Z4}
\end{figure}

Firstly, a baseline architecture is proposed and evaluated to determine how different spectrogram types contributed to the performance of different classes.
To do this, three C-DNN \textit{Encoder} networks, comprising CNN and DNN-02 blocks, each \textit{Encoder} for an individual spectrogram input, are trained.
Meanwhile, another C-DNN encoder network, the entire network as in Figure \ref{fig:c04_encoder} for spectrogram combination, is also trained.
These four trained systems are subsequently used as high-level feature extractors to train the \textit{Decoder} and then to test the overall system. 
Four different \textit{Encoders}, using the MLP \textit{Decoder} architecture from Section~\ref{c04_decoder} to assess individual spectrogram performance are used to combine with four \textit{Encoders}.
These extensive experiments were conducted using the DCASE 2018 Task 1B Dev. set.
To compare performance, class-wise accuracies for the three spectrograms and their combinations are shown in Figure \ref{fig:Z4}, with overall average performance shown at the bottom.
Clearly, the combined features performed best overall, with the log-mel and Gamma performing similarly, and both being better than CQT.
However, a glance at the per-class accuracy shows some interesting variation. 
For example, the CQT spectrogram was particularly good at discriminating the \textit{Bus} and \textit{Metro} classes, compared to the other spectrograms.
Also, while log-mel and Gamma performances were similar, the former excelled on \textit{Airport} and \textit{Public Square} classes, whereas the latter tended to be slightly better for classes containing vehicular sounds (with the exception of the \textit{Metro} class).
It can be concluded that the three spectrograms represent sounds in  ways that have affinity for certain types of sounds (mirroring a conclusion in~\cite{ivm_jr}, albeit on very different types of sound data). It is therefore unsurprising that intelligently combining the three spectrograms into a high-level feature vector can achieve significant performance gaining over single spectrograms.

\subsection{Spectrogram Performance for Each Device}
\label{c04_result_device}
\begin{figure}[h]
    \centering
    \includegraphics[width=0.85\linewidth]{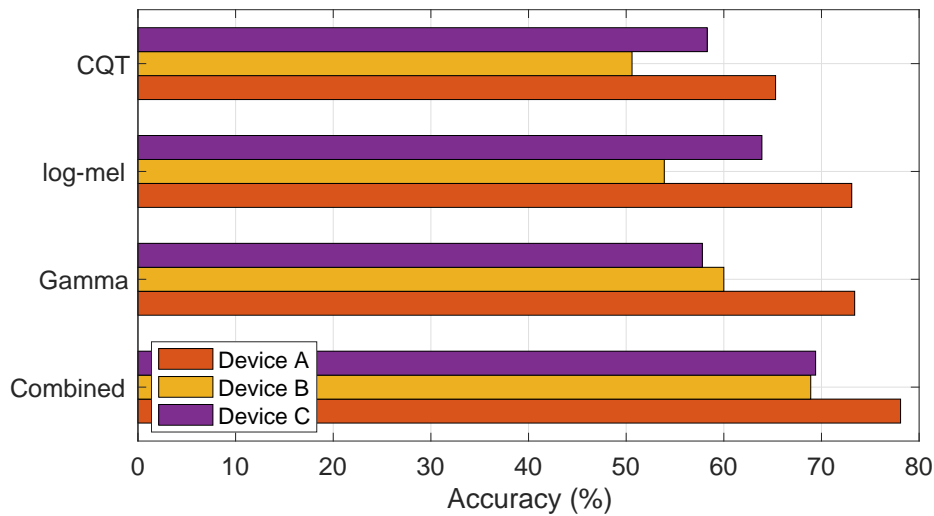}
    	\vspace{-0.2cm}
	\caption{\textit{Performance comparison for different recording \\ devices within the DCASE 2018 Task 1B Dev. set.}}
    \label{fig:Z3}
\end{figure}

DCASE 2018 Task 1B  includes highly unbalanced data recordings from three different devices as described in Section~\ref{c03_dataset}.
The performance of different spectrograms for those three devices is analysed next, results reported as plot in Figure \ref{fig:Z3}.
The device with the largest amount of training data (Device A) obviously scored best, achieving the accuracy around 9.0\% better than devices B and C.
Again, the Gamma and log-mel results were similar, but each `preferred' a different minority device. 
Although there were not enough devices included in the test for the evidence to be conclusive, this variability indicates that spectrograms differ in their affinity for different devices (or device locations, or channels). Again, the combined features effectively leveraged the advantages of each spectrogram type.

\subsection{Spectrogram Performance by Segment Length}
\label{c04_result_length}
\begin{figure}[h]
    \centering
    \includegraphics[width=0.85\linewidth]{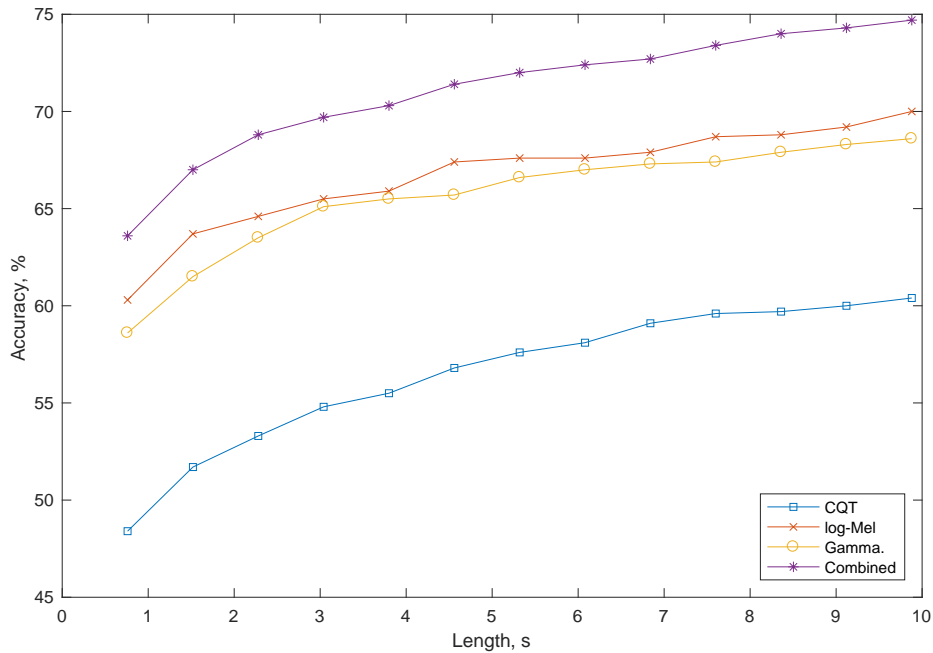}
    	\vspace{-0.2cm}
	\caption{\textit{Classification performance as a function of the length of \\ the test signal (second - s) over DCASE 2018 Task 1B Dev. set - all devices.}}
    \label{fig:Z5}
\end{figure}
%
\begin{figure}[h]
    \centering
    \includegraphics[width=0.85\linewidth]{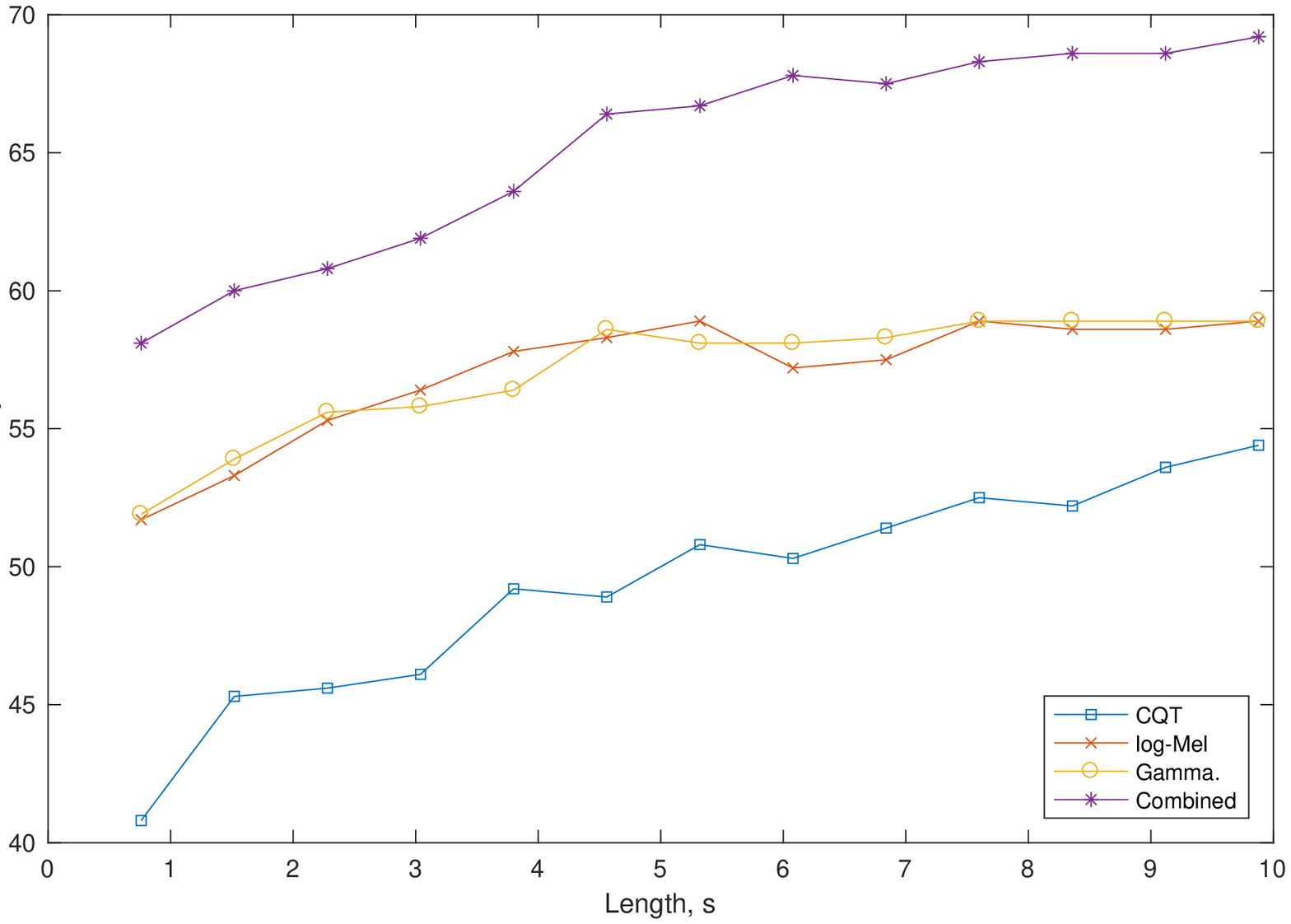}
    	\vspace{-0.0 cm}
	\caption{\textit{Classification performance as a function of the length of \\ the test signal (second - s) over DCASE 2018 Task 1B Dev. set - devices B\&C.}}
    \label{fig:Z6}
\end{figure}

Inspired by some recent research considering the ability of systems to recognise a sound class early (or using partial data)~\cite{huy_early_2018,ivm_early_2018}, this ability for the different spectrogram types is also evaluated.
Figures \ref{fig:Z5} and \ref{fig:Z6} plot early classification accuracy for DCASE 2018 Task 1B for all devices and for devices B$+$C, respectively. 
Early classification means that class assignment is only performed on the first part of the audio recording, rather than the entire duration (i.e. on cropped audio). Performance is plotted for a number of cropped segment lengths between 1 second and the full 10 seconds.
From both plots, immediate observations are that the combined high-level features performed much better than the individual spectrogram types. 
The CQT performed worst while the other two spectrograms had similar performance (as in the experiments above).
Looking closer at Figure~\ref{fig:Z5} (accuracy for all devices), the score for all features continued to climb as duration progressed towards the full 10 seconds. 
This provides a strong indication that the system was data-constrained and is likely to perform better with longer duration recordings.
By contrast, Figure~\ref{fig:Z6} contains indications that the performance of the log-mel and Gamma spectrograms began to plateau as duration exceeded to 5 seconds, indicating that performance might not substantially increase if longer duration recordings were available. 
However the continued improvement of the CQT representation as  length increased gave the combined features an ability to gain higher accuracy from longer recordings: The strength of CQT may lie in the analysis of longer recordings.
However, in these experiments, CQT performance lagged the combined features by around 15.0\% absolute, with the other spectrograms lagging by only around 5.0\% absolute -- apart from the area in Figure \ref{fig:Z6} where they plateaued.
Most remarkable is the one with just 2 seconds of input data from a recording, our proposed combined high-level feature was able to match and outperform any of the individual spectrograms operating with the full 10 seconds of input data.
This clearly demonstrates a major advantage of the proposed system. 
It effectively captures the advantages of the individual spectrogram features, which vary in their affinity for different classes and devices, and yields extremely good performance even when a restricted amount of data is available for classification.
\subsection{Performance of Different Classifiers in The \textit{Decoder}}
\label{c04_result_decoder}
Three methods were proposed in Section~\ref{c04_decoder} to incorporate the three high-level spectrogram features into a combined high-level feature in the \textit{Encoders} network.
These methods were namely ``sum-comb'', ``max-comb'' and ``lin-comb''. 
To make use of the combined features, three back-end classifier methods for the \textit{Decoders} block are introduced, namely RFC \textit{Decoder}, MLP \textit{Decoder} and MoE \textit{Decoder} in Section~\ref{c04_decoder}.
In total, the three classifiers and three combiners yield 9 models to evaluate.
In this section, performance among these 9 models are compared, evaluated on the DCASE 2018 Task 1B Dev. dataset.
It is noted that the accuracy of the \textit{Encoders} network (i.e. the feature extractor, alone) and the full system accuracy (i.e. incorporating the decoder) are separately reported.
Results are presented in Table \ref{table:c04_re_model}, again split into Device A and Devices B \& C performance.
Best performance for both device sets, highlighted in bold, was achieved by the MoE \textit{Decoder} classifier with the ``lin-comb'' combiner. 
However some interesting trends were evident. 
Firstly, MLP \textit{Decoder} was only very slightly inferior to MoE \textit{Decoder} for all combiners and device types.
Secondly, looking at the \textit{Encoders} network results for the Device A evaluation, the  ``max-comb'' combiner actually outperformed the accuracy of ``lin-comb'', although the latter performed best for most of the full systems. 
This means that the optimal high-level feature combiner for the full system was not the best combiner for loss computation when training the \textit{Encoder} network.
However the situation reverses when looking at Devices B \& C -- an indication that the performance gain of ``lin-comb'' may have been due to better generalisation.
\begin{table}[tb!]
    \caption{\textit{Performance of \textit{Encoder/Decoder} (\%) over DCASE 2018 Task 1B Dev. set.}} 
        	\vspace{-0.2cm}
    \centering
    \scalebox{0.85}{
    \begin{tabular}{|l | c |c |c |}
        \hline		            
\textbf{Device A }   &RFC \textit{Decoder}     &MLP \textit{Decoder}      &MoE \textit{Decoder}    \\                     
        \hline 
        sum-comb  &71.5/75.6              &71.5/72.2               &71.5/71.9    \\
        max-comb   &74.1/75.3             &74.1/74.7                &74.1/75.5     \\
        lin-comb     &73.7/75.2             &73.7/75.5                &73.7/\textbf{75.9}  \\
\hline 
\hline 
\textbf{Devices  B \& C:}    &RFC \textit{Decoder}         &MLP \textit{Decoder}      &MoE \textit{Decoder}       \\                     
\hline 
        sum-comb    &63.9/64.4            &63.9/65.6     &63.9/63.9   \\
        max-comb  &61.4/65.3            &61.4/63.9     &61.4/63.9   \\
        lin-comb     &64.2/68.9    &64.2/69.2      &64.2/\textbf{70.6}   \\
                \hline 

    \end{tabular}
    }
    \label{table:c04_re_model} 
\end{table}
\subsection{Per-class Performance of Different \textit{Decoders}}
\label{c04_result_class}
\begin{table}[th!]
    \caption{\textit{Performance comparison (Acc. \%) to DCASE 2018 baselines \\ for Task 1B Dev. set on Device A  (using ``lin-comb'' \\ for extracting high-level features in \textit{Encoder}).}} 
   	\vspace{-0.2cm}
   	\centering
     \resizebox{0.85\textwidth}{!}{

    \begin{tabular}{|l | c |c |c |c  |}
       \hline
\textbf{Categories}     	&\textbf{D.2018}      &\textbf{RFC \textit{Decoder}}      &\textbf{MLP \textit{Decoder}}      &\textbf{MoE \textit{Decoder}}  \\                     
	 \hline                                                            
             Airport            &73.4        &67.5     &60.4     &66.8   \\
             Bus                &56.7        &78.5     &80.2     &80.2   \\
             Metro              &46.6        &67.0     &72.8     &69.3   \\
             Metro station      &52.9        &84.6     &82.6     &80.3   \\
             Park               &80.8        &89.7     &86.8     &88.4   \\
             Public square      &37.9        &47.7     &52.8     &50.9   \\
             Shopping Mall      &46.4        &74.6     &75.3     &73.8   \\
             Street Pedestrian  &55.5        &65.6     &72.5     &71.3   \\
             Street Traffic     &82.5        &91.1     &90.7     &92.3   \\
             Tram               &56.5        &83.1     &79.3     &83.1   \\
       \hline                                                                             
	     \textbf{Average}    &\textbf{58.9}  &\textbf{75.2} &\textbf{75.5} &\textbf{75.9}  \\   
       \hline 
    \end{tabular}    }
    \label{table:comp_dcase_a} 
\end{table}

\begin{table}[th!]
    \caption{\textit{Performance comparison (Acc. \%) to DCASE 2018 baselines \\ for Task 1B Dev. set on Devices B+C (using ``lin-comb'' \\ for extracting high-level features in \textit{Encoder}).}} 
   	\vspace{-0.2cm}
   	\centering
     \resizebox{0.85\textwidth}{!}{%
    \begin{tabular}{|l | c |c |c |c  |}
       \hline
\textbf{Categories}				&\textbf{D.2018}      &\textbf{RFC \textit{Decoder}}      &\textbf{MLP \textit{Decoder}}      &\textbf{MoE \textit{Decoder}} \\                     
	 \hline                                                            
             Airport             &72.5        &55.6        &69.4     &75.0      \\
             Bus                 &78.3        &88.9        &86.1     &88.9      \\
             Metro               &20.6        &75.0        &63.9     &66.7      \\
             Metro station       &32.8        &50.0        &61.1     &50.0      \\
             Park                &59.2        &91.7        &91.7     &94.4      \\
             Public square       &24.7        &52.8        &47.2     &47.2      \\
             Shopping Mall       &61.1        &80.6        &80.6     &80.6      \\
             Street Pedestrian   &20.8        &66.7        &75.0     &77.8      \\
             Street Traffic      &66.4        &75.0        &77.8     &77.8      \\
             Tram                &19.7        &52.8        &38.9     &47.2      \\
       \hline                                                                             
	     \textbf{Average}    &\textbf{45.6}     &\textbf{68.9}     &\textbf{69.2}     & \textbf{70.6}    \\   
       \hline 
    \end{tabular}    }
    \label{table:comp_dcase} 
\end{table}
%
Given that the results presented so far indicate that the \textbf{lin-comb} combiner performed best, these high-level features are fed into the three alternative decoders to explore class-by-class performance.
Table \ref{table:comp_dcase} presents results for DCASE 2018 Task 1B (Dev. set). Device A and Device B \& C results are again shown separately, and the ``D.2018'' column is the DCASE 2018 baseline.
Results show that the three classifiers all outperformed the baseline -- with the mixture of experts system improving accuracy by 17.0\% and 25.0\% absolute, for Device A and Devices B \& C, respectively.

\subsection{Performance Comparison to State-of-the-art Systems}
\label{c04_result_stateoftheart}

While performance against the baseline score of DCASE 2018 is good, 
the same model configuration (i.e. ``lin-comb'' combiner and MoE \textit{Decoder} back-end classifier) is evaluated  on various datasets and competitions, to compare the performance against the state of the art at the time of writing.
The results, listed in Table \ref{table:state_of_the_art}, show that the system proposed achieves the highest accuracy for two datasets -- achieving 70.6\% and 98.9\% for DCASE 2018 Task 1B Dev. and LITIS Rouen, respectively.
For DCASE 2016, an accuracy of 88.2\% was achieved, holding second position on the challenge table, and ranked top-four among state-of-the-art systems. 
DCASE 2017 performance is a little less competitive at 72.6\%.
DCASE 2018 Task 1A performance was 77.5\%, taking third place on the challenge table.
Participating in the recent DCASE 2019 challenge, this system achieved 76.8\% and 72.8\% for DCASE 2019 Task 1A and 1B, respectively. 
It should be noted that there is some inconsistency between the accuracies reported in the DCASE 2018 technical reports and those published on the DCASE 2018 challenge website~\footnote{http://dcase.community/challenge2018/}. 
Therefore, Table \ref{table:state_of_the_art} carefully reports the accuracies stated in the peer-reviewed published papers and technical reports submitted to the challenge, rather than the figures advertised on the websites, which may differ slightly in some cases.

\section{Conclusion}
\label{c04_conclusion}
This chapter has presented a novel \textit{Encoder-Decoder} deep learning framework applied for ASC.
The framework addresses three main factors: low-level feature input, high-level feature extraction, and output classification that affects the final accuracy. 
\begin{enumerate}
\item Firstly, inspired by the belief that low-level features each contain valuable and complementary information, three different spectrograms (log-mel, Gamma, and CQT) were evaluated, as well as their combination.
\item Hence, the \textit{Encoder}  network was proposed to effectively combine three different spectrograms (log-mel, Gamma, and CQT), extracting high-performance high-level feature vectors.
\item Then the  \textit{Decoder} was proposed as a final classifier. Three different models were explored and evaluated, the random forest classification, an MLP-based network, and a Mixture of Expert (MoE). 
\end{enumerate}
The final combined system achieved very competitive results on various datasets, and has been evaluated against state-of-the-art systems to prove that the proposed \textit{Encoder-Decoder} framework is both powerful in terms of performance and also general in terms of particular ASC task.

\begin{landscape}

\begin{table*}[!bt]
    \caption{\textit{Performance comparison (Acc. - \%) of the proposed Encoder-Decoder (E-D) Framework \\ (``lin-comb'' + MoE  \textit{Decoder}) to state-of-the-art results, with best performance in \textbf{bold} \\ (Upper part: Dataset; Middle part: top-ten DCASE challenges; Lower part: State-of-the-art papers) }} 
    \centering
    \vspace{-0.2cm}
    \resizebox{1.5\textwidth}{!}{%
    \begin{tabular}{| l c  |  l c  | l c  | l c |  l c | l c || l c|} 
        \hline 
\textbf{D.2016}              &\textbf{Acc. (\%)}    &\textbf{D.2017}              &\textbf{Acc. (\%)}   &\textbf{D.2018-1A}              &\textbf{Acc. (\%)}        &\textbf{D.2018-1B}                  &\textbf{Acc. (\%)}     &\textbf{D.2019-1A}         &\textbf{Acc. (\%)}       &\textbf{D.2019-1B}                       &\textbf{Acc. (\%)}     &\textbf{LITIS}                &\textbf{Acc. (\%)}\\ [0.5ex] 
                                                                                                                                                                                                                                                                                                                                                               
\multicolumn{2}{l|}{ \textbf{(Eva. set)}}     & \multicolumn{2}{l|}{ \textbf{(Eva. set)}}     & \multicolumn{2}{l|}{ \textbf{(Dev. set)}}             &\multicolumn{2}{l|}{ \textbf{(Dev. set)}}              &\multicolumn{2}{l|}{ \textbf{(Eva. set)}}       &\multicolumn{2}{l||}{ \textbf{(Eva. set)}}                   &\multicolumn{2}{l |}{\textbf{(20-fold Ave.)} }    \\
                                                                                                                                                                                                                                                                                                                                                               
\hline                                                                                                                                                                                                                                                                                                                                            
Wei~\cite{dc_16_t10}          &$84.1$          &Zhao~\cite{dc_17_t10}        &$70.0$          &Li~\cite{dc_18_t10}           &$72.9$               &Baseline~\cite{dc_18_bsl}         &$45.6$            &Mingle~\cite{dc_19_t10}  &$79.9$              &Baseline~\cite{data_dc_18}          &$61.6$            &Bisot~\cite{lit_bisot_eus}    &$93.4$  \\     
Bae~\cite{dc_16_t09}          &$84.1$          &Jung~\cite{dc_17_t09}        &$70.6$          &Jung~\cite{dc_18_t09}         &$73.5$               &Li~\cite{dc_18_tb07}              &$51.7$            &Wu~\cite{dc_19_t09}      &$80.1$              &Kong~\cite{dc_19_tb09}                 &$61.6$            &Ye~\cite{lit_ye_acm}          &$96.0$  \\     
Kim~\cite{dc_16_t08}          &$85.4$          &Karol~\cite{dc_17_t08}       &$70.6$          &Hao~\cite{dc_18_t08}          &$73.6$               &Tchorz~\cite{dc_18_tb06}          &$53.9$            &Gao~\cite{dc_19_t08}     &$80.5$              &Waldekar~\cite{dc_19_tb08}             &$62.1$            &Huy~\cite{huy_lit_acm}        &$96.4$  \\     
Takahasi~\cite{dc_16_t07}     &$85.6$          &Ivan~\cite{dc_17_t07}        &$71.7$          &Christian~\cite{dc_18_t07}    &$74.7$               &Kong~\cite{dc_18_tb05}            &$57.5$            &Wang~\cite{dc_19_t07}    &$80.6$              &Wang~\cite{dc_19_tb07}                 &$70.3$            &Yin~\cite{dc_16_yin}          &$96.4$  \\     
Elizalde~\cite{dc_16_t06}     &$85.9$          &Park~\cite{dc_17_t06}        &$72.6$          &Zhang~\cite{dc_18_t06}        &$75.3$               &Wang~\cite{dc_18_tb04}            &$57.5$            &Jung~\cite{dc_19_t06}    &$81.2$              &Jiang~\cite{dc_19_tb06}                &$70.3$            &Huy~\cite{huy_lit_jr}         &$96.6$  \\     
Valenti~\cite{dc_16_t05}      &$86.2$          &Lehner~\cite{dc_17_t05}      &$73.8$          &Li~\cite{dc_18_t05}           &$76.6$               &Waldekar~\cite{dc_18_tb03}        &$57.8$            &Huang~\cite{dc_19_t05}   &$81.3$              &Song~\cite{dc_19_tb05}                 &$72.2$            &Ye~\cite{lit_ye_jr}           &$97.1$  \\     
Marchi~\cite{dc_16_t04}       &$86.4$          &Hyder~\cite{dc_17_t04}       &$74.1$          &Dang~\cite{dc_18_t04}         &$76.7$               &Zhao~\cite{dc_18_zhao_dcase}      &$58.3$            &Haocong~\cite{dc_19_t04} &$81.6$              &Primus~\cite{dc_19_tb04}               &$74.2$            &Huy~\cite{huy_lit_aes}        &$97.8$  \\     
Park~\cite{dc_16_t03}         &$87.2$          &Zhengh~\cite{dc_17_t03}      &$77.7$          &Octave~\cite{dc_18_oct_dcase} &$78.4$               &Truc~\cite{dc_18_tb01}            &$63.6$            &Hyeji~\cite{dc_19_t03}   &$82.5$              &Hamid~\cite{dc_19_tb03}                &$74.5$            &Zhang~\cite{lit_zang_int_03}  &$97.9$  \\  
Bisot~\cite{dc_16_t02}        &$87.7$          &Han~\cite{dc_17_t02}         &$80.4$          &Yang~\cite{dc_18_yang_dcase}  &$79.8$               &                                  &                  &Koutini~\cite{dc_19_t02} &$83.8$              &Gao~\cite{dc_19_tb02}                  &$74.9$            &Zhang~\cite{lit_zang_int}    &$98.1$  \\     
Hamid~\cite{dc_16_t01}        &$89.7$          &Mun~\cite{dc_17_t01}         &$\textbf{83.3}$ &Golubkov~\cite{dc_18_t01}     &$\textbf{80.1}$      &                                  &                  &Chen~\cite{dc_19_t01}    &$\textbf{85.2}$     &Kosmider~\cite{dc_19_tb01}             &$\textbf{75.3}$   &Huy~\cite{huy_lit_int_02}     &$98.7$  \\ 
                                                                                                                                                                                                                                                                                                                                   
\cmidrule{1-8}                                                                                                                                                                                                                                                                                                            
Mun~\cite{dc_16_mun_ica}      &$86.3$          &Zhao~\cite{dc_17_zhao_jr}     &$64.0$         &Bai~\cite{dc_18_bai_int}      &$66.1$               &Zhao~\cite{dc_18_zhao_ica}         &$63.3$            &                           &                    &                                         &                  & & \\   
Li~\cite{dc_16_li_ica}        &$88.1$          &Yang~\cite{dc_17_kl_ica}      &$69.3$         &Gao~\cite{dc_18_gao_jr}       &$69.6$               &Truc~\cite{dc_18_truc_int}         &$64.7$            &                           &                    &                                         &                  & & \\   
Hyder~\cite{dc_16_hyder_int}  &$88.5$          &Waldekar~\cite{dc_17_int_699} &$69.9$         &Zhao~\cite{dc_18_zhao_ica}    &$72.6$               &Truc~\cite{dc_18_truc_icme}        &$66.1$            &                           &                    &                                         &                  & & \\   
Song~\cite{dc_16_song_int}    &$89.5$          &Wu~\cite{dc_17_wu_ica}        &$75.4$         &Phaye~\cite{dc_18_phaye_ica}  &$74.1$               &Yang~\cite{dc_18_yang_jr}          &$67.8$            &                           &                    &                                         &                  & & \\   
Yin~\cite{dc_16_yin}          &$\textbf{91.0}$ &Chen~\cite{dc_17_chen_ica}    &$77.1$         &Heo~\cite{dc_18_heo_axv}      &$77.4$               &                                   &                  &                           &                    &                                         &                  & & \\    
                                                                                                                                                                                                                                                                                                                                                                    
\hline				                                                                                                                                                                                                                                                                                                                            
\textit{E-D Framework}        &$88.2$          &\textit{E-D Framework}        &$72.6$         &\textit{E-D Framework}        &$77.5$               &\textit{E-D Framework}                          &$\textbf{70.6}$   &\textit{E-D Framework}                 &$76.8$              &\textit{E-D Framework}                               &$72.8$            &\textit{E-D Framework}                    &$\textbf{98.9}$    \\
   \hline 
    \end{tabular} }   
    \label{table:state_of_the_art} 
\end{table*}
\end{landscape}

%

\chapter{Two-level Hierarchical Classification}
\label{c05}

Inspired by the high cross-correlation between sound scenes mentioned in Section \ref{c03_baseline_compare}, the original ``flat''ASC task, i.e. classification of all categories at once, might be better structured into multiple hierarchical sub-tasks operating in a divide-and-conquer manner.
This chapter further explores the high cross-correlation between sound scenes, then based on that, it develops a two-level classification scheme for ASC.

In particular, sound scenes, which are expected to be acoustically similar, are firstly grouped into meta categories.
The meta-categories constitute the first level of the classification hierarchy. 
Next, each category within the meta categories is classified by the second level of the two-level classification scheme.
The two levels could also be referred to as coarse and fine grained classification.
\begin{figure}[h]
	\centering
	\centerline{\includegraphics[width=0.85\linewidth]{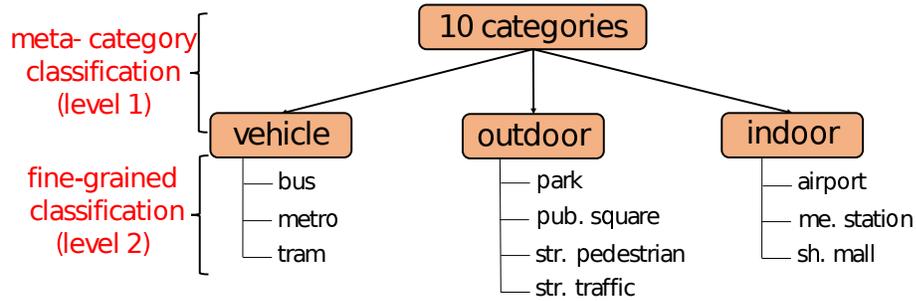}}
	\vspace{-0.2cm}
	\caption{\textit{The two-level hierarchy of scene categories constructed by \\ examination of the categories used in the DCASE 2018 dataset.}}
	\label{fig:c05_two_level_scheme}
\end{figure}
As experiments in this Chapter are conducted on DCASE 2018 Tasks 1A and 1B, the proposed hierarchy scheme is constructed based on them, as shown in Figure~\ref{fig:c05_two_level_scheme} (note that meta categories, such as \textit{Indoor, Outdoor} and \textit{Vehicle}, are selected based on analysis of the incorrect classification cases in Section \ref{c03_baseline_compare}).

The hierarchical classification is performed in a top-down fashion. 
Firstly, the meta-categories are classified, followed by the fine-grained classification of the scene categories within each individual meta-category. 
As a result, four classifiers are trained: one for meta-category classification (referred to as meta-category classifiers), then three are trained for classification of categories within the three meta-categories (namely a ``vehicle'' classifier, ``indoor'' classifier, and ``outdoor'' classifier, respectively). 
An unseen example will be then be deemed to have classified correctly only if it is correctly classified at both levels of the hierarchy.
For example, a ``on bus'' scene example is correctly classified if it is both correctly classified as ``vehicle'' by the meta-category classifier and as ``bus'' by the ``vehicle'' classifier. 
Any misclassifcation by one or both of the classifiers will result in the example being wrongly classified overall.  

\section{The Proposed System}
\label{c06_system}

\subsection{High-level Architecture}
\label{c06_baseline}
\begin{figure}[h]
	\centering
	\includegraphics[width=\linewidth]{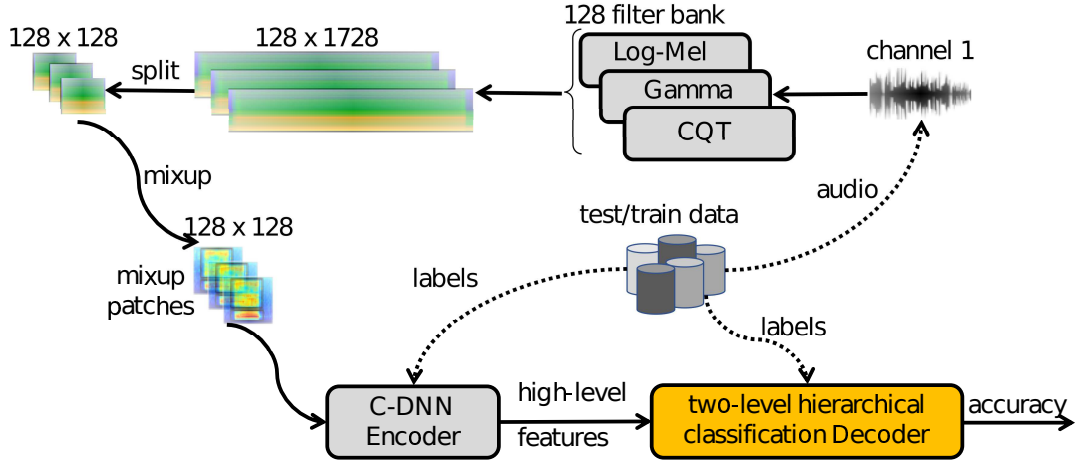}
	\vspace{-0.5cm}
	\caption{\textit{High-level system architecture applying\\ a two-level hierarchical classification scheme.}}
	\label{fig:c05_high_level_arc}
\end{figure}
The high-level architecture of an ASC system applying the hierarchical scheme is described in Figure~\ref{fig:c05_high_level_arc}.
As regards front-end feature extraction, three types of spectrogram (log-mel, Gamma, and CQT) are used to extract spectrogram information from channel 1. 
The entire spectrograms are then split into image patches of $128\times128$ before applying mixup data augmentation. 

Other settings for the front-end feature extraction such as filter number, window size, hop size, etc.,  are re-used from the baseline proposed in Section \ref{c03_baseline}.
The resulted mixup data is used to train a network for high-level feature extraction, referred as to C-DNN \textit{Encoder}. 
The high-level features extracted from the C-DNN \textit{Encoder} are fed into the two-level hierarchical classification scheme as described in Figure \ref{fig:c05_two_level_scheme}, which in turn then report the final classification accuracy.
Compared to the novel \textit{Encoder-Decoder} framework architecture introduced in Chapter \ref{c04}, the two-level hierarchical scheme proposed in this Chapter takes the role of an MLP \textit{Decoder}, RF \textit{Decoder}, or MoE \textit{Decoder} .

\subsection{C-DNN \textit{Encoder} Architecture}
\label{c05_encoder}
The high-level feature extractor in this Chapter uses a deep C-DNN as described in Table \ref{table:c05_CDNN}, comprising batch normalization (BN), convolutional (Cv [kernel size] $@$ kernel number), rectified linear unit (ReLU), average pooling (AP [kernel size]), dropout (Dr) and fully connected (FC) layers.

To clarify it, C-DNN \textit{Encoder} is separated into two parts: the CNN part (the upper of Table \ref{table:c05_CDNN}) for feature learning and the DNN part (the lower of Table \ref{table:c05_CDNN}) for classification. 
Both are based on the previous architectures in Chapter~\ref{c04}, but instead of using a Global Average Pooling layer at the output of the CNN as in the \textit{Encoder} of Section \ref{c04_encoder}, an additional convolutional layer with kernel size of [$8{\times}8$] is incorporated. 
This equates to the time-frequency resolution of the output from the previous layer, and is included to capture the interaction across the convolutional channel dimension. 
In the other words, the final convolutional layer helps to scale temporal and frequency dimensions into one value with trainable parameters learning all pixels of temporal-frequency images.
\begin{table}[tb]
    \caption{\textit{The C-DNN architecture used for high-level feature extraction.}} 
        \vspace{-0.2cm}
    \centering
    \scalebox{0.85}{

    \begin{tabular}{|l |c|} 
        \hline 
            \textbf{Layers}   &  \textbf{Output}  \\
        \hline 
         Input layer (image patch) & $128{\times}128$          \\
         BN - Cv [$3{\times}3$] $@$ 32 - ReLU - BN - AP [$2{\times}2$] - Dr (10\%)      & $64{\times}64{\times}32$\\
         BN - Cv [$3{\times}3$] $@$ 64 - ReLU - BN - AP [$2{\times}2$] - Dr (15\%)      & $32{\times}32{\times}64$\\
         BN - Cv [$3{\times}3$] $@$ 128 - ReLU - BN - Dr (20\%)      & $32{\times}32{\times}128$ \\
         BN - Cv [$3{\times}3$] $@$ 128 - ReLU - BN - AP [$2{\times}2$] - Dr (20\%)       & $16{\times}16{\times}128$\\
         BN - Cv [$3{\times}3$] $@$ 256 - ReLU - BN  - Dr (25\%)      & $16{\times}16{\times}256$ \\
         BN - Cv [$3{\times}3$] $@$ 256 - ReLU - BN  - AP [$2{\times}2$] - Dr (25\%)  & $8{\times}8{\times}256$ \\
         BN - Cv [$8{\times}8$] $@$ 256 - ReLU - BN - Dr (30\%) & $256$ \\  
         \hline 
         \hline 
         Input layer (vector)          & 256          \\
         FC - ReLU - Dr (30\%)        &  512         \\
	     FC - ReLU - Dr (30\%)        &  1024           \\
	     FC - Softmax                     &  10          \\                    
       \hline 
    \end{tabular}
    }
    \label{table:c05_CDNN} 
\end{table}

Once the network has been trained, the feature-learning CNN part of the network is used as a feature extractor and its last convolutional layer is considered to provide high-level features. 
In this way, when presented with a new input, the high-level feature extractor will process the input starting from the first convolutional layer, through to the final convolutional layer and produce a high-level feature vector of dimension 256.

\subsection{Two-level Hierarchical Classification as \textit{Decoder}}
\label{c05_decoder}

Most existing works follow a ``flat'' classification scheme in which all scene categories are classified at once. 
By contrast, this Section proposes performing the classification hierarchically, as recently introduced in Figure~\ref{fig:c05_two_level_scheme}. 
The classifiers involving in the hierarchical classification are realized by Multilayer Perceptron (MLP) based networks. 
The 256 dimensional high-level features presented in Section \ref{c05_encoder} are obtained from the mixup image patches and used to train the MLPs. 
There are in total four MLPs (one for the meta-categories at the first level and three for fine classification within the meta-category groups of \textit{Indoor, Outdoor} and \textit{Vehicle} at the second level), each comprises four fully connected layers and is parametrized as summarised in Table \ref{table:c05_MLP}.
Note that the {MLPs share a common architecture but are trained separately depending on their respective sub-tasks in the hierarchical classification.
\begin{table}[tb]
	\caption{\textit{MLP-based architecture used \\in the two-level hierarchical scheme}} 
	\vspace{-0.2cm}
    \centering
    \scalebox{0.85}{
	\begin{tabular}{|l |c|} 
		\hline 
		\textbf{Layers}   &  \textbf{Output Shape}  \\
		\hline 
		Input layer & $256$          \\
		FC - ReLU - Dr (30\%)        &$512$         \\
		FC - ReLU - Dr (30\%)        &$1024$           \\
		FC - ReLU - Dr (30\%)        &$1024$           \\
		FC - Softmax         &$C$          \\                    
		\hline 
	\end{tabular}
	}
	\label{table:c05_MLP} 
	\vspace{-0.1cm}
\end{table}

The number of categories classified  \(C\) depends on the specific task in the hierarchical scheme. For example \(C\) is 3 for meta-category classification and for the \textit{Vehicle} and \textit{Indoor} group categories. It is 4 for classification of the categories in the \textit{Outdoor} group.
\section{Experimental Setting}
\label{c05_setup}
\subsection{Datasets}
\label{c05_dataset}
To evaluate the two-level hierarchical scheme, DCASE 2018 Task 1A and 1B development datasets~\cite{data_dc_18} are used to conduct experiments.
For consistency, the relevant settings used for these datasets are mentioned and reused from Section \ref{c03_dataset} and Section \ref{c04_dataset}.
\subsection{Setting Hyperparameters and Training Process}
\label{c05_hyper}
Since the labels of the mixup data input are no longer one-hot, the network is trained with Kullback-Leibler (KL) divergence loss~\cite{kl_loss} rather than the standard cross-entropy loss over mixup training image patches:
\begin{align}
    \label{eq:loss_func}
    LOSS_{KL}(\Theta) = \sum_{c=1}^{C}y_{c}\log \left\{ \frac{y_{c}}{\hat{y}_{c}(\Theta)} \right\}  +  \frac{\lambda}{2}||\Theta||_{2}^{2},
\end{align}
where $\Theta$ denotes the trainable network parameters and $\lambda$ denotes the $\ell_2$-norm regularization coefficient set to 0.001. 
$y_{c}$ and $\hat{y}_{c}$  denote the ground-truth and the network output of class $c$, respectively. 

In addition to the KL-divergence loss, the triplet loss function~\cite{tripletloss} is additionally employed to train the MLPs in the second-level classifiers to encourage the networks to improve their discrimination power. 
The motivation is that the triplet loss function has been shown to be efficient in learning a discriminative metric which simultaneously minimises same-category distance while maximising between-category distances. In this way, it enhances Fisher's criterion~\cite{tripletloss} (i.e. the ratio of the between- class distance to the within-class variance in the feature space). 

Suppose that there are two samples from different categories presented to an MLP. 
The ground-truth label of the first sample is the anchor $\mathbf{a}$, the prediction for the first sample is positive $\mathbf{p}$, and the prediction for a second sample is positive $\mathbf{n}$, then the triplet loss is given as:
\begin{align}
    \label{eq:triplet_loss}
    LOSS_{T} = \max \{ d(\mathbf{a},\mathbf{p}) - d(\mathbf{a},\mathbf{n}) + margin, 0  \}, 
\end{align}
where \(d\) is the squared Euclidean distance and the \(margin\) is set to $0.3$.

The final loss function is then a combination of the KL-divergence loss and the triplet loss as follows:
\begin{equation}
    \label{eq:final_loss}
    LOSSES = \gamma LOSS_{KL} + (1-\gamma)LOSS_{T}
\end{equation}
The networks are implemented using the Tensorflow framework. The coefficient $\lambda$ in (\ref{eq:loss_func}) is set to 0.001, and $\gamma$ in (\ref{eq:final_loss}) is experimentally set to $0.2$. The network training is accomplished with the Adam optimiser~\cite{kingma2014adam} with an initial learning rate of $10^{-4}$, a batch size of $100$, and a fixed termination after 100 epochs.

\subsection{Multi-spectrogram Ensemble}
\label{c05_ensemble}

As for the comprehensive analysis of low-level features in Chapter \ref{c03}, using multiple input types, provided a rule of thumb for performance ASC, now all three time-frequency input types are used:  log-mel~\cite{librosa_tool}, gammatone filter (Gamma)~\cite{auditory2009_tool}, and Constant Q Transform (CQT)~\cite{librosa_tool}. Together these three will enable construction of an ensemble of three systems. 
The final decision of each classification task (meta-category classification at the first level or the fine-grained second level classification. shown in Figure \ref{fig:c05_two_level_scheme}) is obtained by aggregating the individual decisions of the three classifiers (each with one type of spectrogram) in an ensemble. In particular, if $\mathbf{\bar{p}}_{\text{log-mel}}$, $\mathbf{\bar{p}}_{\text{Gamma}}$, $\mathbf{\bar{p}}_{\text{CQT}}$ are probabilities corresponding  log-mel, Gamma, and CQT spectrogram input, sum of three probability $\mathbf{\bar{p}}[\bar{p}_1, \bar{p}_2,...,\bar{p}_C]$ is computed by
\begin{equation}
  \label{eq:c05_final_res}
 \mathbf{\bar{p}} = \mathbf{\bar{p}}_{\text{log-mel}} + \mathbf{\bar{p}}_{\text{Gamma}} + \mathbf{\bar{p}}_{\text{CQT}}
\end{equation}
where $C=10$ denotes the number of categories classified in DCASE 2018 Task 1A \& 1B.
Thus, the final classification label is determined as,
\begin{equation}
    \label{eq:c05_label_determine}
   \hat{y} = \argmax_{c \in \{1,2,\ldots,C\}}\bar{p}_c.
\end{equation}
where $\hat{y}$ denotes the final label. 
\section{Experimental Results}
\label{c05_results}

\subsection{Performance comparison to DCASE 2018 baseline}
\label{c05_compare_baseline}
\begin{table}[h]
	\caption{\textit{Performance comparison (in percentage accuracy) \\ between the proposed system (with and without triplet loss), \\ the DCASE 2018 baseline, and the C-DNN \textit{Encoder} baseline.}} 
	\vspace{-0.2cm}
    \centering
    \scalebox{0.85}{
	\begin{tabular}{|l |c |c|} 
		\hline 
		\textbf{Systems Compared}                   & \textbf{Task 1A}  & \textbf{Task 1B}  \\ [0.5ex] 
		\hline 
		DCASE 2018 baseline ~\cite{data_dc_18}      & $59.7$  & $45.6$ \\
		The C-DNN \textit{Encoder}                                & $70.9$  & $61.1$ \\	    
		The proposed w/o triplet loss                                                     & $73.3$ &  $\textbf{62.2}$ \\
		The proposed w/ triplet loss                             & $\textbf{75.3}$   & $58.9$ \\        
		\hline 
	\end{tabular}    
	}
	\label{table:c05_all_compare} 
\end{table}
\begin{figure*}[h]
	\centering
	\includegraphics[width=0.9\linewidth]{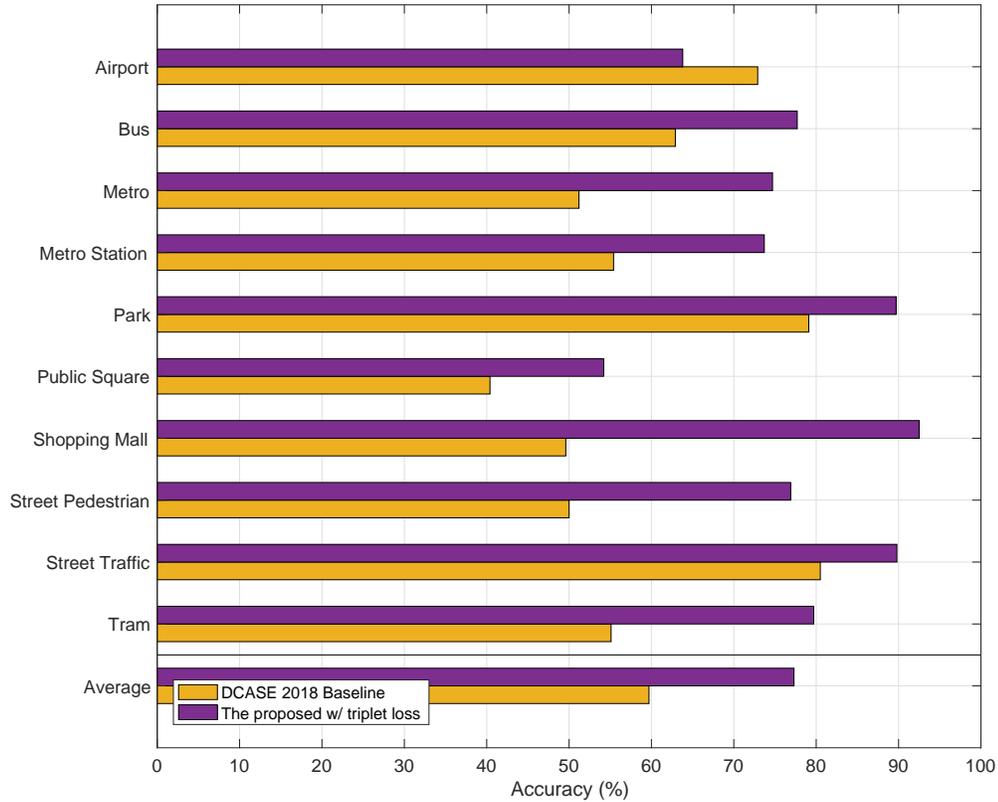}
	\vspace{-0.8cm}
	\caption{\textit{Category-wise performance comparison between the proposed system with triplet loss and the DCASE 2018 baseline on Task 1A.}}
	\label{fig:c05_all_compare}
\end{figure*}
To evaluate the hierarchical scheme, the C-DNN \textit{Encoder} which uses only the Gamma spectrogram is referred as to the baseline.
The performance of the DCASE 2018 baseline, C-DNN \textit{Encoder} baseline, the entire system (applying the two-level hierarchical scheme without ensemble) are compared in Table \ref{table:c05_all_compare}.
As can be seen, the proposed system outperforms the DCASE 2018 baseline by a large margin, around $15.6$\% absolute (with triplet loss) on Task 1A and $16.6$\% absolute on Task 1B (without triplet loss). 
Improvements of the individual categories can also be visualised in Figure~\ref{fig:c05_all_compare}, which compares the proposed system with triplet loss against the DCASE 2018 baseline on Task 1A. 
It is notable that several categories enjoy a significant gain of more than $20.0\%$, such as \textit{shopping mall, tram, metro} and \textit{street-pedestrian}. 
\begin{figure}[h]
	\centering
	\includegraphics[width=0.85\linewidth]{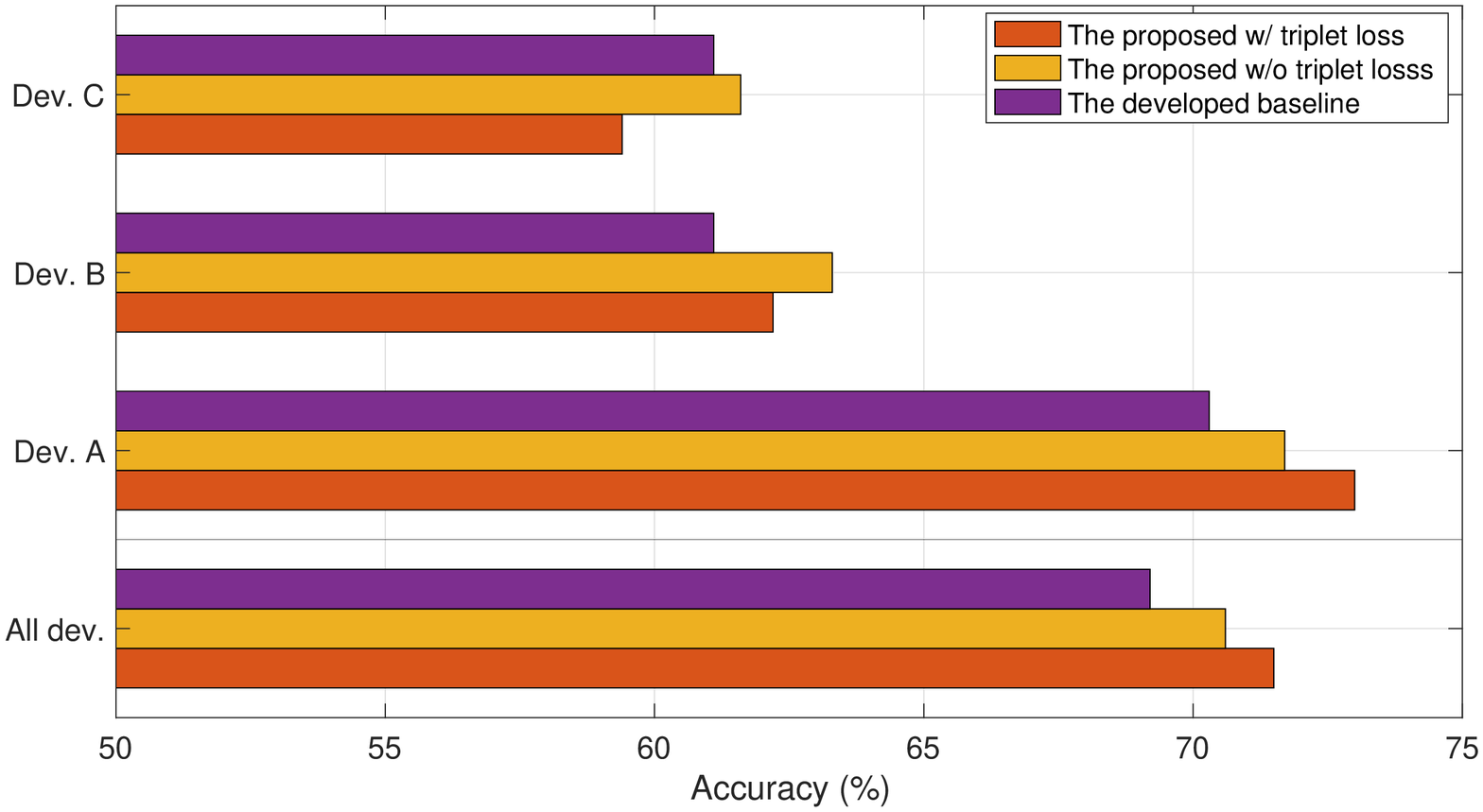}
	\caption{\textit{Accuracy obtained by the systems developed in this work on different devices of Task 1B.}}
	\label{fig:c05_device}
\end{figure}
Compared to C-DNN \textit{Encoder}, the proposed system gains an accuracy of $2.4\%$ and $1.1\%$ on Task 1A and Task 1B, respectively, when the triplet loss is not used. 
When the triplet loss is used, a significant accuracy improvement is seen on Task 1A: $2.4\%$ absolute compared to that without triplet loss and $4.4\%$ compared to the developed baseline thanks to the proposed hierarchical classification scheme. 
However, using triplet loss seems to be counter-productive on Task 1B as the accuracy is reduced by $3.3\%$ absolute compared to the system without  triplet loss. 
This is presumably due to the device mismatch or the lack of training data on the target devices (device B \& C) or both. 
However, averaging over all the devices, the proposed system with triplet loss outperforms all other counterparts, as shown in Figure \ref{fig:c05_device}. 
To further shed light on the performance of the classifiers in the proposed hierarchical classification scheme, their confusion matrices are presented in Figure~\ref{fig:c05_meta_category}. 
Overall, the meta-categories are discriminated very well  by the meta-category classifier, with an average accuracy of $94$\%. 
Given the good performance of the meta-category classifier, the test examples are expected to be directed to the correct groups in the lower level. 
Even though the fine-grained classifiers' performance are not as good as that of the meta-category classifier, this is to be expected since the categories in a group tend to be similar acoustically, however in each group, the fine classification network is able to avoid confusion between its categories and those in other groups. 
\begin{figure*}[h]
	\centering
	\includegraphics[width=0.9\linewidth]{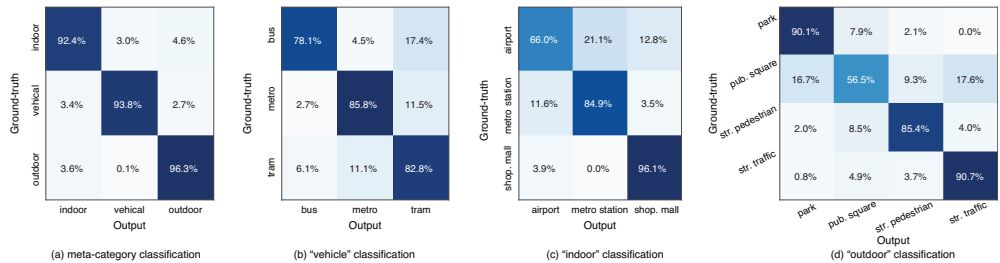}
	\caption{\textit{Confusion matrices obtained by different classifiers in the proposed hierarchical classification scheme on Task 1A.}}
	\label{fig:c05_meta_category}
\end{figure*}

Further compare to \textit{Encoder-Decored} systems proposed in Chapter \ref{c04}, while the best \textit{Encoder-Decored} system using lin-comb, MoE Decoder, and three spectrogram inputs achieves the accuracy of 75.9\% on DCASE 2018 Taks 1A as shown in Table \ref{table:c04_re_model}, the two-level hierarchical classification system proposed achieves the competitive result of 75.3\% on the same task with only using Gammatone spectrogram input.

\subsection{Results of Multi-spectrogram Ensemble}
\label{c05_ensemble}
\begin{figure}[h]
	\centering
	\includegraphics[width=0.85\linewidth]{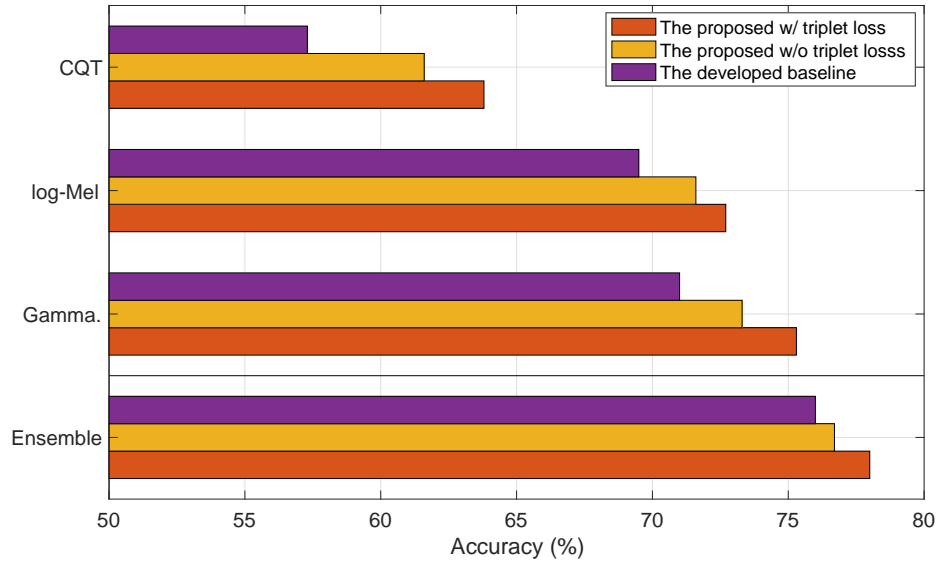}
	\caption{\textit{Performance of individual time-frequency representations \\ and their ensemble on Task 1A.}}
	\label{fig:c05_spec}
\end{figure}
Further experiments are conducted over individual time-frequency inputs (i.e. Gamma, log-mel, and CQT spectrograms). The gammatone spectrogram seems to perform best as shown in Figure \ref{fig:c05_spec} while the CQT spectrogram  performs the worst. 
However, aggregation of the classification outputs of all three, results in significant improvements over the individual ones. 
This is observed over all systems; the proposed system with triplet loss, the proposed system without triplet loss, and the developed baseline. 
It is expected as different time-frequency representations have been shown to be good for different scene categories, and their individual strength is leveraged in the ensemble to improve the performance gain.

The obtained results are further compared  with the previous works (both the DCASE 2018 challenge submission systems and the recent works), providing a comprehensive performance comparison on Task 1A and Task 1B in Table \ref{table:c05_sot}. 
It should be noted that there are inconsistencies between the accuracies reported in the DCASE 2018 technical reports and those published in DCASE 2018 challenge website \footnote{http://dcase.community/challenge2018/}. 
The results in Tables \ref{table:c05_sot} are collated from the technical reports which are the original sources of the reported accuracies. 
For clarity, only top 10 DCASE 2018 challenge submissions are presented in the tables. 
In the one hand, the proposed system outperforms the recent works (i.e. after the DCASE 2018 challenge) on Task 1A while retaining as top-4 performer in the context of the DCASE 2018 submission systems.
 In the other hand, our proposed system achieves very competitive results on Task 1B, achieving an accuracy of 66.9\% and outperforming the DCASE 2018 submission systems.
\begin{table}[tb]
    \caption{\textit{Comparison between the top-10 DCASE 2018 challenge (top), recent papers (middle), and the proposed system (bottom)}}         \vspace{-0.2cm}
    \centering
    \scalebox{0.85}{

    \begin{tabular}{| l c  | l c |} 
        \hline 
            \textbf{D.2018-1A}            &\textbf{Acc. (\%)}   &\textbf{D.2018-1B}                  &\textbf{Acc. (\%)}   \\ [0.5ex] 
        \hline 
            Li~\cite{dc_18_t10}           &$72.9$               &Baseline~\cite{dc_18_bsl}         &$45.6$            \\     
            Jung~\cite{dc_18_t09}         &$73.5$               &Li~\cite{dc_18_tb07}              &$51.7$              \\     
            Hao~\cite{dc_18_t08}          &$73.6$               &Tchorz~\cite{dc_18_tb06}          &$53.9$              \\     
            Christian~\cite{dc_18_t07}    &$74.7$               &Kong~\cite{dc_18_tb05}            &$57.5$              \\     
            Zhang~\cite{dc_18_t06}        &$75.3$               &Wang~\cite{dc_18_tb04}            &$57.5$              \\     
            Li~\cite{dc_18_t05}           &$76.6$               &Waldekar~\cite{dc_18_tb03}        &$57.8$              \\     
            Dang~\cite{dc_18_t04}         &$76.7$               &Zhao~\cite{dc_18_zhao_dcase}      &$58.3$              \\     
            Octave~\cite{dc_18_oct_dcase} &$78.4$               &Truc~\cite{dc_18_tb01}            &$63.6$              \\  
            Yang~\cite{dc_18_yang_dcase}  &$79.8$               &                                  &                   \\     
            Golubkov~\cite{dc_18_t01}     &$\textbf{80.1}$      &                                  &                    \\ 
         \hline 
            Bai~\cite{dc_18_bai_int}      &$66.1$               &Zhao~\cite{dc_18_zhao_ica}         &$63.3$            \\   
            Gao~\cite{dc_18_gao_jr}       &$69.6$               &Truc~\cite{dc_18_truc_int}         &$64.7$            \\   
            Zhao~\cite{dc_18_zhao_ica}    &$72.6$               &Truc~\cite{dc_18_truc_icme}        &$66.1$            \\  
            Phaye~\cite{dc_18_phaye_ica}  &$74.1$               &Yang~\cite{dc_18_yang_jr}          &$\textbf{67.8}$   \\   
            Heo~\cite{dc_18_heo_axv}      &$77.4$               &                                   &                  \\    
          \hline 
            \textit{The proposed w/ triplet loss}        &$78.0$               &\textit{The proposed w/o triplet loss}             &$66.9$   \\
          \hline 
    \end{tabular}
    }
    \label{table:c05_sot} 
\end{table}
\section{Conclusion}
\label{c05_conclusion}
This chapter has presented an approach that trains deep feature embedding networks to extract high-level features for audio scene signals via a C-DNN based \textit{Encoder} and proposed a novel hierarchical classification scheme to accomplish scene classification.
In the classification hierarchy, the similar scene categories are first grouped into meta-categories. 
Meta-category classification is carried out first, followed by the fine-grained classification within the meta groups. 
MLPs were trained, with the contribution of triplet loss, to play the role of the classifiers in the classification hierarchy. 
Experiments on the DCASE 2018 Task 1A and 1B datasets demonstrated that the proposed methods outperform DCASE baseline.

\chapter{Respiratory Disease Detection}
\label{c06}

According to the World Health Organization (WHO)~\cite{who}, respiratory illness, which comprises lung cancer, tuberculosis, asthma, chronic obstructive pulmonary disease (COPD), and lower respiratory tract infection (LRTI), accounts for a significant percentage of mortality worldwide. 
Indeed, records indicate that around 10 million people currently have tuberculosis (TB), 65 million have COPD, and 334 million have asthma.
Notably, the WHO estimates that about 1.4, 1.6, and 3 million people die from TB, lung cancer or COPD annually, respectively. 

To deal with respiratory diseases,  early detection is the key factor in enhancing the effectiveness of intervention, including treatment and limiting spread.
During a respiratory examination, lung auscultation (listening to the sounds of breathing through a stethoscope) is an important aspect of respiratory disease diagnosis.
By listening to respiratory sounds during lung auscultation, experts can recognise adventitious sounds (including \textit{Crackles} and \textit{Wheezes}) during the respiratory cycle. 
These often occur in those who have pulmonary disorders.
If automated methods can be developed to detect such anomalous sounds, it will improve the early detection of respiratory disease and enable screening of a wider population than manual screening.

Inspired by the deep learning techniques that I had developed for effective Acoustic Scene Classification in Chapters \ref{c03}, \ref{c04} and \ref{c05}, I decided to apply these advances on a real world problem: respiratory sound analysis.
Thus, this chapter introduces a robust deep learning framework aiming to classify anomalies in respiratory cycles for detecting disease from respiratory sound recordings.
It evaluates using a standard benchmark, the 2017 International Conference on Biomedical Health Informatics (ICBHI)~\cite{ic_dataset} dataset.
The framework proposed for this task is derived from the baseline mentioned in Chapter \ref{c03}. It begins with front-end feature extraction to transform input sound into a spectrogram representation. 
Then, a back-end deep learning network is used to classify the spectrogram features into categories of respiratory anomaly cycles or disease classes.

The framework proposed confirms three main contributions towards respiratory-sound analysis.
Firstly, it allows an extensive exploration of the effects of spectrogram type, spectral-time resolution, overlapped/non-overlapped windows, and data augmentation, thus indicating which feature has the greatest effect on final prediction accuracy.
This leads to a proposal for a novel deep learning system, developing the proposed framework further, which is shown to significantly outperform current state-of-the-art methods.
However the complexity of that structure is quite high, and so a Teacher-Student scheme is developed and applied with the aim of achieving a trade-off between model performance and complexity. This additionally helps to increase the potential of the proposed framework for building real-time applications, such as in mobile devices which are constrained in terms of processing power.

Before discussing the framework architecture, state-of-the-art systems used for respiratory sound analysis will be analysed, and thus a benchmark dataset used to conduct experiments as well as specific task definition over this dataset are presented. 

\section{State-of-the-art Respiratory Sound Analysis}
\label{c06_intro}

\subsection{Literature Review}
\label{c06_state}

Research into the niche domain of automated detection or analysis of respiratory sounds has some precedents~\cite{sound_early_01, sound_early_02, sound_early_03}, but has drawn attention in recent years as robust machine hearing methods have been developed, leveraging on ever more capable deep learning techniques.
Like traditional ASC systems, most existing respiratory sound analysis systems tend to rely upon frame-based feature representations such as Mel-Frequency Cepstral Coefficients (MFCC)~\cite{sl_hmm_ano_14_embc03, ic_tree_19_embc04}, borrowed from the Automatic Speech Recognition (ASR) and Speaker Recognition (SR) fields. 
However, Gr{\o}nnesby \textit{et al.}~\cite{sl_svm_ano_17_nw} found that MFCCs did not represent crackles well. 
They thus replaced them with five-dimensional feature vectors, comprising four time domain features (variance, range, and sum of simple moving average (coarse and fine)), and one frequency domain feature (spectrum mean). 
Meanwhile, Hanna \textit{et al.}~\cite{ic_tree_18_cbmi} firstly extracted spectral information from bark-bands energy, Mel-bands energy, MFCCs, rhythm features from beat loudness, harmonicity and inharmonicity features, as well as tonal features such as chords strength and tuning frequency. 
Next,  they computed statistical features including standard deviation, variance, minimum, maximum, median, mean, first derivative, second derivative from those features in addition to mean and variance of the raw signal.
This extensive list aimed to maximize the chance of achieving a discriminative feature set.
To further explore audio features, Mendes \textit{et al.}~\cite{sl_lg_ano_16_embc02} went further to propose 35 different types of feature, mainly coming from Music Information Retrieval research.
Inspired by the finding that only some features contributed to the final result, Datta \textit{et al.}~\cite{sl_svm_ano_17_embc01} firstly assessed features such as power spectral density (PSD), FFT and Wavelet spectrograms, MFCCs, and Linear Frequency Cepstral Coefficients (LFCCs). 
Next, they applied a Maximal Information Coefficient (MIC)~\cite{mic_tech} to score each feature, selecting only the most influencing before feeding into a classifier to improve performance and reduce complexity. 
Similarly, Kok \textit{et al.}~\cite{ic_tree_19_embc04} applied the Wilcoxon Sum of Rank test to indicate which features among MFCCs, Discrete Wavelet Transform (DWT) and a set of time domain features (namely power, mean, variance, skewness and kurtosis of audio signal) mainly affected final classification accuracy.
Image processing techniques were then employed by Sengupta \textit{et al.}~\cite{sl_lbsvm_17_fl}, who applied Local Binary Pattern (LBP) analysis on Mel-frequency spectral coefficients (MFSCs) to capture texture information from the MFSC spectrogram, thus obtained an LBP spectrogram.
The LBP spectrogram was converted into a histogram presentation before feeding it into a back-end classifier, which was shown to outperform the previous MFCC-based methods.
In these systems, the time stream of audio feature vectors is classified by a range of traditional machine learning techniques. These include Logistic Regression (LR)~\cite{sl_lg_ano_16_embc02}, $k$-Nearest Neighbour (KNN)~\cite{sl_svm_ano_17_nw, sl_lbsvm_17_fl}, Hidden Markov Models (HMM)~\cite{sl_hmm_ano_14_embc03,  sl_hmm_19_sas, ic_hmm_18_sp}, Support Vector Machines (SVM)~\cite{sl_svm_ano_17_nw, sl_svm_ano_17_embc01, sl_lbsvm_17_fl, ic_pca_svm_18_springer01, ic_svm_18_sp} and Decision Trees (DT)~\cite{ic_tree_19_embc04, sl_svm_ano_17_nw, ic_tree_18_cbmi, ic_baseline}. 
\begin{table}[t]
    \caption{\textit{The state-of-the-art frame-based frameworks}} 
     \vspace{-0.1cm}
    \centering
    \scalebox{0.8}{
    \begin{tabular}{| l | c | c |} 
        \hline 
	    \textbf{Author}  & \textbf{Front-end} & \textbf{Back-end}  \\ [0.5ex] 
	               &  \textbf{Feature Extraction}      & \textbf{Classification} \\
        \hline 
         Okubo \emph{et al.}~\cite{sl_hmm_ano_14_embc03}        &MFCC                               &HMM \\
         Kok \emph{et al.}~\cite{ic_tree_19_embc04}             &MFCC, DWT, \& Handcrafted features  &RUSBoost \\
         Gr{\o}nnesby \textit{et al.}~\cite{sl_svm_ano_17_nw}   & MFCC \& Handcrafted features            &KNN, DT, SVM    \\
         Hanna \textit{et al.}~\cite{ic_tree_18_cbmi} & MFCC \& handcrafted features            &DT    \\

         Mendes \textit{et al.}~\cite{sl_lg_ano_16_embc02}      &MFCC \& Music based features              & LR    \\
         Datta \textit{et al.}~\cite{sl_svm_ano_17_embc01}      &PSD, FFT, Wavelet, MFCC, LFCC             & SVM     \\
         Sengupta \textit{et al.}~\cite{sl_lbsvm_17_fl}         &MFCC \& LBP             &KNN, SVM     \\
        \hline 
    \end{tabular}
    }
    \label{table:c06_ml} 
\end{table}

As we know from earlier chapters, deep learning techniques have achieved strong and robust detection performance for general sound classification~\cite{ivmCNNsounddet},~\cite{ivmDNNsounddet}.
Feature extraction in state-of-the-art deep learning based systems typically involves generating two-dimensional time-frequency spectrograms that are able to capture both fine grained temporal and spectral information as well as present a much wider time context than single frame analysis.
While a variety of spectrogram transformations have been utilised, Mel-based methods such as log-mel spectra~\cite{sl_rnn_19_ieeeopen, ic_cnn_19_ici, ic_cnn_20_ieee_bs} and stacked MFCC features~\cite{sl_rnn_19_ieeeopen, sl_cnn_17_eurasip, ic_cnn_18_bibm, ic_rnn_19_cbms, sl_rnn_18_embc05, ic_rnn_18_ann} are the most popular ones.
Some researchers combined different types of spectrogram, e.g. STFT and Wavelet as proposed by Minami \textit{et al.}~\cite{ic_cnn_19_iccas} or optimized S-Transformations in~\cite{ic_trip_19_ieeeopen}.
Although extracting good quality representative spectrograms is very important for a back-end classifier in general, researchers to date have not yet extensively explored the settings used in this step. This applies to both traditional sound classification, as well as to respiratory sound classification.

\begin{table}[t]
    \caption{\textit{The state-of-the-art spectrogram based frameworks}} 
     \vspace{-0.1cm}
    \centering
    \scalebox{0.8}{
    \begin{tabular}{| l | c | c |} 
        \hline 
	    \textbf{Author}  & \textbf{Front-end} & \textbf{Back-end}  \\ [0.5ex] 
	               &  \textbf{Feature Extraction}      & \textbf{Classification} \\
        \hline 
         Perna \emph{et al.}~\cite{ic_cnn_18_bibm}                   &MFCC             & LeNet (CNN)      \\
         Aykanat \emph{et al.}~\cite{sl_cnn_17_eurasip}                &MFCC             & LeNet (CNN)     \\
         Liu \emph{et al.}~\cite{ic_cnn_19_ici}                    & log-Mel            & VGG (CNN)       \\
         Minami \emph{et al.}~\cite{ic_cnn_19_iccas}                           &STFT \& Wavelet             & Parallel VGGs (CNN)   \\
         Chen \emph{et al.}~\cite{ic_trip_19_ieeeopen}                           & Optimized S-transform    &ResNet50 (CNN)     \\
         Perna \emph{et al.}~\cite{ic_rnn_19_cbms}                  & MFCC            & LSTM (RNN)  \\
         Kochetov \textit{et al.}~\cite{ic_rnn_18_ann}        &MFCC             & LSTM \& GRU (RNN)    \\
         Acharya \emph{et al.}~\cite{ic_cnn_20_ieee_bs}                &log-Mel             & Hybrid (CNN \& RNN)   \\
        \hline 
    \end{tabular}
    }
    \label{table:c06_dl} 
\end{table}
The most recent deep learning classifiers used with spectrogram input for research into respiratory sound analysis are mainly based on Convolutional Neural Networks (CNN), Recurrent Neural Networks (RNN), or hybrid architectures.
These CNN-based systems span some diverse architectures such as LeNet6~\cite{ic_cnn_18_bibm, sl_cnn_17_eurasip}, VGG5~\cite{ic_cnn_19_ici}, two parallel VGG16s~\cite{ic_cnn_19_iccas}, and ResNet50~\cite{ic_trip_19_ieeeopen}.
Inspired by the fact that respiratory indicative sounds such as \textit{Crackle} and \textit{Wheeze} present certain sequential characteristics, RNN-based networks have been developed in order to capture the sequential information. For example, Perna and Tagarelli~\cite{ic_rnn_19_cbms} analysed the use of a Long Short-term Memory (LSTM) network for two tasks of classifying anomalous respiratory sounds and classifying respiratory diseases.
By using LSTM and Gated Recurrent Unit (GRU) cells in a RNN-based network, Kochetov \textit{et al.}~\cite{ic_rnn_18_ann} proposed a novel architecture, namely the Noise Masking Recurrent Neural Network, which aimed to distinguish both noise and anomalous respiratory sounds.
In hybrid architectures proposed in~\cite{ic_cnn_20_ieee_bs}, a CNN was firstly used to map a spectrogram input to a temporal sequence. 
Then, LSTM was used to learn sequence structures before classification takes place via fully connected layers.
Compared to traditional machine learning approaches, state-of-the-art respiratory sound detection performance comparisons presented in~\cite{ic_rnn_19_cbms, ic_cnn_19_iccas, ic_trip_19_ieeeopen} indicate that deep learning classifiers are robust and effective.

\subsection{Exiting Issues and  Proposed Solution}
\label{c06_issue}

As the above literature review of respiratory sound analysis systems shows, both frame-based and spectrogram-based systems deployed for respiratory sound classification are very similar to the state-of-the-art ASC systems mentioned in Chapter \ref{c02}. 
Basically, the current systems used for respiratory analysis makes use of both machine learning in general, and deep learning techniques in particular, to achieve good results.
However, there exist specific issues that differ from the ASC research field.   
Firstly, the state-of-the-art systems involve ever-increasing model complexity, especially for those employing deep learning models, limiting their potential implementation within mobile or wearable real-time devices. 
While choosing a platform for ASC applications is flexible and less affected by efficiency in general, the practicalities of implementing respiratory sound analysis on edge devices is an important aspect of the benefit of such systems.
For instance,  if the function of respiratory sound analysis can be integrated into mobile phones, patients could self-check their situation at home, to track disease or recovery progression, for example.
Additionally,  analysing respiratory sound on real-time embedded devices helps to reduce the cost of manufacturing devices significantly, thus potentially increasing the ability of observation of lung disease, and the application on a much larger scale.
A more serious issue with this research field has been the difficulty of comparing between techniques due to the lack of standardised datasets used by authors for evaluation.
Most publications evaluated on proprietary datasets that are unavailable to others~\cite{sl_lg_ano_16_embc02, sl_svm_ano_17_embc01, sl_hmm_19_sas, sl_rnn_19_ieeeopen, sl_rnn_18_embc05}.
Comparing this to ASC, it is much easier to record an ASC dataset (no ethics approvals required, no access to patients, no infection control issues etc.) and so a number of good ASC dataset have been published for research and academic activities. Having more data, and more diversity of data available, is usually beneficial when building deep learning based systems, but for respiratory sound detection data is lacking.

To tackle these main issues, this chapter proposes a deep learning framework in the following way;
\begin{itemize}

\item Firstly to ensure repeatability and ease of comparison, the 2017 International Conference on Biomedical Health Informatics (ICBHI)~\cite{ic_dataset} dataset is used for all experiments.
The ICBHI dataset is one of the largest currently available which includes audio recordings. 
Using this resource, factors can be comprehensively analysed. This includes investigating different types of spectrogram, the use of overlapped or non-overlapped windowing, variable spectrogram patch sizes, the use of data augmentation techniques. In each case, the standard database allows their effects on performance to be precisely pinpointed.

\item From this analysis, a deep learning framework is proposed to target two related tasks of anomaly sound classification and respiratory disease detection. 
Two methods of train/test splitting are used in the literature (namely random 5-fold cross validation and 60/40 splitting as per the ICBHI challenge's recommendation). Both are evaluated here, and compared directly to state-of-the-art systems.

\item To aid in the trade-off between performance and complexity, a Student-Teacher scheme is proposed.
Specifically, the best deep learning framework, which is used for the task of respiratory disease detection and requires a large number of trainable parameters, is referred to as the Teacher.
Classification information from the Teacher model is extracted and distilled to train another network architecture with fewer trainable parameters, referred to as the Student.
Finally, a reduced-size Student network results, and when evaluated, is shown to achieve similar performance to the Teacher, but with significantly lower complexity.
\end{itemize}

\section{ICBHI Dataset and Tasks Proposed}
\label{c06_icbhi_tasks}

\subsection{ICBHI Dataset}
\label{c06_icbhi}

The 2017 ICBHI dataset~\cite{ic_dataset}, which was collected from School of Health Sciences, University of Aveiro (ESSUA) an Aristotle University of Thessaloniki (AUTH) as shown in Table \ref{table:c06_dataset}, provides a large database of labelled respiratory sounds comprising 920 audio recordings with a combined duration of 5.5 hours. 
\begin{table}[t]
    \caption{ICBHI dataset~\cite{ic_dataset}} 
        	\vspace{-0.2cm}
    \centering
    \scalebox{0.9}{
    \begin{tabular}{|l |c c c |c c c |} 
        \hline 
             \textbf{Database}          &  &\textbf{Testing Set} &  & &\textbf{Traing Set} & \\
                               &  \textbf{ESSUA}  &  \textbf{AUTH} &  \textbf{All}  &  \textbf{ESSUA} &  \textbf{AUTH} &  \textbf{All} \\
        \hline 
             \textbf{Patients}  	       & 38             & 11          & 49          & 72      & 7         & 79 \\        
             \textbf{Recordings}           & 617            & 64          & 381         & 507     & 32         & 539\\
             \textbf{Wheezes}		       & 588            & 61          & 649         & 459     & 42         & 501\\
             \textbf{Crackles}             & 273            & 112         & 385         & 1140    & 111         & 1215\\
             \textbf{Crackles \& Wheezes}  & 106            & 37          & 143         & 335     & 28         & 363\\
             \textbf{Normal}               & 1216           & 363         & 1579        & 1740    & 323         & 2063\\

       \hline 
    \end{tabular}
    }
    \label{table:c06_dataset} 
\end{table}
The recording lengths are uneven, ranging from from 10 to 90 seconds, and were recorded with a wide range of sampling frequencies from 4000\,Hz to 44100\,Hz. 
In total, the dataset contains recordings from 128 patients, who are identified in terms of being healthy or exhibiting one of the following respiratory diseases or conditions: COPD, Bronchiectasis, Asthma, upper and lower respiratory tract infection, Pneumonia, Bronchiolitis.
These respiratory condition labels are linked to audio recording files.
Within each audio recording, four different types of respiratory cycle are presented -- called \textit{Crackle}, \textit{Wheeze}, \textit{Both (Crackle \& Wheeze)}, and \textit{Normal}.
These cycles, labelled by experts, include identified onset and offset times.
The cycles have various recording lengths ranging from 0.2 up to 16.2 seconds, with the number of cycles being unbalanced (i.e. 1864, 886, 506 and 3642 cycles respectively for \textit{Crackle}, \textit{Wheeze}, \textit{Both}, and \textit{Normal}).

\subsection{Main Tasks Proposed for ICBHI Dataset}
\label{c06_tasks}

Given the ICBHI recordings and metadata, this chapter targets performance over two main tasks.

\textbf{Task 1}, respiratory anomaly classification, is separated into two sub-tasks.
The first aims to classify four different cycles (\textit{Crackle}, \textit{Wheeze}, \textit{Both}, and \textit{Normal}).
The second is to classify the four types of cycle into two groups of \textit{Normal} and \textit{Anomaly} sounds (the latter group consisting of \textit{Crackle}, \textit{Wheeze}, and \textit{Both}). 
For convenience, these are named Task 1-1 and Task 1-2, respectively in this thesis.

\textbf{Task 2}, respiratory disease prediction, also comprises two sub-tasks.
The first aims to classify audio recordings into three groups of disease conditions: \textit{Healthy}, \textit{Chronic Disease} (i.e. COPD, Bronchiectasis and Asthma) and \textit{Non-Chronic Disease} (i.e. upper and lower respiratory tract infection, Pneumonia, and Bronchiolitis) (Note that individual disease condition is not directly classified as the dataset is unbalanced, especially only 1 patient with Asthma disease and 2 patients with LRTI disease).
The second sub-task is for classification into two groups of \textit{Healthy} and \textit{Unhealthy} (comprising the \textit{Chronic} and \textit{Non-Chronic} disease groups combined).
These sub-tasks are referred to as Tasks 2-1 and Task 2-2, respectively in this thesis.
While Tasks 1-1 and 1-2 are evaluated over individual respiratory cycles, Task 2-1 and 2-2 are evaluated over entire audio recordings. 

State-of-the-art published systems that used the ICBHI dataset follow two different approaches to split the database into training and testing portions.
The first~\cite{ic_hmm_18_sp, ic_svm_18_sp, ic_baseline, ic_cnn_19_iccas} followed the ICBHI challenge recommendations~\cite{ic_dataset} to divide the dataset into non-overlapping 60\% and 40\% portions for training and test subsets, respectively.
Notably, this avoids a situation in which audio recordings from one subject are found in both of the subsets.
Meanwhile, the second~\cite{ic_tree_19_embc04, ic_tree_18_cbmi, ic_cnn_19_ici, ic_cnn_18_bibm, ic_rnn_19_cbms} randomly separated the entire dataset into training and test subsets, with different ratios. 

To evaluate our proposed framework on each task in this chapter, the ICBHI dataset is firstly separated (6898 respiratory cycles for Task 1 and 920 entire recordings for Task 2) into five folds for cross validation.
Next, a baseline system is introduced to evaluate the effects of a number of settings and influencing factors as noted above. 
Due to extensive training times, this initial exploration evaluates on one cross-validation fold.
Then, following the initial exploration, two systems are proposed; one for anomaly cycle detection (Tasks 1-1 and 1-2) and the other for respiratory disease detection (Tasks 2-1 and 2-2).
Each of those systems is then trained and evaluated with both the full 5-fold cross validation and 60/40 split as the ICBHI challenge's recommendation. Each system is then compared against state-of-the-art methods. 

\subsection{Evaluation Metrics}
\label{tasks}
The baseline and proposed framework variants are assessed using the metrics of \textit{Sensitivity} (Sen.), \textit{Specitivity} (Spec.), and \textit{ICBHI score}~\cite{ic_rnn_19_cbms, ic_dataset}.
To understand these scores, a confusion matrix for Task 1 as presented in Table \ref{table:c06_cycle_tab} is considered. 
In this case, \textit{C, W, B,} and \textit{N} denote the numbers of cycles of  \textit{Crackle}, \textit{Wheeze}, \textit{Both}, and \textit{Normal} respectively, whereas \textit{c, w, b,} and \textit{n} subscripts indicate the classification results. The sums $C_{t}$, $W_{t}$, $B_{t}$ and $N_{t}$ are the total numbers of cycles. 
\textit{Sensitivity} is computed for Task 1-1 (4-class anomaly classification) as:
 \begin{equation}
     \label{eq:task_1_1_sen}
     Sensitivity =  \frac{C_{c} + W_{w} + B_{b}}{C_{t} + W_{t} + B_{t}}
 \end{equation}
and for Task 1-2 (binary anomaly classification) as:
 \begin{equation}
     \label{eq:task_1_2_sen}
     Sensitivity =  \frac{C_{c+w+b} + W_{c+w+b} + B_{c+w+b}}{C_{t} + W_{t} + B_{t}}
 \end{equation}
where $C_{c+w+b} = C_c + C_w + C_b$, $W_{c+w+b} = W_c + W_w + W_b$, and $B_{c+w+b} = B_c + B_w + B_b$. Then, \textit{Specificity} can be defined 

 \begin{equation}
Specificity =  \frac{N_{n}}{N_{t}}
 \end{equation}
%
\begin{table}[t]
    \caption{Confusion matrix of anomaly cycle classification.} 
        	\vspace{-0.2cm}
    \centering
    \scalebox{0.85}{
    \begin{tabular}{|c |c |c |c |c  |} 
        \hline 
                               &  \textbf{Crackle}  &  \textbf{Wheeze} &  \textbf{Both}  &  \textbf{Normal} \\
        \hline 
             \textbf{Crackle}  	& $C_{c}$            & $W_{c}$          & $B_{c}$         & $N_{c}$ \\        
             \textbf{Wheeze}   & $C_{w}$            & $W_{w}$          & $B_{w}$         & $N_{w}$ \\
             \textbf{Both}		& $C_{b}$            & $W_{b}$          & $B_{b}$         & $N_{b}$ \\
             \textbf{Normal}   & $C_{n}$            & $W_{n}$          & $B_{n}$         & $N_{n}$ \\
       \hline 
             \textbf{Total}    	& $C_{t}$            & $W_{t}$          & $B_{t}$         & $N_{t}$ \\
       \hline 
    \end{tabular}
    }
    \label{table:c06_cycle_tab} 
\end{table}
%
\begin{table}[t]
    \caption{Confusion matrix of respiratory disease detection.} 
        \vspace{-0.2cm}
    \centering
    \scalebox{0.85}{
    \begin{tabular}{| c |c  |c  |c  |} 
        \hline 
              &  \textbf{Chronic}  &  \textbf{Non-chronic} &  \textbf{Healthy}       \\
        \hline 
             \textbf{Chronic}     & $C_{c}$    & $NC_{c}$    & $H_{c}$        \\ 
	     \textbf{Non-chronic} & $C_{nc}$ & $NC_{nc}$ & $H_{nc}$     \\
	     \textbf{Healthy}	  & $C_{h}$     & $NC_{h}$     & $H_{h}$         \\ 
       \hline 
	     \textbf{Total}	  & $C_{t}$     & $NC_{t}$     & $H_{t}$         \\            
       \hline 
    \end{tabular}
    }
    \label{table:c06_disease_tab} 
\end{table}
Similarly, Task 2's confusion matrix as shown in Table \ref{table:c06_disease_tab} is considered. 
In this case, \textit{C, NC} and \textit{H} are the numbers of recordings of the three Task 2 classes. \textit{c, nc} and \textit{h} subscripts indicate the classification results. As before, $C_{t}$, $NC_{t}$, and $H_{t}$ are the total numbers of \textit{Chronic}, \textit{Non-chronic}, and \textit{Healthy} recordings, respectively.

For Task 2-1,  \textit{Sensitivity} is defined as follows:
\begin{equation}
     \label{eq:task_2_1_sen}
     Sensitivity =  \frac{C_{c} + NC_{nc}}{C_{t} + NC_{t}}
 \end{equation}
and for Task 2-2 as:
  \begin{equation}
     \label{eq:task_2_2_sen}
     Sensitivity =  \frac{(C_{c} + C_{nc}) + (NC_{c} + NC_{nc})}{C_{t} + NC_{t}}
 \end{equation}
\textit{Specificity} is then defined as 
  \begin{equation}
Specificity =  \frac{H_{h}}{H_{t}}
 \end{equation}
Regarding the composite \textit{ICBHI score}, this represents an equal trade-off between the two metrics and is computed in the same way for each task -- namely averaging the \textit{Sensitivity} and the \textit{Specificity} scores. \\
\begin{equation}
ICBHI score =  \frac{1}{2}(Specificity + Sensitivity)
 \end{equation}

\section{High-level Framework Architecture}
\label{c06_architecture}

\subsection{High-level Description}
\label{c06_high_level}

The high-level architecture of the proposed system is described in Figure~\ref{fig:c06_high_level_arc}.
The architecture is divided into two main parts: front-end feature extraction (the upper part) and back-end deep learning models (the lower part).
In general, respiratory cycles in Task 1 or entire audio recording in Tasks 2 are transformed into one or more spectrogram representations.
The spectrograms are then split into equal-sized image patches.
During training, mixup data augmentation~\cite{aug_mixup_s01, aug_mixup_s02} is applied to the patches to generate an expanded set of training data that is fed into a deep learning classifier.
\begin{figure}[t]
    \centering
    \includegraphics[width =\linewidth]{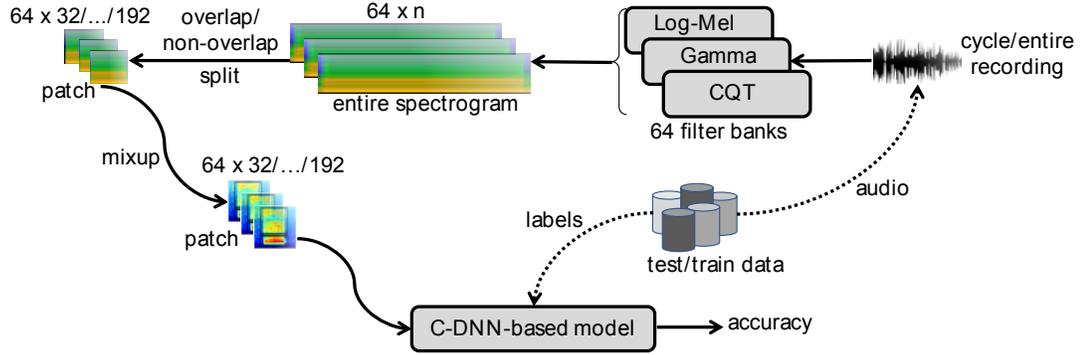}
	\caption{The high-level architecture and\\  processing pipeline of the proposed framework.}
    \label{fig:c06_high_level_arc}
\end{figure}
 
\subsection{Baseline System}
\label{c06_baseline}

From the high-level architecture shown in Figure~\ref{fig:c06_high_level_arc}, it can be seen that a variety of factors in front-end feature extraction could affect the performance of the classifier. 
These include the type of spectrogram used, the size of image patches and their degree of overlap, and the use of data augmentation. 
These factors are investigated, thus indicate the most influencing factors among those listed above. 
To limit the investigation scope to manageable proportions, the deep learning architecture assessed is constrained, thus a C-DNN baseline like VGG-7~\cite{vgg_net} is proposed, defined below.

The main characteristics and settings of this baseline architecture are listed in Table \ref{table:c06_baseline}, while the network architecture is presented in Table~\ref{table:c06_C-DNN}.
During processing, all audio recordings are re-sampled (they were, as aforementioned, recorded with various sample rates) to 16000 Hz mono.
Since respiratory cycle lengths differ quite widely, short cycles are repeated to ensure that inputs for Task 1 have a minimum length of 5 seconds or longer. This is of course unnecessary for Task 2 which uses entire recordings.
Next, each cycle (for Task 1) or recording (for Task 2) is transformed into a spectrogram with 64 features per analysis frame.
For example, the log-mel spectrogram is extracted with a window size of 1024 samples, a hop size of 256 samples, and 2048-point FFT, followed by average pooling in the frequency direction to yield a spectrogram with 64 frequency bins.
Whichever type of spectrogram is used, the resulting time-frequency output is split into square non-overlapping patches of size $64{\times}64$.
Since data augmentation is one of factors evaluated, this technique is not applied to the baseline system.

As can be seen from Table~\ref{table:c06_C-DNN}, the network architecture consists of seven blocks -- six are convolutional and one is a dense block. The former blocks comprise batch normalization (BN) layers, convolutional (Cv [kernel size] $@$ kernel number) layers, rectified linear units (ReLU), average pooling (AP [kernel size]) and global average pooling (GAP) layers, and use dropout (Dr (dropout percentage)). The dense block comprises a fully connected (FC), and a final Softmax layer for classification.
\textit{C} refers to the number of classes, which depend on the specific task being evaluated (i.e. two separate C-DNN models are trained and test with \textit{C} set to 4 and 3 for Tasks 1 and 2, respectively).
\begin{table}[t]
    \caption{Baseline system settings.} 
        	 \vspace{-0.2cm}
    \centering
    \scalebox{0.85}{
    \begin{tabular}{|l |c |} 
        \hline 
            \textbf{Factors}   &  \textbf{Setting}  \\
        \hline 
             Re-sample  & 16kHz \\         
             Cycle duration (only for Task 1) & 5s \\         
             Spectrogram & log-mel \\         
             Patch splitting & non-overlapped \\         
             Patch size & $64\times64$ \\         
             Data augmentation & None \\         
             Deep learning model & C-DNN based architecture\\
      \hline 
    \end{tabular}
    }
    \label{table:c06_baseline} 
\end{table}
\begin{table}[t]
    \caption{Baseline C-DNN network architecture} 
         \vspace{-0.2cm}
    \centering
    \scalebox{0.85}{
    \begin{tabular}{|l |l |c |} 
        \hline 
            \textbf{Architecture} & \textbf{Layers}   &  \textbf{Output}  \\
        \hline 
             & Input layer (image patch)  &          $64{\times}64$   \\
         Conv. Block 01   & BN - Cv [$3{\times}3$] $@$ 64 - ReLU - BN - AP [$2{\times}2$] - Dr ($10\%$)      & $32{\times}32{\times}64$\\
         Conv. Block 02  & BN - Cv [$3{\times}3$] $@$ 128 - ReLU - BN - AP [$2{\times}2$] - Dr ($15\%$)      & $16{\times}16{\times}128$\\
         Conv. Block 03  & BN - Cv [$3{\times}3$] $@$ 256 - ReLU - BN - Dr ($20\%$)      & $16{\times}16{\times}256$ \\
         Conv. Block 04  & BN - Cv [$3{\times}3$] $@$ 256 - ReLU - BN - AP [$2{\times}2$] - Dr ($20\%$)       & $8{\times}8{\times}256$\\
         Conv. Block 05  & BN - Cv [$3{\times}3$] $@$ 512 - ReLU - BN  - Dr ($25\%$)      & $8{\times}8{\times}512$ \\
         Conv. Block 06  & BN - Cv [$3{\times}3$] $@$ 512 - ReLU - BN -  GAP - Dr ($25\%$) & $512$ \\           
         Dense Block & FC - Softmax  & $C$ \\         
       \hline 
    \end{tabular}
    }
    \label{table:c06_C-DNN} 
\end{table}
\subsection{Experimental Settings for The Baseline System}
\label{c05_baseline}

All the systems are implemented using TensorFlow. Network training makes use of the Adam optimiser~\cite{kingma2014adam} with $100$ training epochs, a mini batch size of $100$, and cross entropy loss:
\begin{equation}
    \label{eq:entro_loss}
    LOSS_{EN}(\Theta) = -\sum_{c=1}^{C}y_{c}\log\left\lbrace\hat{y}_{c}(\Theta)\right\rbrace + \frac{\lambda}{2}||\Theta||_{2}^{2},
\end{equation}
where \(\Theta\) are all trainable parameters, \(C\) is the number of categories classified, and constant \(\lambda\) is empirically set to 0.001.  
$y_{c}$ and $\hat{y}_{c}$ denote ground truth and predicted results of class $c$, respectively.

An entire spectrogram or cycle is separated into smaller patches and applied patch-by-patch to the C-DNN model which then returns the probability computed over each patch. 
The probability of an entire spectrogram is the average of all patches' probabilities.
Let us consider $\mathbf{p}^{n} = (p_{1}^{n}, p_{2}^{n},\ldots,p_{C}^{n})$ the probability obtained from the \(n^{th}\) out of \(N\) patches. Then, the mean probability of a test sound instance is denoted as \(\mathbf{\bar{p}} = (\bar{p}_{1}, \bar{p}_{2},\ldots, \bar{{p}}_{C})\) where
\begin{equation}
\label{eq:mean_stratergy_patch}
\bar{p}_{c} = \frac{1}{N}\sum_{n=1}^{N}p_{c}^{n}  
\end{equation}
The predicted label  \(\hat{y}\) is then determined as 
\begin{equation}
\hat{y} = \argmax_{c \in \{1,2,\ldots,C\}}\bar{p}_c.
\end{equation}

\section{Analysis of Influencing Factors}
\label{c06_analyse}

By using the baseline system mentioned above, the impact of various factors on performance is investigated below.

\subsection{Influence of Spectrogram Type}
\label{c06_spec}
\begin{figure}[h]
	\centering
	\includegraphics[width =0.85\linewidth]{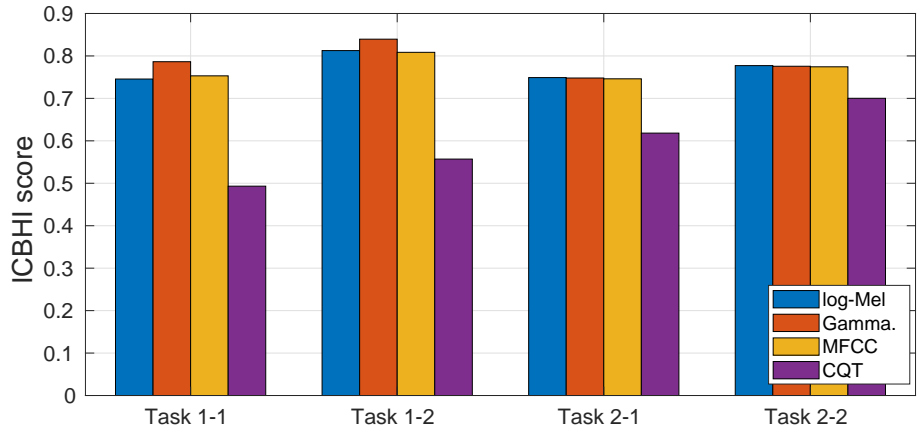}
	\vspace{-0.1 cm}
	\caption{Comparison of baseline performance using different spectrograms.}
	\label{fig:c06_spec_01}
\end{figure}
From previous work on natural sound datasets~\cite{dc_16_lam_int, dc_18_lam_aes}, it is clear that the choice of spectrogram is one of the most important factors that affects final classification accuracy.
Therefore, the effect of spectrogram types on ICBHI performance for each task is evaluated.
To this end, all settings as described in Table \ref{table:c06_baseline} are maintained, but four spectrogram types: log-mel spectrogram, Gamma spectrogram, stacked Mel-Frequency Cepstral Coefficients (MFCC), and rectangular Constant Q Transform (CQT) spectrogram are used. 
Each of the spectrogram types is evaluated on all four subtasks.
Just as in the experiments presented in Chapter \ref{c03}, while log-mel, MFCC, and CQT spectrograms are generated by using the Librosa toolbox~\cite{librosa_tool}, and the Gamma spectrogram by \cite{auditory2009_tool}  (note that detailed computation of these spectrograms are described in the Appendix).

The obtained results in terms of ICBHI Score are shown in Figure~\ref{fig:c06_spec_01}, revealing that MFCC, log-mel, and Gamma spectrogram perform competitively, and are much better than CQT for all subtasks. 
Compared to log-mel, Gamma spectrogram results achieve an improvement of 0.04 for Task 1-1 and 0.03 for Task 1-2. 
However log-mel slightly outperforms its Gamma counterpart for Task 2.
MFCC is, meanwhile, better than log-mel in Task 1-1 (0.01) but the opposite is seen for all other subtasks.

These results suggest that the Gamma spectrogram is optimal for anomaly cycle classification (Task 1) while the log-mel spectrogram works best for detection of respiratory diseases (Task 2). As a result these two spectrograms are adopted in the following experiments for those respective tasks.

\subsection{Influence of the overlapping degree}
\label{overlap}
\begin{table}[h]
    \caption{Baseline performance loss or gain on each subtask \\ when overlapping spectrogram patches are used (ICBHI Score).}
        	\vspace{-0.1cm}
    \centering
    \scalebox{0.85}{
    \begin{tabular}{|l |c |c |c |c |} 
        \hline 
          \textbf{Patches} & \textbf{Task 1-1} & \textbf{Task 1-2}   &  \textbf{Task 2-1} &  \textbf{Task 2-2}  \\
        \hline 
         No overlap & \textbf{0.79} &\textbf{0.84}  &0.75                &0.77\\
        Overlap          &  0.78              & 0.83                &\textbf{0.77} & \textbf{0.79} \\
         \hline 
    \end{tabular}
    } 
    \label{table:c06_overlap} 
\end{table}
As the spectrogram of an entire cycle or audio recoding is large in temporal dimension and is of variable length, they are split into smaller patches of $64\times64$ for presentation before feeding to the back-end deep learning models. 
In traditional signal processing systems, overlapping analysis windows are used to prevent occlusion of important features in the original data by edge effects. 
Therefore, the effect of overlapped or non-overlapped patches on ICBHI performance is examined in this section.
Specifically, the baseline with non-overlapping patches (the settings in Table \ref{table:c06_baseline}) is contrasted to the system with patches overlapped by 50\% (note that Gamma and log-mel are applied on Task 1 and Task 2, respectively).

Results shown in Table \ref{table:c06_overlap} reveal that Task 1 performs better with non-overlapped patches (subtask scores of 0.79 and 0.84, respectively) while those results for Task 2 performs better with overlapped patches (subtask scores of 0.77 and 0.79, respectively).
These results can be explained by two potential factors: firstly different spectrogram types were used in the two tasks, and secondly Task 1 repeats respiratory cycles, whereas Task 2 classifies unrepeated recordings.

\subsection{Influence of Time Resolution}
\label{resolution}
\begin{figure}[h]
	\centering
	\includegraphics[width =0.85\linewidth]{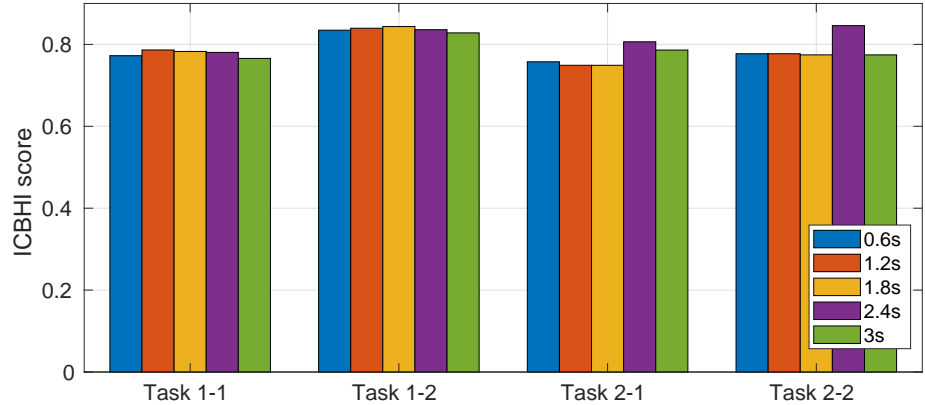}
	\vspace{-0.1 cm}
	\caption{Performance comparison between different time resolutions on each task.}
	\label{fig:c06_time_res}
\end{figure}
The baseline network operates on fixed-size patches where the time span encoded in each patch is defined by its horizontal dimension and sampling rate.
Features are presented sequentially, and therefore the time span also defines the temporal resolution of features presented to the classifier.
In this section, the effect of different temporal resolution is explored by adjusting patch widths to 0.6\,s, 1.2\,s, 1.8\,s, 2.4\,s, and 3.0\,s. 
This is achieved  by changing the patch size to be $64{\times}32$, $64{\times}64$, $64{\times}96$, $64{\times}128$, and $64{\times}160$, respectively, then repeat the experiments for each of them.
Note that all settings are reused from Table \ref{table:c06_baseline} with exception that Gamma and log-mel spectrograms are used for Task 1 and Task 2, respectively. The frequency resolution (vertical dimension) remains unchanged in each case. The dimension of the network input layer is increased or decreased to accommodate the differing time resolution.

The obtained results are shown in Figure~\ref{fig:c06_time_res} for the four subtasks. 
As can be seen, patch size of $64\times64$ (i.e. 1.2\,s time resolution) as in the baseline system performs best for Task 1-1 and second best for Task 1-2 (achieving 0.79 and 0.84, respectively). 
However, a double sized patch, $64\times128$ (i.e. 2.4\,s time resolution) is clearly the best for Tasks 2-1 and 2-2 (achieving 0.81 and 0.85, respectively).

\subsection{Influence of Data Augmentation}
\label{c06_augmentation}
\begin{table}[h]
    \caption{Performance (ICBHI Score) with and without mixup data augmentation.} 
        	\vspace{-0.3cm}
    \centering
    \scalebox{0.85}{
    \begin{tabular}{|l |c |c |c |c |} 
        \hline 
           & \textbf{Task 1-1} & \textbf{Task 1-2}   &  \textbf{Task 2-1} &  \textbf{Task 2-2}  \\
        \hline 
         Non-mixup & 0.79              &0.84               &0.75                &0.77\\
         mixup         & \textbf{0.80} &\textbf{0.85} & \textbf{0.84} & \textbf{0.85} \\
         \hline 
    \end{tabular}
    } 
    \label{table:c06_augmentation} 
\end{table}
Data augmentation (DA) has been shown useful to improve the learning ability of deep learning models in tasks involving natural sound classification~\cite{dc_18_lam_aes, dc_16_lam_int}.
Therefore, DA in the form of mixup~\cite{aug_mixup_s01, aug_mixup_s02} is applied and evaluated its effect on respiratory sound classification.

By using two types of Uniform or Beta Distribution to generate mixing coefficient $\alpha$, this doubles the data size and hence, the training time. Note that in Task 1, the DA mixes the \textit{Normal} class with one of the other classes (since there is already one mixed class in the dataset, i.e. \textit{Crackle \& Wheeze}), whereas it randomly mixes samples of all classes for Task 2.
After mixup, the generated patches are shuffled and fed into the C-DNN baseline. 
Since the labels $\mathbf{y}_{mp1}$ and $\mathbf{y}_{mp2}$ of the resulting patches are no longer one-hot encoded, it is, therefore, necessary to replace the cross-entropy loss by the Kullback-Leibler (KL) divergence loss~\cite{kl_loss}:
\begin{align}
   \label{eq:kl_loss}
   LOSS_{KL}(\Theta) = \sum_{c=1}^{C}y_{c}\log \left\{ \frac{y_{c}}{\hat{y}_{c}(\Theta)} \right\}  +  \frac{\lambda}{2}||\Theta||_{2}^{2}.
\end{align}
Again, $\Theta$ denotes the trainable network parameters and $\lambda$ denote the $\ell_2$-norm regularization coefficient, set to 0.001. \(C\) is the number of categories classified,
$y_{c}$ and $\hat{c}_{n}$  denote the ground-truth and the network output at class $c$, respectively.

Using the settings in Table \ref{table:c06_baseline} with Gamma spectrogram in Task 1 and the log-mel spectrogram in Task 2, the improvement over the baseline ICBHI score for each subtask due to mixup data augmentation can be assessed.
Results shown in Table \ref{table:c06_augmentation} indicate that mixup data augmentation substantially improves the ICBHI score in Task 2 by 0.09 and 0.08 on Tasks 2-1 and 2-2, respectively. However, modest improvements are seen for Task 1.

\section{Enhanced Deep Learning Framework}
\label{c06_enhance_framework}
\begin{table}[h!]
    \caption{Deep learning frameworks for Tasks 1 and 2.}  
        	\vspace{-0.2cm}
    \centering
    \scalebox{0.85}{
    \begin{tabular}{|l |c |c|} 
        \hline 
            \textbf{Factors}   &  \textbf{Anomaly cycle}  &  \textbf{Respiratory disease}  \\
                                         &  \textbf{classification}  &  \textbf{detection} \\
        \hline 
             Resample  & 16000 Hz & 16000 Hz\\         
             Cycle duration & 5s & N/A \\         
             Spectrogram & Gamma & log-mel \\         
             Patch splitting & non-overlapped & overlapped \\         
             Patch size & $64\times64$ & $64\times128$ \\         
             Data augmentation & Yes & Yes \\      
     \hline 
    \end{tabular}
    }
    \label{table:c06_framework} 
\end{table}
From the analysis of influencing factors presented above, two systems are proposed. 
One for Task 1 anomaly cycle classification, and the other for Task 2 respiratory disease detection, both summarised in  Table \ref{table:c06_framework}.
In this section, the performance of the C-DNN architecture is enhanced by incorporating a mixture-of-experts (MoE) technique into the DNN part of the network, leading to a CNN-MoE architecture, similar to that in Chapter \ref{c05}.

\subsection{CNN-MoE Network Architecture}
\label{c06_framework_architecture}
According to the C-DNN architecture entailed in Table \ref{table:c06_C-DNN}, the first six convolutional blocks are used to map the image patch input to condensed and discriminative embeddings, often referred to as high-level features.
The features are then classified by a dense block comprising a fully-connected layer and Softmax.
On the basis that the embedding may contain more information than a single fully connected layer can unlock, the dense block in replaced by a mixture-of-experts (MoE) block as shown in Figure~\ref{fig:c06_framework}.
The MoE block architecture is reused and mentioned in Section \ref{c04_decoder} 

The proposed systems, as defined in Table \ref{table:c06_framework}, are trained with KL-divergence loss~\cite{kl_loss} (due to the use of mixup data augmentation) and use the same training settings as the previous experiments with the C-DNN baseline.
\begin{figure}[t]
    \centering
    \includegraphics[width =\linewidth]{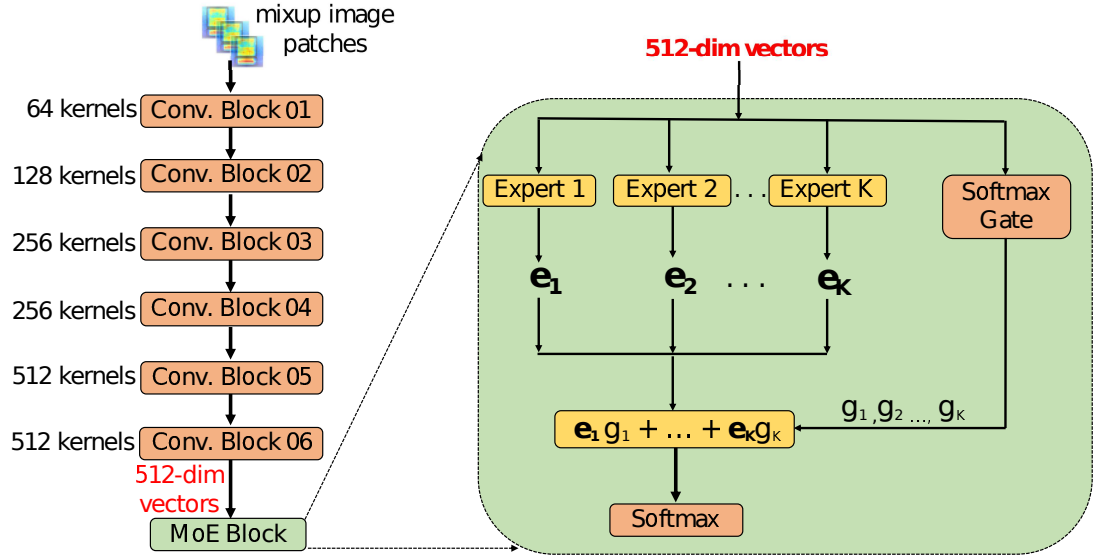}
    	\vspace{-0.2 cm}
    	\caption{The proposed CNN-MoE architecture.}
    \label{fig:c06_framework}
\end{figure}

\subsection{Performance Comparison}
\label{c06_compare}
This section firstly compares the performance of using C-DNN and CNN-MoE, analyses if MoE technique is effective to improve the classification accuracy. Next, the best systems proposed are compared to the state of the art.
 
\textbf{Comparing C-DNN to CNN-MoE}: The efficiency of the MoE technique (experimentally using $K$=10 experts) is evaluated and compared to the C-DNN system, reporting the performance of both in Table \ref{table:c06_comp_moe} (note that both the systems follows the settings in Table \ref{table:c06_framework}, with the back-end classifier being either C-DNN or CNN-MoE – there are thus eight systems in total, two C-DNNs and two CNN-MoEs for each kind of data split).
The results in Table \ref{table:c06_comp_moe} clearly indicate that the CNN-MoE systems perform best overall. Although only marginal gains is seen over the C-DNN for Task 1, results in improvement with a margin as large as 0.06 absolute in terms of ICBHI score with both the data splits, 5-fold cross validation and ICBHI challenge's data split, in Task 2.  
\begin{table}[t]
    \caption{ICBHI score comparison between the C-DNN and CNN-MoE frameworks over 5-fold cross validation and ICBHI challenge splitting (highest scores in \textbf{bold}).} 
        	\vspace{-0.2cm}
    \centering
    \scalebox{0.85}{

    \begin{tabular}{|c  |c |c |c |c |} 
        \hline            
           & \textbf{C-DNN} &   \textbf{CNN-MoE} &  \textbf{C-DNN} &   \textbf{CNN-MoE} \\
                \textbf{Tasks}      & (5-fold) &  (5-fold) & (ICBHI) &  (ICBHI)\\

        \hline 
        1-1, 4-category                         &0.77              &\textbf{0.79}     &0.43 & \textbf{0.47}\\        
	    1-2, 2-category                             &0.84    &  \textbf{0.84}      &0.53  & \textbf{0.54} \\
	    2-1, 3-category                                   &84.7    &\textbf{0.91}    &0.79 &   \textbf{0.84}         \\
	    2-2, 2-category                        &0.86    &\textbf{0.92}  & 0.79 &  \textbf{0.84} \\
       \hline 
    \end{tabular}
    }
    \label{table:c06_comp_moe} 
\end{table}

\textbf{Comparing to state-of-the-art systems:} Next, the proposed framework is contrasted to state-of-the-art systems.
For each task, challenge's data split is evaluated twice -- once with the ICBHI challenge train/test split, and once with random splitting (as described in Section~\ref{tasks}).
Considering the first splitting method specified in the ICBHI challenge, Table \ref{table:c06_comp_sta_icb} presents scores obtained by the proposed framework and state-of-the-art published systems (where available). 
It is noted that the proposed framework lies second in terms of Task 1-1 evaluation. 
Our results for other subtasks were listed in Table \ref{table:c06_comp_moe}. Only Task 2-2 is found in the literature (for the ICBHI data split) achieved 0.72~\cite{ic_cnn_20_ieee_bs}, which is surpassed by 0.84 obtained by our system.
\begin{table}[t!]
    \caption{Comparison against state-of-the-art systems with \\ ICBHI challenge splitting (highest scores in \textbf{bold}).} 
        	\vspace{-0.2cm}
    \centering
    \scalebox{0.85}{

    \begin{tabular}{|c |l |l |c |c |c |} 
        \hline 
	    \textbf{Tasks}   &\textbf{Method}                        &\textbf{Spec.}   &\textbf{Sen.}   &\textbf{Score}  \\
        \hline 
        1-1, 4-category      &DT~\cite{ic_baseline}                    &0.75             &0.12           &0.43  \\        
        1-1, 4-category      &HMM~\cite{ic_hmm_18_sp}                   &0.38             &\textbf{0.41}           &0.39  \\        
        1-1, 4-category      &SVM~\cite{ic_svm_18_sp}                   &0.78             &0.20           &0.47  \\
        1-1, 4-category     &CNN-RNN~\cite{ic_cnn_19_iccas}          &\textbf{0.81} &0.28 & \textbf{0.54}    \\ 
	    1-1, 4-category      &\textbf{Our system}                                &0.68    &0.26   &0.47    \\
       \hline 
    \end{tabular}
    }
    \label{table:c06_comp_sta_icb} 
\end{table}

Table \ref{table:c06_comp_sta_rand} compares the performance obtained by our system with previously published results that use the random train/test splitting method. For Tasks 1-1 and 1-2, the proposed framework clearly outperforms other systems quite consistently. 
Meanwhile for Task 2-1 and 2-2 the proposed method also outperforms other systems in terms of overall ICBHI score, but not necessarily simultaneously for both subcomponents of specificity or sensitivity.
\begin{table}[t!]
    \caption{Performance comparison between the proposed system and state-of-the-art systems following random splitting (highest scores are highlighted in \textbf{bold}).} 
        	\vspace{-0.2cm}
    \centering
    \scalebox{0.85}{

    \begin{tabular}{|c |l |l |c |c |c |c |} 
        \hline 
	    \textbf{Tasks}   &\textbf{Methods}                        &\textbf{train/test}      &\textbf{Spec.}   &\textbf{Sen.}   &\textbf{Score}  \\
        \hline 
	    1-1, 4-category      &Boosted DT~\cite{ic_tree_18_cbmi}  &60/40                    &0.78             &0.21            &0.49  \\
	    1-1, 4-category      &CNN~\cite{ic_cnn_18_bibm}            &80/20                    &0.77             &0.45            &0.61  \\
    	    1-1, 4-category      &CNN-RNN~\cite{ic_cnn_20_ieee_bs}  &5 folds      &0.84             &0.49           &0.66  \\
	    1-1, 4-category      &LSTM~\cite{ic_rnn_19_cbms}           &80/20                    &0.85             &0.62            &0.74  \\
	    1-1, 4-category      &\textbf{Our system}                  &5 folds               &\textbf{0.90}    &\textbf{0.68}   &\textbf{0.79}    \\
       \hline 
	    1-2, 2-category      &Boosted DT~\cite{ic_tree_18_cbmi}  &60/40                    &0.78             &0.33            &0.56  \\
	    1-2, 2-category      &LSTM~\cite{ic_rnn_19_cbms}           &80/20                    &-                &-               &0.81  \\
   	    1-2, 2-category      &CNN~\cite{ic_cnn_19_ici}           &75/25                    &-                &-               &0.82  \\
	    1-2, 2-category      &\textbf{Our system}                  &5 folds               &\textbf{0.90}    &\textbf{0.78}   &\textbf{0.84}    \\
       \hline 
       \hline
	    2-1, 3-category      &CNN~\cite{ic_cnn_18_bibm}            &80/20                    &0.76             &0.89             &0.83  \\
	    2-1, 3-category      &LSTM~\cite{ic_rnn_19_cbms}           &80/20                    &0.82             &\textbf{0.98}    &0.90   \\
	    2-1, 3-category      &\textbf{Our system}                  &5 folds               &\textbf{0.86}    &0.95             &\textbf{0.91}      \\
       \hline 
   	    2-2, 2-category      &Boosted DT~\cite{ic_tree_18_cbmi}  &60/40                    &0.85             &0.85            &0.85  \\
	    2-2, 2-category      &CNN~\cite{ic_cnn_18_bibm}            &80/20                    &0.78             &0.97             &0.88  \\
   	    2-2, 2-category      &RUSBoost DT~\cite{ic_tree_19_embc04}            &50/50                    &\textbf{0.93}             &0.86             &0.90  \\
	    2-2, 2-category      &LSTM~\cite{ic_rnn_19_cbms}           &80/20                    &0.82             &\textbf{0.99}    &0.91 \\
	    2-2, 2-category      &\textbf{Our system}                  &5 folds               &0.86    &0.98    &\textbf{0.92} \\
       \hline 
    \end{tabular}
    }
    \label{table:c06_comp_sta_rand} 
\end{table}

\subsection{Discussion}
\label{c06_discussion}

Comparing Tables \ref{table:c06_comp_sta_icb} and \ref{table:c06_comp_sta_rand}, it is notable that those systems following the ICBHI data split (i.e. recordings from the same patient are never found in both train/test subsets) exhibit considerably lower performance over all tasks than those following random splitting.
This indicates that the ICBHI dataset presents a very high dependence on patient characteristics, which is likely make respiratory cycle classification challenging in practice. 

However, all the results obtained by the proposed framework for Tasks 2-1 and 2-2 (with both splitting methods) exceed 84\%. These results for recording-based classification of lung disease -- which is highly related to the overall aim of lung disease detection -- provide a strong indicator of the robustness of the proposed framework. As does the fact that the same proposed  framework is capable of performing well for all subtasks.

\section{Student-Teacher Scheme for Respiratory Disease Detection}

\subsection{The Proposed Student-Teacher Arrangement}
\begin{figure}[th]
    \centering
    \includegraphics[width =\linewidth]{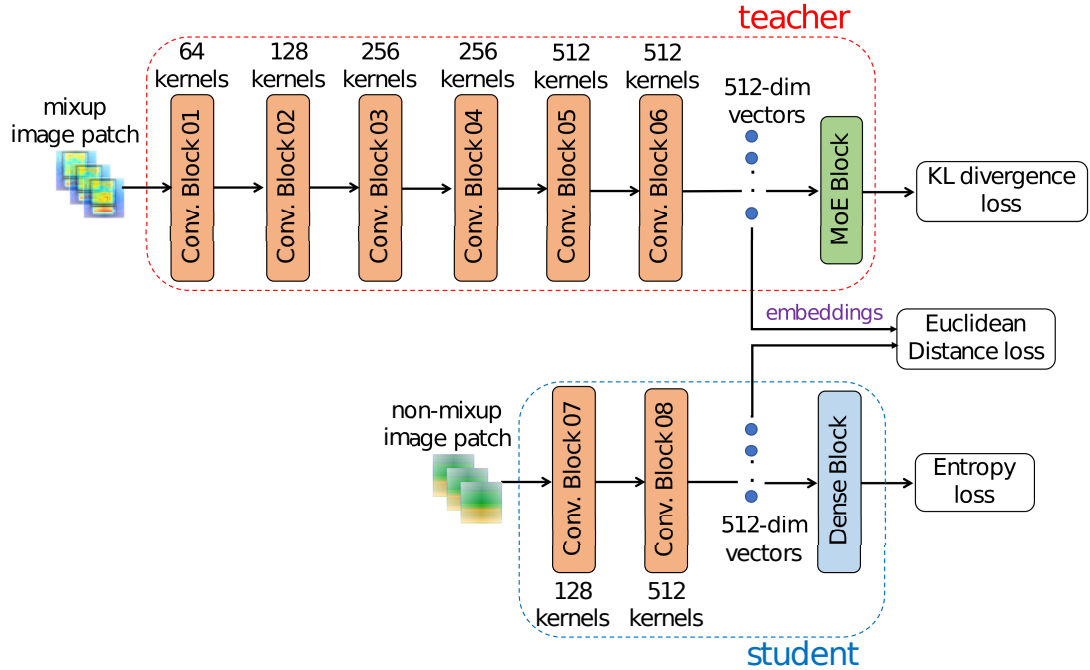}
    \vspace{-1cm}
	\caption{Architecture of the Student-Teacher scheme.}
    \label{fig:c06_tea_stu}
\end{figure}
Recent works on sound scene and sound event detection reported the effectiveness of Teacher-Student learning schemes~\cite{teacher_student_01, teacher_student_02}.
Among other advantages, these schemes offer a trade-off between model size and performance. 
Since the complexity of our best model based on the proposed MoE framework may be a barrier to future real-time implementation, it is explored whether a student-teacher scheme can be used to train a network with much lower complexity and perform well on the task of respiratory disease detection (Task 2).

\begin{table}[t]
    \caption{The Student network architecture.} 
        	\vspace{-0.2cm}
    \centering
    \scalebox{0.85}{
    \begin{tabular}{|l |l |c |} 
        \hline 
            \textbf{Architecture} & \textbf{Layers}   &  \textbf{Output}  \\
        \hline 
             & Input layer (image patch)  &          $64{\times}128$   \\
         Conv. Block 07   & Cv [$3{\times}3$] $@$ 128 - ReLU - AP [$4{\times}4$]  & $16{\times}32{\times}128$\\
         Conv. Block 08  & Cv [$3{\times}3$] $@$ 512 - ReLU - GAP       & $512$\\
         Dense Block & FC - Softmax & 3 \\         
       \hline 
    \end{tabular}
    }
    \label{table:c06_student} 
\end{table}

The proposed solution, as shown in Figure \ref{fig:c06_tea_stu}, comprises two networks, namely the Teacher and the Student.
The teacher network re-uses the high-performance CNN-MoE architecture introduced in Section \ref{c06_framework_architecture}. 
The student network features a compact architecture, comprising two convolutional blocks (identified \textit{Conv. Block 07} and \textit{Conv. Block 08} in the figure), and a dense block whose configuration is the same as the one in Table \ref{table:c06_student} (note that the student network does not apply batch normalisation, dropout or mixup data augmentation).

Training the Teacher-Student network is separated into two phases.
First, the Teacher is trained as usual. 
Afterwards, the Teacher's embedding is distilled to the Student's embedding to assist in the Student's learning process. 
The influence of this knowledge distillation on the student network's performance is empirically investigated. 
With the presence of this knowledge distillation, training the student network, therefore, aims to minimize two losses: (1) the Euclidean distance $LOSS_{EU}$ between the teacher and student embedding, and (2) the standard cross-entropy loss $LOSS_{EN} $ on the student's classification output. 
The combined loss function is therefore,
\begin{equation}
    \label{eq:final_loss}
        LOSS = (1-\gamma)LOSS_{EN} + \gamma LOSS_{EU}
\end{equation}
Here, the hyperparameter \(\gamma\) is empirically set to 0.5 to balance the two constituent losses. Other hyper-parameters and settings are inherited from Section \ref{c06_framework_architecture}. 

\subsection{Results From the Teacher-Student Scheme}

The experimental results obtained by the student network in comparison with the teacher network are shown in Table \ref{table:c06_student_res}.
On the one hand, it can be seen that without knowledge distillation from the teacher network, the small-footprint student network obtains a substantially low specificity score, although it maintains a very good sensitivity. This observation is consistent with the overall ICBHI score and can be explained by the simplicity of the network which results in low learning capacity.
On the other hand, distilling knowledge from the teacher significantly boosts the student performance,  yielding specificity, sensitivity, and ICBHI scores that are very competitive to those of the teacher network -- even though the student network is much smaller and simpler. 
\begin{table}[t]
    \caption{Performance comparison between Teacher \\ and Student with and without knowledge distillation.}
        	\vspace{-0.2cm}
    \centering
     \resizebox{0.98\textwidth}{!}{%

    \begin{tabular}{|c  |l  |c c c |c c c |} 
        \hline 
	        &                 &   \multicolumn{3}{c|}{\textbf{Five-fold random split}}  &  \multicolumn{3}{c|}{\textbf{ICBHI split}}  \\

	    \textbf{Tasks}        & \textbf{Models}                                  &\textbf{Spec.}  &\textbf{Sen.}  &\textbf{ICBHI Score}     &\textbf{Spec.}  &\textbf{Sen.}  &\textbf{ICBHI Score} \\
        \hline
				 &Teacher                            &0.86	&0.95	&0.91	                           &0.71	&0.98	&0.84                    \\
	    2-1, 3-category      &Student w/o knowledge distill            &0.43	&0.94	&0.68	                            &0.41	&0.97	&0.69                    \\
	                         &\textbf{Student w/ knowledge distill}    &0.86	&0.90	&0.88	                           &0.71	&0.98	&0.84                    \\
       \hline 
       \hline 
				 &Teacher                            &0.86	&0.98	&0.92                                &0.71	&0.98	&0.84  \\
	    2-2, 2-category      &Student w/o knowledge distill            &0.43	&0.99	&0.71                              &0.41	&0.99	&0.70  \\
	                         &\textbf{Student w/ knowledge distill}    &0.86	&0.96	&0.91                             &0.71	&0.98	&0.84  \\
       \hline 
    \end{tabular}
    }
    \label{table:c06_student_res} 
\end{table}
\begin{table}[t]
    \caption{Model footprint comparison between Teacher and Student} 
        	\vspace{-0.2cm}
    \centering
    \scalebox{0.85}{

    \begin{tabular}{| l | c | c |} 
        \hline 
            \textbf{Features} & \textbf{Teacher}   &  \textbf{Student}  \\
        \hline 
         Trainable Convolutional Layers & 6 & 2\\
         Trainable Fully-connected Layers & 11 & 1\\
         Batch normalization & 12 & 0 \\
         Number of trainable parameters & $4.5\times 10^6$ & $0.6\times 10^6$ \\      
         Number of MAC operations  & 44,886 K   &9,513 K \\   
       \hline 
    \end{tabular}
    }
    \label{table:c06_mem} 
\end{table}

Details of the model footprint are shown in Table~\ref{table:c06_mem}, it can be seen that the Teacher uses six convolutional layers, eleven fully-connected layers and twelve batch normalization layers that together contribute to a large model size with $4.5\times106$ trainable parameters. Meanwhile, the Student only uses two convolutional layers and one fully-connected layer, requiring only $0.6\times106$ parameters, approximately one-seventh of the Teacher’s. The model footprints also scale in terms of computational cost of multiply-accumulate (MAC) operations during inference. While an inference process on the Teacher costs 44,886 kMAC operations, the Student only costs 9,513 kMAC (the MAC operation computation for a deep learning network is presented in~\cite{mac_comp}). The inference process for a 20-second long recording in Task 1-1, conducted by a Tesla P100 GPU, takes 0.5 second; nearly ten times longer than the 0.045 second required for the Student’s inference process.

\section{Conclusion}
This chapter has presented a robust deep learning framework for the analysis of respiratory anomalies and detection of lung diseases from lung auscultation recordings.
Extensive experiments were conducted with different architectures and system settings using the ICBHI dataset, and two defined tasks related to that.
The proposed system is evaluated against existing state-of-the-art methods, outperforming them for most of the challenge tasks.
Furthermore, to facilitate implementation in real-time systems, a Teacher-Student learning scheme was employed to significantly reduce model complexity while still achieving very high accuracy.
The final experimental results validate the application of deep learning for the timely diagnosis of respiratory diseases, bringing this research area one step closer to clinical applications.

\chapter{Conclusion and Future Work}
\label{c07}

\section{Summary}
\label{c07_conclusion}

Concretely, this thesis has focused on dealing with the task of acoustic scene classification (ASC), and then applied the techniques developed for ASC to a real-life application of detecting respiratory disease. 

To deal with ASC challenges, this thesis addresses three main factors that directly affect the performance of an ASC system.
Firstly, this thesis explores input features by making use of multiple spectrograms (log-mel, Gamma, and CQT) for low-level feature extraction to tackle the issue of insufficiently discriminative or descriptive input features.
Next, a novel \textit{Encoder} network architecture is introduced. 
The \textit{Encoder} firstly transforms each low-level spectrogram into high-level intermediate features, or embeddings, and thus combines these high-level features to form a very distinct composite feature.
The composite or combined feature is then explored in terms of classification performance, with different \textit{Decoders} such as Random Forest (RF), Multilayer Perception (MLP), and Mixture of Experts (MoE).
By using this \textit{Encoder-Decoder} framework, it helps to reduce the computation cost of the reference process in ASC systems which make use of multiple spectrogram inputs.
Inspired by high-cross correlation among sound categories, and the potentially useful information that this might yield, the architecture is further explored, and a hierarchical classification, referred as to \textit{Decoder}, is proposed to make use of that. 
The scheme helps to structure the original ``flat'' ASC task into multiple hierarchical sub-tasks that operates in a divide-and-conquer manner. 
Since each sub-task is only suitable for some sound categories, a combination of triplet loss and cross entropy loss proves effective to enhance the classification accuracy. 
To evaluate the \textit{Encoder-Decoder} framework, recently published datasets (Litis Rouen and DCASE 2016 Task 1A, DCASE 2017 Task 1A, DCASE 2018 Task 1A \& 1B, DCASE 2019 Task 1A \& 1B) are used, and demonstrate very competitive results compared to the state-of-the-art systems.
Additionally, the obtained results also indicate that the framework is effective for early detection of sound scenes, which is potentially very useful for real-time applications on edge devices.
The results strongly demonstrate that the proposed \textit{Encoder-Decoder} framework is robust for ASC tasks.

Since the proposed techniques applied for general ASC tasks were shown to be highly effective, this inspired an application to a specific real-life application. 
This was namely the 2017 Internal Conference on Biomedical Health Informatics (ICBHI) respiratory sound dataset. 
Building upon the proposed ASC framework, the ICBHI tasks were tackled with a deep learning framework, and the resulting system shown to be capable at detecting respiratory anomaly cycles and diseases.  
The experimental results obtained validated the deep learning techniques used in the general ASC task for the timely diagnosis of respiratory diseases, thus potentially for a wide ranges of clinical applications.

\section{Future Work}
\label{c07_future_work}

Applying deep learning techniques developed for ASC tasks for specific applications, like the respiratory disease detection task mentioned in Chapter \ref{c06}, has promise for a number of audio detection problems.
However, the challenge to implement complicated deep learning frameworks on edge devices (e.g. mobile and low powered hardware) is the high degree of complexity of the computation required.
To deal with such challenges, model compression techniques have drawn increasing attention in recent years.
Two main approaches of compression are quantization and pruning. 
Recently, the Tensorflow framework 2.0 provides a complete guide for both the compression methods mentioned in ~\cite{google_opt}.
The toolbox is very usable and is quick to compress an originally complicated model to a more simple Tensorflow lite model that can be suitable for embedding on a wide range of embedded operating systems.
Therefore, one item of future work proposed in this thesis is to implement the deep learning framework applied for respiratory disease detection in Chapter \ref{c06} on an embedded or edge hardware platform. 
Successfully achieving a real-time system for detecting respiratory disease will help patients to self-observe their situation, reduce the cost of fabrication and possibly increase the scale of respiratory disease detection.

As mentioned in Chapter \ref{c02}, acoustic scene classification (ASC) and acoustic event detection (AED) are two main tasks of the emerging `machine hearing' research field~\cite{bk_lyon_17_human}. 
Currently, these two tasks are considered to be separate.
In particular, while ASC datasets are easily recorded in nature, it is hard to collect AED datasets in real-life environments.
Therefore, current AED datasets were synthesised which enable AED and ASC tasks in challenges to be independent yet share the same underlying database.
Furthermore, sound events follow certain structures, but this is not always true for sound scenes. 
Techniques applied to detect sound events and sound scenes are therefore basically different.
While CNN-based architectures are explored for sound scenes, analysis of sound events tend to use RNN-based networks which are robust for time sequence information.
However, sound context awareness abilities integrated in certain devices in the future should be base on both ASC and AED techniques to achieve high performance.
For instance, quite environments such as \textit{in park} or \textit{in home} are very challenging to detect if only based on static sound scene information. 
However, such quite environments are easier to recognize if specific sound events can be detected, such as \textit{bird song} in a park or sound of \textit{tap water} inside a home.
Therefore, if both sound events and scenes can be integrated into sound-based systems, it would be effective at improving sound context awareness.
From the analysis above, another idea for future work is exploring the high-cross correlation between ASC and AED to improve the sound context awareness of future sound-based systems.

\appendix 
\chapter{Mathematical definitions} 
\label{c08}

\renewcommand{\chaptername}{Appendix}
\makeatletter
\renewcommand{\chapter}[1]{\markboth{\sffamily\normalsize\bfseries\@chapapp\ 
\thechapter.\ #1}{}} 
\makeatother

This appendix provides the detailed derivation of the spectrogram transformations, namely the log-mel, CQT and Gamma used in experiments in Chapter \ref{c03}, \ref{c04}, \ref{c05}, and MFCC mentioned in Chapter \ref{c06}.
Additionally, this appendix details the computation of network layers used in this thesis such as convolutional (Cv), batch normalization (BN), rectify linear unit (ReLU), dropout (Dr), fully connected (FC), and Softmax layers.
Note that ``$\times$'' is used for matrix product.

\section{Spectrogram Computation}
\label{c08_secp}

\subsection{STFT spectrogram}
\label{c08_STFT}

The Short-Time Fourier Transform (SFFT) applies a Fourier Transform to a frame of the time series signal to extract the frequency content of the local section of analysed input.
If $\mathbf{s}$($n$) is considered as the digital audio signal, then the STFT spectrogram, denoted by $ \mathbf{STFT}[F,T]$, is computed as,
\begin{equation}
      \label{eq:stft}
      \mathbf{STFT} = \sum_{n=1}^{N} \mathbf{s}[n]\mathbf{w}[n]e^{-j2{\pi}Fn/N} 
\end{equation}
where $\mathbf{w}$[$n$] is a window function, typically Hamming.
While time resolution ($T$) of the STFT spectrogram is set by the hop size, the frequency ($F$) resolution depends on window length and the sample rate of the audio signal. 

\subsection{log-mel spectrogram}
\label{c08_log_mel}

To generate a log-mel spectrogram, the section of time series audio being analysed is first transformed into an STFT spectrogram as noted above.
Next, a Mel filter bank, which simulates the overall frequency selectivity of the human auditory system is applied. The filter bank uses the frequency warping $F_{mel} = 2595.log_{10}(1 + {F}/{700})$~\cite{bk_ian_speech} to generate a Mel spectrogram $\mathbf{MEL}[F_{mel}, T]$ (note that frequency resolution $F_{mel}$ depends on the number of Mel filters).
Logarithmic scaling is then applied to obtain the log-mel spectrogram. 

If $\mathbf{COE_{MEL}}[F_{mel}, F]$ is considered as a matrix storing coefficients of the Mel filters, then log-mel spectrogram $\mathbf{LOG\_MEL}[F_{mel}, T]$ is a matrix computed by a multiplication of the two matrices as follows, 
\begin{equation}
      \label{eq:log-mel}
      \mathbf{LOG\_MEL}[F_{mel},T] = log_{10}\left(\mathbf{COE_{MEL}}[F_{mel}, F]{\times}\mathbf{STFT}[F, T]\right) 
\end{equation}
\subsection{MFCC spectrogram}
\label{c08_mfcc}

From log-Mel spectrogram, Discrete Cosine Transform (DCT) is used to extract a sequence of uncorrelated coefficients crossing frequency dimension, reducing log-Mel frequency resolution into smaller space.  A pixel value $dct[f_{dct}, t_{dct}]$ of DCT matrix $\mathbf{DCT}[F_{dct}, T_{dct}]$, where $F_{dct}$ and $T_{dct}$ are frequency and time resolutions, is computed by:
\setlength{\arraycolsep}{0.0em}
\begin{eqnarray}
\label{eq:dct}
dct[f_{dct}, t_{dct}]&{}={}&\left(   \frac{2}{F_{mel}} \right)^{\frac{1}{2}} \left( \frac{2}{T} \right)^{\frac{1}{2}}  \sum_{f_{mel}=0}^{F_{mel}-1}\sum_{t=0}^{T-1}\nonumber\\
&&\Lambda(f_{mel})cos\left[ \frac{{\pi}f_{dct}}{F_{mel}}(2f_{mel}+ 1) \right]\nonumber \\
&& \Lambda(t)cos\left[ \frac{{\pi}t_{dct}}{T}(2t+1)   \right]\nonumber \\
&& log{-}mel[f_{mel}, t]
\end{eqnarray}
\setlength{\arraycolsep}{5pt}
%
where
\begin{equation}
\label{eq:lamda}
\Lambda(x) = 
\begin{cases}
\frac{1}{\sqrt{2}} & \text{if } x = 0  \\
1 & \text{otherwise}
\end{cases}
\end{equation}
$T$, $F_{mel}$, and $log{-}mel[f_{mel}, t]$ are time resolution, frequency resolution, and magnitude of a pixel of log-mel spectrogram, respectively. 

\subsection{Gamma spectrogram}
\label{c08_GAM}

Gammatone filters are designed to model the frequency-selective cochlea activation response of the human inner ear~\cite{aud_model_pat}, in which filter output simulates the frequency response of the basilar membrane.
The impulse response is given by
\begin{equation}
    \label{eq:gammaton}
    g[k] = k^{P-1 } T^{P-1}  e^{-2b\pi kT} cos(2\pi f kT + \theta)  
\end{equation}
where \(P\) is the filter order, \(\theta\) is the phase of  the carrier, \(b\) is filter bandwidth, and \(f\) is central frequency, and \(T\) is sampling period.
The filter bank is then formulated on the equivalent rectangular bandwidth (ERB) scale~\cite{aud_model_gla} as
\begin{equation}
    \label{eq:gammaton}
    ERB = 24.7(4.37.10^{-3}f + 1)
\end{equation}
To quickly and conveniently generate the gammatonegram, Ellis \emph{et al.}~\cite{auditory2009_tool} introduced a toolbox which first transforms the audio signal into STFT spectra as mentioned above. Then, a matrix of gammatone weighting $\mathbf{COE_{GAMMA}}[F_{gamma}, F]$ is applied to the STFT to obtain the Gamma spectrogram.
\begin{equation}
      \label{eq:log-mel}
      \mathbf{GAMMA}[F_{gamma},T] = \mathbf{COE_{GAMMA}}[F_{gamma}, F]{\times}\mathbf{STFT}[F, T]
\end{equation}
where $F_{gamma}$ is frequency resolution the depends on the number of gammatone filters used.

\subsection{Constant Q Transform (CQT)}
\label{c08_CQT}

The CQT applies a bank of filters corresponding to tonal spacing, where each filter is equivalent to a subdivision of an octave, with central frequencies given by, 
\begin{equation}
    \label{eq:cqt-0}
     F_{k} = (2^{\frac{1}{b}})^{k}f_{min}
\end{equation} 
where $F_{k}$ denotes the frequency of the $k$th spectral component, $f_{min}$ is the minimum frequency, and $b$ is the number of filters per octave.
As the name suggests, the Q value (which is commonly known to be the ratio of central frequency to bandwidth in electrical and control systems), is set to a constant as in,
\begin{equation}
    \label{eq:cqt-1}
     Q = \frac{F_{k}}{\Delta F_{k}} = \frac{F_{k}}{F_{k+1}-F_{k}} = \left (2^{\frac{1}{b}} -1 \right )^{-1}
\end{equation}
Like the STFT, the CQT spectrogram $ \mathbf{CQT}[F_{k}, T] $ is extracted using Fourier-based transformation, 
\begin{equation}
    \label{eq:cqt-2}
     \mathbf{CQT} = \frac{1}{N(k)}\sum_{n=0}^{N(k)-1}\mathbf{s}[n]\mathbf{w}[k, n]e^{-i2\pi \frac{nQ}{N(k)}}
\end{equation}
where
\begin{equation}
    \label{eq:cqt-3}
     N(k) = Q\frac{f{s}}{F_{k}}
\end{equation}
\begin{equation}
    \label{eq:cqt-4}
     \mathbf{w}[k,n] = \alpha + (1-\alpha)cos\frac{2{\pi}n}{N(k)-1}
\end{equation}
and $f{s}$ is the sample rate of the input signal.

\section{Computation of Network Layers}
\label{c08_layer}

if $\mathbf{X[F,T,C]} \in \mathbb{R^{F{\times}T{\times}C}}$ and $\mathbf{x} \in \mathbb{R^{N}}$ are considered as an input tensor and input vector, where $F, T$, $C$ are frequency, time, channel dimensions of the input tensor and $N$ is the dimension of input vector, the computation of network layers used in thesis is described in detail as below,

\subsection{Rectify Linear Unit (ReLU) Layer}
\label{c08_relu}

Relu function takes each pixel $x_{f,t,c}$ of input tensor $\mathbf{X}$ and thus returns the output $ f_{Relu}(x_{f,t,c})$ as,

\begin{equation}
       f_{ReLU}(x_{f,t,c})=\begin{cases}
     x_{f,t,c}, & \text{if } x_{f,t,c}>0 \\
     0, & \text{if } x_{f,t,c} <=0
   \end{cases}
\end{equation}

As regards input vector $\mathbf{x}[x_1,..., x_N]$ , this function takes $x_{n}$ and returns

\begin{equation}
       f_{ReLU}(x_{n})=\begin{cases}
     x_{n}, & \text{if } x_{n}>0 \\
     0, & \text{if } x_{n} <=0
   \end{cases}
\end{equation}

\subsection{Dropout Layer}
\label{c08_relu}

Dropout function $f_{Drop}$ for input tensor $\mathbf{X}$ is computed by,
\begin{equation}
              f_{Drop}(\mathbf{X})= \mathbf{XD}
\end{equation}
and for input vector $\mathbf{x}$ as
\begin{equation}
              f_{Drop}(\mathbf{x})= \mathbf{xd}
\end{equation}
where $\mathbf{D} \in \mathbb{R^{F{\times}T{\times}C}}$  is a matrix with similar dimension of input tensor $\mathbf{X}$ and $\mathbf{d} \in \mathbb{R^{N}}$  is a vector with similar dimension of input vector $\mathbf{x}$.
$\mathbf{D}$ and $\mathbf{d}$ are generated by $\mathbf{D/d} \sim$ Bernoulli($1-p$) where $p$ is the percentage of input dropped.

\subsection{Batch Normalization Layer}
\label{c08_batch}
If a batch of $B$ tensor is described by $\mathbf{X=\{X_{1}, X_{2}, ..., X_{B}\}}$, where $\mathbf{X_{b}}$ is a tensor input, the batch normalization function takes a tensor $\mathbf{X_b}$, and thus returns an output $f_{Batch}(\mathbf{X_b})$ as

\begin{equation}
f_{Batch}(\mathbf{X_b}) = \lambda\frac{\mathbf{X_b-\mu}}{\sqrt{\sigma + \epsilon}}
\end{equation}
where $\mu$ and $\sigma$ are defined by
\begin{equation}
\mu = \frac{1}{B} \sum_{b=1}^{B}\mathbf{X_{b}}\\
\end{equation}

\begin{equation}
\sigma = \frac{1}{B} \sum_{b=1}^{B}\mathbf{(X_{b} - \mu)^2}\\
\end{equation}
and $\lambda$ and $\epsilon$ are scale and ship parameters that are learned during training process.
In this thesis, the batch normalization function is only applied over input tensors and across the channel dimension. 

\subsection{Fully Connected Layer}
\label{c08_fully}
In this thesis, the fully-connected layer is applied for input vector. 
The output vector $\mathbf{y}[y_{1}, y_{2}, ..., y_{M}]$ with output dimension of $M$ is computed by 
\begin{equation}
              y_{m}= \sum_{n=0}^{N}w_{m,n}x_{n} + b_m
\end{equation}
where pixel $w_{m,n}$  in the coefficient matrix $\mathbf{W[M,N]}$ and $b_m$ in bias vector $\mathbf{b}[b_1,..., b_M]$ are trainable parameters. 

\subsection{Softmax Layer}
\label{c08_softmax}

In this thesis, the Softmax layer is applied on input vectors. The output vector $\mathbf{y}=[y_{1}, y_{2}, ..., y_{N}]$ of this layer is computed by 

\begin{equation}
              y_{n}= \frac{\exp(x_n)}{\sum_{n=1}^{N}\exp(x_n) }
\end{equation}
Note that the output and input vectors of Softmax layer have same dimension of $N$.

\subsection{Convolutional Layer}
\label{c08_conv}
In this thesis, convolutional layer is applied on input tensors.
if $C'$ kernels with size of $[K,P]$ are applied on the  input tensor $\mathbf{X[F,T,C]}$ in a convolutional layer, the output tensor has size of $\mathbf{Y[F,T,C']}$ (note that the frequency $F$ and time $T$ dimensions are remained by adding zero padding), a pixel $y_{f,t,c'}$ of output tensor $\mathbf{Y}$ is computed by,
 
\begin{equation}
              y_{f,t,c'}= \sum_{k=1}^{K}\sum_{p=1}^{P}\sum_{c=1}^{C}w_{k,p,c,c'}x_{k,p,c} + b_{c'}
\end{equation}
where pixel $x_{k,p,c}$ is in tensor input $\mathbf{X[F,T,C]}$; pixel $w_{k,p,c,c'}$  in coefficient matrix $\mathbf{W[K,P,C,C']}$ and $b_{c'}$ in bias vector $\mathbf{b}=[b_1,..., b_{C'}]$ are trainable parameters.

\bibliographystyle{ieeetr}
\bibliography{thesis}

\end{document}